\documentclass[12pt]{article}
\usepackage{dcolumn}
\usepackage{bm}
\usepackage{color}
\usepackage{verbatim}
\usepackage{amsmath}
\usepackage{amsthm}
\usepackage{amssymb}
\usepackage{amsfonts}
\usepackage{latexsym}
\usepackage{epsfig}
\usepackage{url}

\newcommand\be{\begin{equation}}
\newcommand\ee{\end{equation}}
\newcommand\bea{\begin{eqnarray}}
\newcommand\eea{\end{eqnarray}}

\newcommand\Bb{{\bf B}}

\newcommand\Eb{{\bf E}}

\newcommand\N{{\mathbb N}}

\newcommand\Z{{\mathbb Z}}

\newcommand\hb{{\bf h}}

\newcommand\rb{{\bf r}}

\newcommand\pb{{\bf p}}

\newcommand\vb{{\bf v}}
\newcommand\eb{{\bf e}}

\newcommand\eps{{\varepsilon}}
\newcommand\R{{\mathbb R}}

\newcommand{\D}{{\mathcal D}}
\newtheorem{theorem}{Theorem}
\newtheorem{definition}{Definition}

\newtheorem{proposition}{Proposition}

\begin{document}
\title{Planar undulator motion excited by a fixed traveling wave: \\Quasiperiodic Averaging, normal forms and the FEL Pendulum}
\author{James A. Ellison$^a$, Klaus Heinemann$^a$, Mathias Vogt$^b$, \\
Matthew Gooden$^c$
\\
$^a$ \small{Department of Mathematics and Statistics, The University of New
Mexico,} \\
\small{Albuquerque, New Mexico 87131, U.S.A. }\\
$^b$ \small{Deutsches~Elektronen--Synchrotron, DESY, ~22607 ~Hamburg,~Germany}
\\
$^c$ \small{Department of Physics, North Carolina State University,} \\
\small{Raleigh, North Carolina 27695, U.S.A.}}
\date{\today}
\allowdisplaybreaks
\maketitle
\newpage

\begin{abstract}
We present a mathematical analysis of planar motion of energetic electrons moving through a planar dipole undulator, excited by a fixed planar polarized plane wave Maxwell field in the X-Ray FEL regime. 
Our starting point is the 6D Lorentz system, which allows planar motions, and we examine this dynamical system as the wave length $\lambda$ of the traveling wave  varies.
By scalings and transformations the 6D system is reduced, without approximation, to a 2D system in a form for a rigorous asymptotic analysis using the Method of Averaging (MoA), a long time perturbation theory. 
The two dependent variables are a scaled energy deviation and a generalization of the so-called ponderomotive phase.
As $\lambda$ varies the system passes through resonant and nonresonant (NR) zones and we develop NR and near-to-resonant (NtoR) MoA normal form approximations.
The NtoR normal forms contain a parameter which measures the distance from a resonance.
For a special initial condition, for the planar motion and on resonance, the NtoR normal form reduces to the well known FEL pendulum system.
We then state and prove NR and NtoR first-order averaging theorems which give explicit error bounds for the normal form approximations. 
We prove the theorems in great detail, giving the interested reader a tutorial on mathematically rigorous perturbation theory in a context where the proofs are easily understood. 
The proofs are novel in that they do not use a near identity transformation and they use a system of differential inequalities. 
The NR case is an example of quasiperiodic averaging where the small divisor problem enters in the simplest possible way.
To our knowledge the planar problem has not been analyzed with the generality we aspire to here nor has the standard FEL pendulum system been derived with associated error bounds as we do here.
We briefly discuss the low gain theory in light of our NtoR normal form.
Our mathematical treatment of the {\it noncollective} FEL beam dynamics problem in the framework of {\it dynamical systems theory} sets the stage for our mathematical investigation of the {\it collective} high gain regime.
\end{abstract}

\newpage
\tableofcontents
\newpage

\setcounter{section}{0}
\setcounter{equation}{0}
\section{Introduction}
\label{1}
We present a normal form analysis of the three-degree-of-freedom Lorentz
force system of six ODE's (ordinary differential equations) governing the planar $(x,y=0,z)$ motion of relativistic electrons moving through a planar dipole undulator along the $z$-axis perturbed by a traveling wave radiation field along the $z$ direction. We are interested in the parameter range for an X-Ray FEL.

Our normal form analysis is based on the Method of Averaging (MoA) at first order. The method has four steps. 
The first step is to put the ODE's into a standard form. 
The second step is to identify the normal form approximations. 
The third step is the derivation of error bounds relating the
exact and normal form solutions. The final step is the transformation
back to the original variables of the Lorentz force system.
In the first step new variables are typically introduced using scalings and transformations. 
In this process we discover that the exact problem can be formulated, without approximation, in terms of two ODE's for the normalized energy deviation and a generalized ponderomotive phase. 
Important in this process is the identification of an appropriate small dimensionless parameter, often denoted by $\eps$, so that the system can be written as $\dot u=\eps f(u,t)+O(\eps^2)$.
In the present context this is the most complicated step.
The normal form approximation is obtained by dropping the $O(\eps^2)$ term and replacing $f$ by its $t-$average.
The third step is often the most difficult, however here the system in standard form is fairly simple and we use this opportunity to give very detailed proofs of two averaging theorems, partly as a tutorial on the methods of proof, rather than applying general theorems from the literature. The latter allows us to obtain quite explicit error bounds which are likely near optimal.

An electron, as a member of an electron bunch, will enter the undulator with a given angle in the $y=0$ plane and a given Lorentz factor. Here the normalized angle will be given by $\Delta P_{x0}$ and the Lorentz factor will be written $\gamma=\gamma_c(1+\eta)$ where $\gamma_c$ is a characteristic value of $\gamma$ for the electron bunch, e.g. the mean, and $\eta$ is the so-called normalized energy deviation. 
We will replace $\eta$ by $\chi$ via the relation $\eta=\eps\chi$, where 
a posteriori $\eps$ will be a measure of the spread of $\eta$ values 
which lead to an FEL pendulum type behavior.
We let $B_u,k_u$ denote the undulator field strength and wave number and let $E_r,\nu k_r$ denote the Maxwell field strength and wave number of the fixed traveling wave radiation field.
Thus our basic parameters are eight, namely  $\Delta P_{x0}, \gamma_c, \eps, B_u, k_u, E_r,k_r,\nu$. 
We will study the electron response to the radiation field as $\nu=O(1)$ varies.
The choice of the parameter $k_r$ will be discussed below.

For an X-Ray FEL, $\eps$ is small, $\gamma_c$ is large
and the undulator parameter, 
\begin{eqnarray}
&& K:=\frac{eB_u}{mck_u}=.934\lambda_u[cm]B_u[T]  \; , 
\label{eq:1.10}
\end{eqnarray}
is $O(1)$.
Also $k_r=O(k_u\gamma_c^2)$ and we define the O(1) constant $K_r$ by
\begin{eqnarray}
K_r:=\frac{k_r}{k_u\gamma_c^2} \; .
\label{eq:1.11}
\end{eqnarray}
In \S\ref{2.3} we will fix $K_r$ (and thus $k_r$) by setting
\begin{eqnarray}
&& \hspace{-8mm} 
K_r=2[1+ \frac{1}{2}K^2+K^2(\Delta P_{x0})^2]^{-1} \; .
\label{eq:1.12}
\end{eqnarray}
For those familiar with FEL theory,  $k_r$ is, for $\Delta P_{x0}=0$, the usual so-called resonant wave
number (See e.g., \cite{BRH}). 
The dependence of $K_r$ on $\Delta P_{x0}$ will be a consequence
of our analysis. 
 For the LCLS (Linac Coherent Light Source) $\lambda_u=3\text{cm}$, $mc^2\gamma_c=15$GeV and 
$B_u=1.32$T so that $K=3.70$ 
(see \url{http://www-ssrl.slac.stanford.edu/lcls/lcls_parms.html}).

Mathematically then, we are interested in an asymptotic analysis of the electron motion for $\eps$ small and $\gamma_c$ large as $\nu$ varies. In particular we are interested in the $(\eps,\gamma_c)$ regime that gives rise to the pendulum type behavior important for the functioning of an X-Ray FEL.
We find that in order to obtain this behavior, in the MoA at first-order, there must be a relation between $\eps$ and $\gamma_c$. Introducing the normalized field strength
\begin{eqnarray}
&& \hspace{-8mm} {\cal E}:=\frac{E_r}{cB_u} \; ,
\label{eq:1.14}
\end{eqnarray}
we show a pendulum type behavior emerges when $\eps=O(\sqrt{\cal E}/\gamma_c)$ for $\gamma_c\gg 1$. Without loss of generality we will take the order constant to be $1$, and choose
\begin{eqnarray}
&& \hspace{-8mm} 
\eps=\sqrt{\cal E}\frac{1}{\gamma_c} \; .
\label{eq:1.13}
\end{eqnarray}
We also show that, for $\eps$ small, the system associated with  (\ref{eq:1.13}) has a resonance structure, such that as $\nu$ varies the system goes through a sequence of nonresonant (NR) and near-to-resonant (NtoR) zones. The associated NtoR approximating normal forms are pendulum like and reduce to the standard FEL pendulum system for $\Delta P_{x0}=0$ and $\nu$ an odd integer.
This behavior is not present for $\eps\ll 1/\gamma_c$ or $\eps\gg 1/\gamma_c$ and so we refer to (\ref{eq:1.13}) as a {\it distinguished case}. This turns out to be a very simple example of the concept of a ``distinguished limit" in the singular perturbation literature. This can be seen in action in the context of our equations (\ref{eq:2.48n}) and (\ref{eq:2.48an}).

In summary, for the distinguished case of (\ref{eq:1.13}), our basic
nondimensional parameters are $K,\Delta P_{x0},{\cal E},\eps,\nu$.
For $\eps$ small we will obtain a sequence of nonresonant (NR) and
near-to-resonant (NtoR) normal form approximations as $\nu$ varies. The NtoR
normal forms can be understood in terms of the simple pendulum system and reduce to the usual FEL
pendulum equations for $\Delta P_{x0}=0$ and $\nu$ an odd integer (See Sections \ref{3.4.2} and \ref{3.4.3}). 
The NtoR normal form allows us to study the effect of $\nu$ being slightly off resonance.
This completes the first two steps in the MoA. In the third step we prove
two theorems which give error bounds, relating the exact and
normal form solutions, which go to zero as $\eps\rightarrow 0+$.
Our goal is to present a mathematically rigorous analysis that is self contained.

Standard derivations of the FEL pendulum equations can be found in \cite{KHL},\cite{SDR},\cite{MP},\cite{SSY}. 
They differ from our approach in that they start from the
ODE for the normalized energy deviation, $\eta$, and 
use physical reasoning to introduce approximations leading to the FEL pendulum normal form for $\Delta P_{x0}=0$.
In contrast, our starting point is the three-degree-of-freedom Lorentz force ODE's
which are clearly more general and we make no approximation in going to the standard 
form for the MoA. Thus our only approximation is in going from the
averaging standard form to the normal form approximations.
Furthermore we obtain error bounds which do not appear to be possible in the standard derivations and 
these bounds are covered by our averaging theorems.
Our definition of resonance is intimately linked to the derivation of our averaging normal forms, whereas in the standard derivations resonance is introduced in the context of maximizing energy exchange. 
We emphasize that we obtain more than the pendulum normal form; we also obtain the more general NtoR normal form as well as the NR normal forms.

We do not intend to minimize the importance of the standard derivations, the physical derivations are certainly important and as is often the case show great physical insight. Here we want to show what can be done in a mathematically rigorous way in the context of dynamical systems theory, but in that we have been guided by and are indebted to the work of  e.g., \cite{KHL},\cite{SDR},\cite{MP},\cite{SSY}.

For ODE's, the MoA is the most robust of the longtime perturbation theories which include e.g., Lindstedt series \cite{M1}, multiple scales \cite{M1}, renormalization group methods \cite{Oono} and Hamiltonian perturbation theory \cite{HPT}. For example, Hamiltonian perturbation theory has the advantage that one is transforming a scalar function, however the MoA is more robust in that transformations and scalings are not restricted to canonical transformations. Central to the MoA, and in contrast to those just mentioned,  is the derivation of error bounds. We emphasize these are true bounds and not just estimates. 
The MoA is a mature subject and there are several good books, see
\cite{M1,LM,SVM} for example as well as the 
Scholarpedia articles
\cite{M2,Sanders}.
We refer to the MoA approximation as a normal form. Generally, a normal form of a mathematical object is a simplified form of the object obtained with the aid of, for example, scalings and transformations such that the essential features of the object are preserved. Here we not only preserve the essential features of the exact ODE's
but bound the errors in the approximation with a bound proportional to the small parameter $\eps$. See \cite{M2} for the use of normal form in a similar context.

This paper has a pedagogical aspect, giving the reader, who may not be familiar with modern long time perturbation theory, an introduction in a context
where the proofs are easily understood.
In addition, we hope that both newcomers to the field and mathematical scientists will find this a good introduction to the noncollective case of an FEL. We also hope that experts will find something of interest.
The reader does not need to be familiar with averaging theory as we give complete proofs including detailed error bounds. 
Furthermore we obtain better results as our theorems are tuned to the problem at hand. In addition, 
to our knowledge, the treatment of the undulator problem in the mathematically  rigorous and self-contained way that we do here has 
not been done before. Our mathematical analysis is not deep, using only undergraduate mathematics as commonly taught in advanced calculus courses, however it is complicated and somewhat intricate in spots. Finally, for us, it sets the stage for our more serious goal of a deep mathematical understanding of the collective high gain FEL theory.

We proceed as follows. In \S\ref{2} we start with 
the three-degree-of-freedom Lorentz equations with a general traveling
wave field in  (\ref{eq:2.20})-(\ref{eq:2.23}) and then introduce $z$ as the independent variable. The system has planar solutions where $0=y=p_y$ and using a conservation law we arrive at a system of two ODE's  
(\ref{eq:2.32}),(\ref{eq:2.33}) for the energy deviation and a precursor to a generalization of the so-called ponderomotive phase. 
By scalings and transformations we discover the distinguished case of (\ref{eq:1.13}) which then leads to a standard form for the method of averaging in (\ref{eq:2.72}),(\ref{eq:2.73}).
The two dependent variables are now a scaled energy deviation and a generalization of the so-called ponderomotive phase. 

In \S\ref{3} we present our main results. We begin by introducing the monochromatic traveling wave field, the case 
of main physical interest.  The system is carefully defined in \S\ref{3.1}. 
In \S \ref{3.2} we define nonresonant, $\Delta$-nonresonant, resonant, and 
near-to-resonant $\nu$ 
in the MoA context. We emphasize that as $\nu$ varies the system passes through resonant and nonresonant zones. The NR case, its first-order averaging normal form and associated solutions are presented in \S\ref{3.3} along with a proposition 
giving an appropriate domain for the associated vector field. \S\ref{3.3}
sets the stage for the more interesting NtoR case of \S\ref{3.4}. The NtoR system is carefully defined along with a proposition giving an appropriate domain for the associated vector field. The first-order averaging normal form is derived and solutions written in terms of solutions of the simple pendulum system. It is unlikely that all $\nu$ values are covered accurately by our normal forms, however we are able to argue in \S \ref{3.4.4} that there is a sense in which the NR case emerges from the NtoR case. The third and fourth steps of the MoA are performed
in \S\ref{3.5} and \S\ref{3.6}. In fact,
the statements of our first-order averaging theorems, which 
give an order $\eps$ bound on the error for long times, i.e., intervals of $O(1/\eps)$, are presented in \S\ref{3.5} and applied to the phase space variables in \S\ref{3.6}. By taking special initial conditions ($\Delta P_{x0}=0$) we recover the result of standard approaches which focus on the energy transfer equations alone and do not consider the phase space variables. Finally in \S\ref{3.7} we use our results in a low gain calculation and compare the result with \cite{KHL}.

The proofs of the two averaging theorems are presented in \S\ref{4}
and they are based on an idea of Besjes (see \cite{ESD,DEH,Besjes})
which leads to proofs without 
using a near-identity transformation, as in usual treatments of, e.g.,
\cite{M1,LM,SVM}.
The NR case is an example of quasiperiodic averaging with a rigorous treatment of a small divisor problem in what is surely the simplest setting. The NtoR case is an example of periodic averaging. A novelty of our approach is that we use a {\it system} of differential inequalities, rather than the usual Gronwall inequality, to obtain better error bounds.

The appendices contain calculations needed in the main text. 
Appendix \ref{A} provides properties of the Bessel expansion of
the function $jj$ which is introduced in Section \ref{3.2}.
In Appendices \ref{B},\ref{C} we study the next-to-leading
order terms $g_1,g_2$ used in Theorem \ref{T1} and
in Appendices \ref{D},\ref{E} we study the next-to-leading
order terms $g_1^R,g_2^R$ used in Theorem \ref{T2}. Appendix \ref{F}
gives an outline of a rigorous approach to regular perturbation theory which 
could be made into a theorem at the level of our averaging theorems. It is applied in \S\ref{3.4.4}. Appendix \ref{G} provides some formulas used in 
Section \ref{3.7}. In Appendix \ref{H} we  discuss ${\cal E}=E_r/cB_u$ in the high gain regime and obtain a crude upper bound estimate of it. Finally, in Appendix \ref{I} we show that the solution of the system of differential inequalities that is used in the proof of both averaging theorems (as well as in Appendix \ref{F}) is indeed a solution.
\setcounter{equation}{0}
\section{General Planar Undulator model}
\label{2}
\subsection{Lorentz force equations}
\label{2.1}
Using SI units, the Lorentz equations for motion of a relativistic 
electron in an electromagnetic field, $(\Eb,\Bb)$, are 
\begin{eqnarray}
&& \dot{\rb}=\vb(\pb) \; , 
\label{eq:2.10} \\
&& \dot{\pb} =-e(\Eb+\vb(\pb) \times \Bb),  \label{eq:2.11}
\end{eqnarray}
with $\dot{}=d/dt$ and where 
\begin{eqnarray}
&& \vb(\pb)=\frac{\pb}{m\gamma} \; , 
\label{eq:2.12}
\end{eqnarray}
is the velocity, $\gamma$ is the Lorentz factor defined by 
\begin{eqnarray}
&& \gamma^2=1+\pb\cdot\pb/m^2c^2  \; , 
\label{eq:2.13}
\end{eqnarray}
and $m$ and $-e$ are the electron mass and charge respectively. 
We introduce Cartesian coordinates as follows: 
\begin{eqnarray}
&&\rb=x\eb_x+y\eb_y+z\eb_z \;, \label{eq:2.15}\\
&&\pb=p_x\eb_x+p_y\eb_y+p_z\eb_z \; , \label{eq:2.16}
\end{eqnarray}
where $\eb_x,\eb_y,\eb_z$ are the standard unit vectors.
Using (\ref{eq:2.10})-(\ref{eq:2.16}) 
the system in Cartesian coordinates is
\bea
&&\hspace{-8mm}\dot{x}=\frac{p_x}{m\gamma} \; , \quad
\dot{y}=\frac{p_y}{m\gamma} \; , \quad
\dot{z}=\frac{p_z}{m\gamma} \; , \label{eq:2.20}\\
&&\hspace{-8mm}\dot{p}_x = -e[E_x + v_yB_z-v_zB_y] \; , \label{eq:2.21}\\
&&\hspace{-8mm}\dot{p}_y = -e[E_y + v_zB_x-v_xB_z] \; , \label{eq:2.22}\\
&&\hspace{-8mm}\dot{p}_z = -e[E_z + v_xB_y-v_yB_x] \; . \label{eq:2.23}
\eea
We denote the undulator magnetic field by $\Bb_u$ and the radiation field by 
$(\Eb_r,\Bb_r)$ whence
\begin{eqnarray}
&& \Eb = \Eb_r \; , \quad  \Bb = \Bb_r + \Bb_u \; . 
\label{eq:2.30}
\end{eqnarray}
A simple planar undulator model magnetic field which satisfies  the Maxwell equations, $\nabla \cdot \Bb_u=0$ and $\nabla  \times \Bb_u=0$,
as in \cite{SDR}, is
\begin{eqnarray}
&&\Bb_u=-B_u[\cosh(k_uy)\sin(k_u z)\eb_y +\sinh(k_uy)\cos(k_u z)\eb_z] \; ,
 \label{eq:2.35}
\end{eqnarray}
where $B_u>0$. 
Since $\nabla  \times \Bb_u=0$ there is a scalar potential $\phi$ such that $\Bb_u=\nabla \phi$. To satisfy $\nabla \cdot \Bb_u=0$, $\phi$ 
must satisfy Laplace's equation. The field (\ref{eq:2.35}) is easily constructed by separation of variables and requiring periodicity in $z$ with period $\lambda_u$ and
then taking the first eigen-mode  
(See, e.g., \cite[p. 145]{Cl}). The scalar field is 
$\phi=-(B_u/k_u)\sinh(k_uy)\sin(k_u z)$.

The traveling wave radiation field we choose is also a Maxwell field and is given by
\begin{eqnarray}
&&\Eb_r=E_rh(\check{\alpha})\eb_x \; , \quad \Bb_r=\frac{1}{c}(\eb_z\times \Eb_r)
=\frac{E_r}{c}h(\check{\alpha})\eb_y \; , 
\label{eq:2.40}
\end{eqnarray}
where $E_r$ is a constant, $h$ is a real valued function on $\R$ and 
\begin{eqnarray}
&& \check{\alpha}(z,t) = k_r(z-ct) \; , 
\label{eq:2.45}
\end{eqnarray}
and $k_r$ is the parameter mentioned in the Introduction.

Our primary emphasis is on the standard monochromatic example where 
\begin{eqnarray}
H(\check{\alpha})=(1/\nu)\sin (\nu \check{\alpha})  \; , \quad 
h(\check{\alpha})=H'(\check{\alpha})=\cos (\nu \check{\alpha}) \; ,
\label{eq:2.170}
\end{eqnarray}
and $\nu\geq 1/2$
thus $h(\check{\alpha}(z,t))=\cos(\nu k_r(z-ct))$.
Note that the prime $'$ always indicates a derivative.
Thus from \S\ref{3} onwards we will use (\ref{eq:2.170}).
However it is easy to carry through the first part of the analysis
with general $H$ and we do want to make a comment on the more general
case. In this monochromatic case $k_r$ will be defined by
(\ref{eq:1.11}),(\ref{eq:1.12}) and the 
$\nu$ will allow for a variable wave number for the traveling wave; 
it will be shown that $\nu=1$ gives the primary resonance with the concomitant pendulum normal form.
The extension to a sum of monochromatic waves
is trivial and won't be discussed. 

Using (\ref{eq:2.12}),(\ref{eq:2.35}),(\ref{eq:2.40}) one can write 
(\ref{eq:2.21})-(\ref{eq:2.23}) as
\bea
&& \hspace{-8mm}\dot{p}_x = -e[\frac{p_z}{m\gamma}B_u
\cosh(k_uy)\sin(k_uz)- \frac{p_y}{m\gamma}B_u
\sinh(k_uy)\cos(k_uz) \nonumber\\
&&\quad +E_r(1-\frac{p_z}{m\gamma c})h(\check{\alpha}(z,t))] \; , \label{eq:2.50}\\
&& \hspace{-8mm}\dot{p}_y = -e\frac{p_x}{m\gamma}B_u
\sinh(k_uy)\cos(k_uz) \; , \label{eq:2.51}\\
&& \hspace{-8mm}\dot{p}_z = -e [-\frac{p_x}{m\gamma}B_u
\cosh(k_uy)\sin(k_uz) +E_r\frac{p_x}{m\gamma c}h(\check{\alpha}(z,t))] \; .
\label{eq:2.52}
\eea
It is easy to check that 
(\ref{eq:2.20}),(\ref{eq:2.50})-(\ref{eq:2.52}) is a Hamiltonian system
with Hamiltonian ${\cal H}$:
\begin{eqnarray}
&& {\cal H}=c\sqrt{({\bf P}_c+ e{\bf A}(\rb,t))^2+m^2c^2} =mc^2\gamma\; ,
\label{eq:2.60}
\end{eqnarray}
where the canonical momentum vector ${\bf P}_c$ is related to $\pb$ by
$\pb={\bf P}_c + e{\bf A}$ and the vector potential
${\bf A}$ is given by
\begin{eqnarray}
&& {\bf A}(y,z,t)=[\frac{B_u}{k_u}\cosh(k_uy)\cos(k_u z)+\frac{E_r}{k_r c} 
H(\check{\alpha}(z,t))]\eb_x \; .
\label{eq:2.65}
\end{eqnarray}
Since ${\bf A}$ is independent of $x$ the $x$-component,
$P_{c,x}$, of the canonical momentum vector ${\bf P}_c$ is conserved, i.e.,
\bea
p_x - e A_x(y,z,t) \; ,
\label{eq:2.70}
\eea
is constant along solutions 
of (\ref{eq:2.20}),(\ref{eq:2.50})-(\ref{eq:2.52}) as is easily
confirmed directly.
We will not make explicit use of the Hamiltonian structure in the following.
The MoA does not rely on a Hamiltonian structure and this frees us from
having to deal only with canonical transformations as we proceed 
to put (\ref{eq:2.20}),(\ref{eq:2.50})-(\ref{eq:2.52}) in an
averaging standard form.
\subsection{Motion in $y=0$ plane with $z$ as the independent 
variable}
\label{2.2}
It is common to take the distance $z$ along the undulator as the independent variable, rather than the time $t$. In fact after unsuccessfully trying to
stay with $t$ we decided to follow the common procedure.
With the usual abuse of notation, we write, from now on 
$x(z),y(z),p_x(z),p_y(z),p_z(z)$ instead of 
$x(t(z)),y(t(z)),p_x(t(z)),p_y(t(z)),p_z(t(z))$  whence the ODE's 
(\ref{eq:2.20}),(\ref{eq:2.50})-(\ref{eq:2.52}) become
\bea
&&\hspace{-8mm}\frac{dx}{dz}=\frac{p_x}{p_z} \; , \quad
\frac{dy}{dz}=\frac{p_y}{p_z} \; , \quad
\frac{dt}{dz}=\frac{m\gamma}{p_z} \; , \label{eq:2.25}\\
&& \hspace{-8mm}\frac{dp_x}{dz}
= -\frac{e}{c}[cB_u\cosh(k_uy)\sin(k_uz)- \frac{p_y}{p_z}cB_u
\sinh(k_uy)\cos(k_uz) \nonumber\\
&&\quad +E_r(\frac{m\gamma c}{p_z}-1)h(\check{\alpha}(z,t))] \; , \label{eq:2.14}\\
&& \hspace{-8mm}\frac{dp_y}{dz} = -\frac{e}{c}\frac{p_x}{p_z}cB_u
\sinh(k_uy)\cos(k_uz) \; , \label{eq:2.17}\\
&& \hspace{-8mm}\frac{dp_z}{dz}= -\frac{e}{c} [-\frac{p_x}{p_z}cB_u
\cosh(k_uy)\sin(k_uz) +E_r\frac{p_x}{p_z}h(\check{\alpha}(z,t))] \; .
\label{eq:2.18}
\eea
The initial conditions at $z=0$ will be denoted by a subscript $0$, e.g., $t(0)=t_0$. Clearly $t_0$ is the arrival time of an electron at the entrance, $z=0$, 
of the undulator. 

Here and in the rest of the paper we consider the initial value problem (IVP) 
with $y_0=p_{y0}=0$. It follows, with no approximation, that $y(z)=p_y(z)=0$ for all $z$ and the six ODE's 
(\ref{eq:2.25})-(\ref{eq:2.18}) reduce to four.
The righthand sides (rhs's) of (\ref{eq:2.25})-(\ref{eq:2.18})
are independent of $x$ and so we do not need to consider the $x$ equation 
until \S\ref{3.6}. It is standard, and also quite convenient, to replace $p_z$
by the energy variable $\gamma$.
With $\gamma(z)$ defined in terms of $p_x(z)$ and $p_z(z)$ by (\ref{eq:2.13}) and using (\ref{eq:2.14}) and (\ref{eq:2.18}), we obtain $\gamma'=(p_xp_x'+p_zp_z')/m^2c^2\gamma=-(eE_r/mc^2)(p_x/p_z)h(\check{\alpha}(z,t))$. 
Finally, we take $\check{\alpha}$ as a dependent variable in place of $t$ and we define 
\begin{equation}
\alpha(z):=\check{\alpha}(z,t(z))=k_r(z-ct(z)) \; .
\label{eq:2.19}
\end{equation}
Later it will be seen
that $\alpha$ is a precursor to a generalization of
the so-called ponderomotive phase
which emerges naturally as we put the ODE's in a standard form
for averaging.

With the above four changes the ODE's for $t,p_x,p_z$ in
(\ref{eq:2.25}),(\ref{eq:2.14}),(\ref{eq:2.18}) become 
\bea
&&\hspace{-8mm}\frac{d\alpha}{dz}=k_r(1-\frac{m\gamma c}{p_z}) \; , 
\label{eq:2.24}
\\
&& \hspace{-8mm}\frac{dp_x}{dz}= -\frac{e}{c}[cB_u\sin(k_uz) +E_r(\frac{m\gamma c}{p_z}-1)h(\alpha)] \; , 
\label{eq:2.26}
\\
&& \hspace{-8mm}\frac{d\gamma}{dz}= -\frac{eE_r}{mc^2} \frac{p_x}{p_z}h(\alpha) \; ,
\label{eq:2.27}
\eea
where the initial conditions are $\alpha(0)= 
\alpha_0:=-k_rct_0, p_x(0)=:p_{x0}, \gamma(0)=:\gamma_0$.
Here $p_z$ must be replaced by
\begin{eqnarray}
&& p_z=
\sqrt{m^2c^2(\gamma^2-1)-p_x^2} \; ,
\label{eq:2.28}
\end{eqnarray}
and it is easy to see that (\ref{eq:2.24})-(\ref{eq:2.27})
are then self contained. From now on we restrict $p_z$ to be positive:
\begin{eqnarray}
&& p_z > 0 \; .
\label{eq:2.29}
\end{eqnarray}
Note that, by (\ref{eq:2.24}), $\alpha$ is
a strictly decreasing function whence, as one expects, $z< c(t(z)-t_0)$.
It is also easy to check that
\bea
\frac{p_x}{mcK} - \cos(k_u z) - \frac{E_r}{cB_u}\frac{k_u}{k_r} 
H(\alpha) \; ,
\label{eq:2.31}
\eea
is conserved along solutions of (\ref{eq:2.24})-(\ref{eq:2.27}).
This conservation law
is identical to (\ref{eq:2.70}) with $y=0$. Recall that $K$
was defined by (\ref{eq:1.10}).

In summary, the solution of the IVP for (\ref{eq:2.25})-(\ref{eq:2.18}) 
with $y_0=p_{y0}=0$, which entails $y=p_y=0$, is given in terms of
the solution of (\ref{eq:2.24}),(\ref{eq:2.27}), i.e., of
\bea
&&\hspace{-8mm}\frac{d\alpha}{dz}=k_r(1-\frac{m\gamma c}{p_z}) \; , 
\quad \alpha(0) = \alpha_0 \; ,
\label{eq:2.32}
\\
&& \hspace{-8mm}\frac{d\gamma}{dz} =- \frac{eE_r}{mc^2} 
\frac{p_x}{p_z}h(\alpha) \; , \quad \gamma(0) = \gamma_0 \; ,
\label{eq:2.33}
\eea
with
\bea
p_x = p_{x0} + mcK \biggl(\cos(k_u z) - 1 + \frac{E_r}{cB_u}\frac{k_u}{k_r} 
[H(\alpha) - H(\alpha_0)]\biggr)
\; ,
\label{eq:2.34}
\eea
and $p_z$ in (\ref{eq:2.28}). To complete the solution of
(\ref{eq:2.25})-(\ref{eq:2.18}) it suffices to note that
$t(z)$ is determined from (\ref{eq:2.19}) in terms of $\alpha(z)$ and
$x(z)$ is determined from (\ref{eq:2.25}) by integration.
\subsection{Standard form for Method of Averaging}
\label{2.3}
We begin by introducing the
normalized energy deviation $\eta$ and its $O(1)$ counterpart $\chi$ via
\begin{eqnarray}
\gamma=\gamma_c(1+\eta)=\gamma_c(1+\eps\chi) \; ,
\label{eq:2.39}
\end{eqnarray}
as mentioned in the Introduction.
Here $\gamma_c$ is a characteristic value of $\gamma$, e.g., its mean
and $\eps$ is a characteristic spread of $\eta$ so that $\chi$ becomes
the new $O(1)$ dependent variable replacing
$\gamma$ in (\ref{eq:2.32}),(\ref{eq:2.33}).
We are interested in an asymptotic analysis for $\gamma_c$ large and
$\eta$ small as in an X-Ray FEL.
Here we determine a relation between $\eps$ and $\gamma_c$ 
which leads to a standard form for the MoA and which will contain the
FEL pendulum system at first order in the case of (\ref{eq:2.170}).

As a first step we introduce new variables, in addition to $\chi$, as follows.
From the conservation law in (\ref{eq:2.31}) we anticipate that the
order of magnitude of $p_x$ will be $mcK$. In addition $\beta_z:=p_z/mc\gamma$ 
will be near $1$ and so $p_z\approx mc\gamma$. Thus we define
dimensionless momenta by
\begin{eqnarray}
p_x=mcKP_x \; , \quad p_z=mc\gamma P_z \; .
\label{eq:2.36}
\end{eqnarray}
Of course, by (\ref{eq:2.29}),
\begin{eqnarray}
&& P_z > 0 \; .
\label{eq:2.37}
\end{eqnarray}
A natural scaling for $z$ is
\begin{eqnarray}
z=\zeta/k_u \;,
\label{eq:2.38}
\end{eqnarray}
so that the undulator period is $2\pi$ in $\zeta$. 

Abbreviating
\begin{eqnarray}
\theta_{aux}(\zeta):=\alpha(\zeta/k_u) \; ,
\label{eq:2.85}
\end{eqnarray}
and with (\ref{eq:1.11}) the system (\ref{eq:2.32}),(\ref{eq:2.33}) becomes
\bea
&&\hspace{-8mm}\theta_{aux}'=K_r\gamma_c^2
(1-\frac{1}{P_z}) \; , 
\label{eq:2.41}\\
&& \hspace{-8mm}\chi'= -K^2\frac{{\cal E}}{\eps\gamma_c^2} 
\frac{1}{1+\eps\chi}\frac{P_x}{P_z}h(\theta_{aux}) \; ,
\label{eq:2.42}
\eea
where $'=d/d\zeta$ 
and ${\cal E}$ is defined in (\ref{eq:1.14}).
The initial conditions are 
$\theta_{aux}(0,\eps)=\theta_0:=\alpha_0,\chi(0,\eps)=\chi_0$.
Moreover $P_z$ must be replaced, due to (\ref{eq:2.28}), by 
\begin{equation}
P_z=\sqrt{1-\frac{1}{\gamma^2}(1+K^2P_x^2)} \quad \text{with} 
\quad \gamma=\gamma_c(1+\eps\chi) \; ,
\label{eq:2.43}
\end{equation}
and $P_x$ must be replaced, due to (\ref{eq:2.34}), by
\begin{eqnarray}
&& \hspace{-8mm} P_x =  \cos\zeta + \Delta P_{x0} 
+ \frac{{\cal E}}{K_r\gamma_c^2} 
[H(\theta_{aux})-H(\theta_0)] \; ,
\label{eq:2.44} 
\end{eqnarray}
where
\begin{eqnarray}
&& \hspace{-8mm} \Delta P_{x0}:=P_{x0} -1 \; , \quad 
P_{x0}:=P_x(0)=\frac{p_{x0}}{mcK} \; .
\label{eq:2.46} 
\end{eqnarray}
Since $p_z>0$ we have $0< P_z<1$. 
We note that most derivations of the FEL
pendulum take $\Delta P_{x0}=0$, see \cite{KHL,SDR,MP,SSY}. 

To expand $P_z$ we need
\begin{eqnarray}
&&\hspace{-12mm}
1 + K^2 P_x^2 =  1 + K^2(\cos\zeta + \Delta P_{x0})^2 
\nonumber\\
&&\hspace{-10mm}
+ \frac{2K^2{\cal E}}{K_r\gamma_c^2}(\cos\zeta + \Delta P_{x0}) 
( H(\theta_{aux}) -H(\theta_0) )
+ \frac{K^2{\cal E}^2}{K_r^2\gamma_c^4}
( H(\theta_{aux}) -H(\theta_0) )^2 \; ,
\nonumber\\
\label{eq:2.57}
\end{eqnarray}
and it is convenient to define
\begin{eqnarray}
&& \hspace{-12mm} q(\zeta):= 1 + K^2(\cos\zeta + \Delta P_{x0})^2 
= \bar{q} + 2K^2\Delta P_{x0} \cos\zeta
+ \frac{K^2}{2}\cos2\zeta 
\; ,
\label{eq:2.59} \\
&&  \hspace{-12mm} \bar{q}:= 1+ \frac{1}{2}K^2+K^2(\Delta P_{x0})^2 \; .
\label{eq:2.61} 
\end{eqnarray}
Clearly $\bar{q}$ is the average of $q(\zeta)$ over $\zeta$.
Now $P_x$ is $O(1)$ so, by (\ref{eq:2.43}),
\begin{eqnarray}
&&  \hspace{-8mm} 
\frac{1}{P_z} = 1 + \frac{1+K^2P_x^2}{2\gamma_c^2(1+\eps\chi)^2}
+ O(\frac{1}{\gamma_c^4}) \nonumber\\
&& = 1 + \frac{q(\zeta)}{2\gamma_c^2}
(1-2\eps\chi + O(\eps^2))  + O(\frac{1}{\gamma_c^4}) 
 \nonumber\\
&& = 1 + \frac{q(\zeta)}{2\gamma_c^2}
(1-2\eps\chi)  + O(\frac{1}{\gamma_c^4})+ O(\frac{\eps^2}{\gamma_c^2})
\; .
\label{eq:2.47}
\end{eqnarray}
Thus using (\ref{eq:2.44}) and (\ref{eq:2.47}), eq.'s
(\ref{eq:2.41}) and  (\ref{eq:2.42}) become
\bea
&&\hspace{-8mm}
\theta_{aux}'=
 -\frac{K_rq(\zeta)}{2} + \eps K_r q(\zeta)\chi 
+ O(\frac{1}{\gamma_c^2})+O(\eps^2)\; ,
\label{eq:2.48} \\
&& \hspace{-8mm} \chi' = -K^2\frac{{\cal E}}{\eps\gamma_c^2}
(\cos\zeta + \Delta P_{x0})h(\theta_{aux}) 
+ O(1/\gamma_c^2) + O(1/\eps\gamma_c^4) \; .
\label{eq:2.48a}
\eea

To transform (\ref{eq:2.48}),(\ref{eq:2.48a}) into a standard form for the
MoA we need to introduce dependent variables that are slowly varying.
We anticipate that $\chi$ will be slowly varying, i.e.,
$\frac{{\cal E}}{\eps\gamma_c^2}$ will be small. To remove the 
$O(1)$ in (\ref{eq:2.48}) we define
\begin{eqnarray}
&& \theta:=\theta_{aux} + Q(\zeta) \; ,
\label{eq:2.66}
\end{eqnarray}
where
\begin{eqnarray}
&& Q(\zeta):= \zeta + \Upsilon_0\sin\zeta +\Upsilon_1\sin 2\zeta \; ,
\label{eq:2.67} \\
&& \Upsilon_0:=\frac{2K^2\Delta P_{x0}}{\bar{q}}, \quad 
\Upsilon_1:=\frac{K^2}{4\bar{q}}\;.
\label{eq:2.68}
\end{eqnarray}
Note that $\Upsilon_0$ and $\Upsilon_1$ depend only on
$K$ and $\Delta P_{x0}$ and that
\begin{eqnarray}
&& Q'(\zeta) = \frac{K_r q(\zeta)}{2} \; .
\label{eq:2.71}
\end{eqnarray}
Thus the system (\ref{eq:2.48}),(\ref{eq:2.48a}) becomes
\bea
&&\hspace{-8mm} \theta'= \eps K_r q(\zeta)\chi
+ O(1/\gamma_c^2)+O(\eps^2) \; , \label{eq:2.48n} \\
&& \hspace{-8mm} \chi' = -K^2\frac{{\cal E}}{\eps\gamma_c^2}
(\cos\zeta + \Delta P_{x0})h(\theta - Q(\zeta)) 
+ O(1/\gamma_c^2) + O(1/\eps\gamma_c^4) \; .
\label{eq:2.48an}
\eea
The initial conditions are 
$\theta(0,\eps)=\theta_0,\chi(0,\eps)=\chi_0$.
To obtain a system where $\theta$ and $\chi$ interact 
with each other in
first-order averaging we must balance the $O(\eps)$ term in 
(\ref{eq:2.48n}) with the $O({\cal E}/\eps\gamma_c^2)$ in (\ref{eq:2.48an}). 
In this spirit we relate $\eps$ and $\gamma_c$ by choosing
\begin{eqnarray}
&& \hspace{-8mm} 
\eps=\frac{{\cal E}}{\eps\gamma_c^2} \; , 
\label{eq:2.49}
\end{eqnarray}
and so we obtain (\ref{eq:1.13}).
It is this balance that will lead to the FEL pendulum equations in \S\ref{3}. 
This is the distinguished case mentioned in the Introduction and the
system (\ref{eq:2.48n}),(\ref{eq:2.48an}) can be written
\bea
&&\hspace{-8mm} \theta'=\eps K_r q(\zeta)\chi
+ O(\eps^2) \; , \label{eq:2.62}\\
&& \hspace{-8mm} \chi'= -\eps K^2
(\cos\zeta + \Delta P_{x0}) h(\theta - Q(\zeta)) +  O(\eps^2) \; ,
\label{eq:2.63}
\eea
which are now in standard form. 
Up to this point $K_r$ has not been fixed but now
it is convenient to take 
\bea
&&\hspace{-8mm}
K_r=2/\bar{q} \; ,
\label{eq:2.64}
\eea
which we do from now on. Using  (\ref{eq:2.61}), (\ref{eq:2.64}) 
is identical to (\ref{eq:1.12}).
Furthermore in the monochromatic case of (\ref{eq:2.170}) and \S \ref{3}, we will see that, with (\ref{eq:2.64}),
the primary resonance appears at $\nu=1$.

With (\ref{eq:2.64}) the ODE's (\ref{eq:2.62}),
(\ref{eq:2.63}) become
\bea
&&\hspace{-8mm} \theta'= \eps
\frac{2 q(\zeta)}{\bar{q}}\chi + O(\eps^2)  \; , 
\label{eq:2.72}\\
&& \hspace{-8mm} \chi'= -\eps K^2
(\cos\zeta + \Delta P_{x0}) h(\theta - Q(\zeta)) +  O(\eps^2) \; .
\label{eq:2.73}
\eea

We now relate $\theta$ to the so-called ponderomotive phase.
We have, from (\ref{eq:2.19}),(\ref{eq:2.85}),
(\ref{eq:2.66}) and (\ref{eq:2.67}),
\begin{eqnarray}
&& \theta(\zeta,\eps) = \frac{k_r}{k_u}(\zeta - k_u ct(\zeta/k_u)) 
+ [ \zeta + \Upsilon_0\sin\zeta +\Upsilon_1\sin 2\zeta] \; .
\label{eq:2.80}
\end{eqnarray}
Using (\ref{eq:2.38}) and (\ref{eq:2.80}) we obtain
\begin{eqnarray}
&& \hspace{-10mm}
\theta(k_u z,\eps) =k_r(z-ct(z)) + k_uz + 
\Upsilon_0\sin k_u z +\Upsilon_1\sin(2k_u z) \; .
\label{eq:2.81}
\end{eqnarray}
For $\Delta P_{x0}=0$ the variable $\theta$ 
is the so-called ponderomotive phase, i.e.,
\bea
&&\hspace{-8mm} 
\theta(k_u z,\eps) = (k_u + k_r)z - k_rct(z) 
+\Upsilon_1\sin(2k_uz) \; ,
\label{eq:2.82}
\eea
where, for $\Delta P_{x0}=0$,
\bea
&&\hspace{-8mm} 
\Upsilon_1=
\frac{k_r K^2}{8k_u\gamma_c^2} = \frac{K_r K^2}{8} =  \frac{K^2}{4\bar{q}} 
=  \frac{K^2}{4+2K^2}
\; .
\label{eq:2.83}
\eea
Thus in our context the ponderomotive phase
 arises naturally in the process of finding
the distinguished relation between $\eps$ and $\gamma_c$
and transforming to slowly varying coordinates.
In standard treatments it is introduced heuristically to maximize energy
transfer.

To make the $O(\eps^2)$ terms in
(\ref{eq:2.72}),(\ref{eq:2.73}) explicit we first rewrite
(\ref{eq:2.41}),(\ref{eq:2.42}) in terms of  
$\eps,K$ and ${\cal E}$ as
\bea
&&\hspace{-8mm}\theta_{aux}'=\frac{2{\cal E}}{\bar{q}\eps^2}
(1-\frac{1}{P_z}) \; , \label{eq:2.53}\\
&& \hspace{-8mm}\chi'= -K^2\eps\frac{1}{1+\eps\chi}
\frac{P_x}{P_z}h(\theta_{aux}) \; ,
\label{eq:2.54}
\eea
where
\begin{eqnarray}
&& \hspace{-8mm}
P_z^2 = 1-\frac{\eps^2}{{\cal E}}(1+\eps\chi)^{-2}(1+K^2 P_x^2) \; , 
\label{eq:2.55} \\
&& \hspace{-8mm} P_x =  \cos\zeta + \Delta P_{x0} 
+ \frac{\eps^2\bar{q}}{2}[H(\theta_{aux}) -H(\theta_0)] \; .
\label{eq:2.56} 
\end{eqnarray}
The initial conditions are 
$\theta_{aux}(0,\eps)=\theta_0,\chi(0,\eps)=\chi_0$.
Under (\ref{eq:2.66}),(\ref{eq:2.64}), the system becomes
(\ref{eq:2.53}),(\ref{eq:2.54}) becomes
\bea
&&\hspace{-8mm} \theta'= 
\frac{2{\cal E}}{\eps^2\bar{q}}(1-\frac{1}{P_z}) 
+ \frac{q(\zeta)}{\bar{q}} \; ,
\label{eq:2.74}\\
&& \hspace{-8mm} \chi'= -\eps K^2\frac{1}{1+\eps\chi}
\frac{P_x}{P_z}
h(\theta -  Q(\zeta)) \; ,
\label{eq:2.75}
\eea
where
\begin{eqnarray}
&& \hspace{-8mm} P_x = \cos\zeta + \Delta P_{x0} 
+ \frac{\eps^2 \bar{q}}{2}[H(\theta - Q(\zeta)) 
-H(\theta_0)] \; .
\label{eq:2.76}  
\end{eqnarray}
The $O(\eps^2)$ terms in (\ref{eq:2.72}),(\ref{eq:2.73})
can now be determined by comparison
with (\ref{eq:2.74}),(\ref{eq:2.75}).
We will do this in the monochromatic case of \S\ref{3}.

\noindent{\bf Remarks:}
\begin{itemize}
\item[(1)]
Note that, by (\ref{eq:1.14}),(\ref{eq:1.13}), 
$\gamma_c=\sqrt{{\cal E}}/\eps$, in particular $\gamma_c>0$ and, by
(\ref{eq:2.39}),
\begin{eqnarray}
&& \gamma=\gamma_c(1+\eps\chi) 
= \sqrt{{\cal E}}(\frac{1}{\eps}+\chi) \; .
\label{eq:2.77}
\end{eqnarray}
Since, by (\ref{eq:2.29}), we have the restriction 
$\gamma >1$ we also have, by (\ref{eq:2.77}),
\begin{eqnarray}
&& 1 + \eps \chi > 0 \; .
\label{eq:2.78}
\end{eqnarray}
Because, by (\ref{eq:2.37}), $P_z>0$, Eq. (\ref{eq:2.55}) 
gives $\frac{\eps}{\sqrt{{\cal E}}}\sqrt{1+K^2 P_x^2}< |1+\eps\chi|$
and (\ref{eq:2.78}) gives
\begin{eqnarray}
&& \chi > -\frac{1}{\eps} + \frac{1}{\sqrt{{\cal E}}}\sqrt{1+K^2 P_x^2} \; .
\label{eq:2.79}
\end{eqnarray}
Note that (\ref{eq:2.79}) defines our maximal domain
of points $(\theta,\chi,\zeta)$, in particular it entails
(\ref{eq:2.37}),(\ref{eq:2.78}).
We will in \S\ref{3.1}
further restrict this domain.

Of course always $\gamma\geq 1$ and, in fact, 
in applications $\gamma_c,\gamma\gg 1$. However for our purposes it
is convenient to base our work on the maximal domain (\ref{eq:2.79}).
\item[(2)] The transformation to the slowly varying $\theta$ in
(\ref{eq:2.66}) works nicely because $\zeta$ (equivalently $z$)
is the independent variable. If we had stayed with $t$ as the
independent variable this step wouldn't work.
\item[(3)] Equations (\ref{eq:2.72}),(\ref{eq:2.73}) are
in the standard form for the MoA. However we did not prove
that the $O(\eps^2)$ are actually bounded by an $\eps$-independent
constant times $\eps^2$.
In the monochromatic case
in \S\ref{3} we will show that the two $O(\eps^2)$ terms are
truly bounded by ${\cal C}\eps^2$ on an appropriate domain for
appropriate constants ${\cal C}$.
\item[(4)]
For the results of this paper the normalized field strength $\cal E$ cannot be too big (or $\eps$ won't be small) and it cannot be too small or another distinguished case will come into play.
Of course for a seeded FEL, $\cal E$ will be set by the seeding field. In Appendix \ref{H}
we present two very crude bounds that have some relevance to the beginning stages of a High Gain FEL. Here we simply note that for ${\cal E} =1000$, $\eps$ is approximately $0.001$.

In an early approach to this problem we built a normal form analysis assuming  $\cal E$ small, so that the radiation field was a small perturbation of the undulator motion. We thus considered $\cal E$ as a small parameter in addition to $1/\gamma_c$. This led to another distinguished case, which also had a resonant structure but with a different pendulum type behavior. Later we realized that ${\cal E}$ is not necessarily small for cases of interest and we were led to the current case of (\ref{eq:1.13}).
\item[(5)] As will become clear
in \S\ref{3} the normal form for (\ref{eq:2.72}) is
$\theta'= \eps 2\chi$. The normal form of (\ref{eq:2.73}) 
depends on $h$. In the monochromatic case
$h(\theta - Q(\zeta))=\cos(\nu[\theta - Q(\zeta)])$ and the nonresonant, 
resonant and near-to-resonant structure will appear as $\nu$ varies.
In particular the primary resonance will appear at $\nu=1$. However it is
curious that if
\begin{eqnarray}
h(\alpha)=\int^\infty_{-\infty}\; \tilde{h}(\xi)\exp(-i\xi\alpha) d\xi \; ,
\label{eq:2.84}
\end{eqnarray}
with $\tilde{h}(\xi)$ smooth and localized near $\xi=\pm 1$ the 
resonance effect is
washed out in first-order averaging. We will explore this
briefly in \S\ref{n5}. We are studying the consequence
of this in the collective case.
\end{itemize}
\setcounter{equation}{0}
\section{Special Planar Undulator Model and averaging theorems}
\label{3}
We have the planar undulator in
a standard form for the MoA in 
(\ref{eq:2.72}),(\ref{eq:2.73}) where the $O(\eps^2)$ terms
can be determined from (\ref{eq:2.74}),(\ref{eq:2.75}).
We now specialize to a
monochromatic radiation traveling wave, write the system in
Fourier form, discuss resonance as a normal form phenomenon, develop the
NR and NtoR normal forms and state two theorems giving 
precise bounds on the normal form approximations. Thus from now on the 
radiation field in (\ref{eq:2.40}) is monochromatic, i.e., 
$h,H$ have the form (\ref{eq:2.170}) with $\nu\geq 1/2$.
\subsection{The basic ODE's for the monochromatic radiation field}
\label{3.1}
In this section we introduce the notation which will allow us
to state and prove our three propositions and two theorems.
With (\ref{eq:2.170}),(\ref{eq:2.55}),
(\ref{eq:2.76}) we show the dependencies of $P_x$ and $P_z$  on $(\theta,\chi,\zeta,\eps,\nu)$ by the
replacement
\begin{eqnarray}
&& \hspace{-8mm} P_x = \Pi_x \; , \qquad
P_z = \Pi_z  \; ,
\label{eq:3.10}
\end{eqnarray}
where 
\begin{eqnarray}
&& \hspace{-8mm} \Pi_x(\theta,\zeta,\eps,\nu):=\cos\zeta + \Delta P_{x0} 
+ \frac{\eps^2 \bar{q}}{2\nu}[\sin(\nu[ \theta - Q(\zeta)]) 
-\sin(\nu\theta_0)] \; ,
\label{eq:3.11} \\ 
&&\hspace{-8mm} 
\Pi_z(\theta,\chi,\zeta,\eps,\nu) 
:= \sqrt{1-\frac{\eps^2}{{\cal E}}(1+\eps\chi)^{-2}(1+K^2 
\Pi_x^2(\theta,\zeta,\eps,\nu)} \; .
\label{eq:3.12}
\end{eqnarray}
Note that, by (\ref{eq:2.79}),(\ref{eq:3.10}),
\begin{eqnarray}
&& \chi > -\frac{1}{\eps} + \frac{1}{\sqrt{{\cal E}}}\sqrt{1+K^2 
\Pi_x^2(\theta,\zeta,\eps,\nu)} \; .
\label{eq:3.13}
\end{eqnarray}
From now on, we restrict $\eps$ to a finite interval $(0,\eps_0]$.
We are of course interested in $\eps$ small, i.e., $0<\eps\ll 1$,
and so, without loss of generality, we take
\begin{eqnarray}
&& 0<\eps \leq \eps_0 \; , \quad  0<\eps_0 \leq 1 \; .
\label{eq:3.14} 
\end{eqnarray}
Using (\ref{eq:3.13}),(\ref{eq:3.14})
we define the open set $\D(\eps,\nu)$, for
$0<\eps\leq \eps_0,\nu\geq 1/2$, by
\begin{eqnarray}
&& \hspace{-8mm}
\D(\eps,\nu):=\lbrace (\theta,\chi,\zeta)\in\R^3:
\chi> -\frac{1}{\eps} + \frac{1}{\sqrt{{\cal E}}}\sqrt{1+K^2 
\Pi_x^2(\theta,\zeta,\eps,\nu)} \rbrace \; ,
\label{eq:3.15}
\end{eqnarray}
which is our maximal domain in extended phase space.
Accordingly we define the domain of $\Pi_x$
to be $\lbrace(\theta,\zeta,\eps,\nu)\in\R^4:0 < \eps\leq \eps_0,\nu\geq 1/2
\rbrace$ and the domain of $\Pi_z$
to be $\lbrace (\theta,\chi,\zeta,\eps,\nu)\in(\D(\eps,\nu)\times\R^2)
:0<\eps\leq \eps_0,,\nu\geq 1/2\rbrace$. 
It is easy to check that on the domain of $\Pi_z$ the argument of the 
square root in (\ref{eq:3.12}) is positive and, 
for $(\theta,\chi,\zeta)\in\D(\eps,\nu)$, we have
(\ref{eq:2.78}) and 
\begin{eqnarray}
&& \hspace{-8mm} 0 < \Pi_z(\theta,\chi,\zeta,\eps,\nu) < 1 \; .
\label{eq:3.16}
\end{eqnarray}
Moreover with (\ref{eq:2.170}) the ODE's 
(\ref{eq:2.74}),(\ref{eq:2.75}) become
\bea
&&\hspace{-8mm} \theta'= 
\frac{2{\cal E}}{\eps^2\bar{q}}(1-\frac{1}{\Pi_z(\theta,\chi,\zeta,\eps,\nu)}) 
+ \frac{q(\zeta)}{\bar{q}}
 \; , 
\label{eq:3.17}\\
&& \hspace{-8mm} \chi'= -\eps K^2\frac{1}{1+\eps\chi}
\frac{\Pi_x(\theta,\zeta,\eps,\nu)}{\Pi_z(\theta,\chi,\zeta,\eps,\nu)}
\cos(\nu[\theta -  Q(\zeta)]) \; ,
\label{eq:3.18}
\eea
where $q$ and $Q$ are defined in (\ref{eq:2.59}),(\ref{eq:2.67}). 
Of course the
initial conditions are $\theta(0,\eps)=\theta_0,\chi(0,\eps)=\chi_0$.

As suggested by (\ref{eq:2.72}),
(\ref{eq:2.73}) we now write
(\ref{eq:3.17}),(\ref{eq:3.18}) as
\bea
&&\hspace{-8mm} \theta'= \eps f_1(\chi,\zeta)
+ \eps^2 g_1( \theta,\chi,\zeta;\eps,\nu) \; , \label{eq:3.19}\\
&& \hspace{-8mm} \chi'=  \eps f_2(\theta,\zeta;\nu)
+  \eps^2 g_2( \theta,\chi,\zeta;\eps,\nu)
\; ,
\label{eq:3.20}
\eea
where $f_1,f_2$ are given by 
\bea
&&\hspace{-8mm}  f_1(\chi,\zeta):= 
\frac{2q(\zeta)\chi}{\bar{q}} \; , 
\label{eq:3.21}\\
&& \hspace{-8mm}  f_2(\theta,\zeta;\nu):= - K^2
(\cos\zeta + \Delta P_{x0}) \cos( \nu[\theta - Q(\zeta)]) 
\; ,
\label{eq:3.22}
\eea
so that $g_1,g_2$ are given by 
\bea
&&\hspace{-8mm}
\eps^2 g_1( \theta,\chi,\zeta;\eps,\nu) 
:=\frac{2{\cal E}}{\eps^2\bar{q}}(1-\frac{1}{\Pi_z(\theta,\chi,\zeta,\eps,\nu)}) 
+\frac{q(\zeta)}{\bar{q}} (1 - 2\eps\chi) \; , 
\label{eq:3.23} \\
&&\hspace{-8mm}
\eps^2 g_2( \theta,\chi,\zeta;\eps,\nu) 
:= \eps K^2 
\cos( \nu[\theta - Q(\zeta)])
[ \cos\zeta + \Delta P_{x0} 
\nonumber\\
&&
-\frac{1}{1+\eps\chi}
\frac{\Pi_x(\theta,\zeta,\eps,\nu) }
{\Pi_z(\theta,\chi,\zeta,\eps,\nu)}] \; .
\label{eq:3.24}
\eea
The ODE's (\ref{eq:3.17}),(\ref{eq:3.18}) and their equivalent
form, (\ref{eq:3.19}),(\ref{eq:3.20}), will be the subject of 
Theorem \ref{T1}, i.e., the averaging theorem for the NR case
(see also Definition \ref{D1} in \S\ref{3.2}). 
They will also be the basis for the NtoR case.

We need an appropriate
domain for the vector field in 
(\ref{eq:3.19}),(\ref{eq:3.20})
when it comes to averaging theorems.
There are two types of singularities in (\ref{eq:3.19}),(\ref{eq:3.20}).
The first involves the $\eps$ dependence of $g_1,g_2$ as $\eps\rightarrow 0+$.
On the surface it appears that the first term on the rhs of (\ref{eq:3.23})
is $O(1/\eps^2)$, however it is $O(1)$.
In fact, when combined with the second term the rhs is $O(\eps^2)$ so that 
$g_1$ is $O(1)$.
Similarly, $g_2$ appears to be $O(1/\eps)$, 
however again there is a cancellation so that $g_2=O(1)$.
This should not come as a surprise since the construction of the distinguished case (see the remarks before (\ref{eq:2.62}))
Proposition 1 makes this precise by finding the limits of $g_1,g_2$ as $\eps\rightarrow 0+$.
Thus the $\eps=0$ singularity is removable.
There are also singularities for $\Pi_z=0,\eps\chi=-1$ which are not removable.
This is reflected in the fact that even though $f_1,f_2$ are
nice, $g_1,g_2$ have these singularities. 
However these singularities are excluded from our maximal 
domain $\D(\eps,\nu)$ (see (\ref{eq:2.78}),(\ref{eq:3.16}))
and so the vector field in 
(\ref{eq:3.19}),(\ref{eq:3.20})
is of class $C^\infty$ on $\D(\eps,\nu)$ for
$0<\eps\leq \eps_0\leq 1,\nu\geq 1/2$. Nevertheless
since $\D(\eps,\nu)$ is dependent on $\eps$ it is
inconvenient to use it in an averaging theorem.
Thus we now restrict $\D(\eps,\nu)$ to an $\eps$-independent 
domain $W(\eps_0)\times\R$.

To motivate $W$ we note that, by (\ref{eq:3.11}) and since $\nu\geq 1/2$,
\begin{eqnarray}
&& 
|\Pi_x(\theta,\zeta,\eps,\nu)| \leq \Pi_{x,ub}(\eps) \; ,
\label{eq:3.25}
\end{eqnarray}
where
\bea
&& \hspace{-8mm} \Pi_{x,ub}(\eps):=1+ |\Delta P_{x0}| 
+2\eps^2 \bar{q} \; .
\label{eq:3.26}
\eea
Clearly, by (\ref{eq:3.25}),(\ref{eq:3.26}),
\begin{eqnarray}
&&\hspace{-8mm} 
-\frac{1}{\eps} + \frac{1}{\sqrt{{\cal E}}}
\sqrt{1+K^2\Pi_x^2(\theta,\zeta,\eps,\nu)}
\leq -\frac{1}{\eps}  + \frac{1}{\sqrt{{\cal E}}}
\sqrt{1+K^2 \Pi_{x,ub}^2(\eps)}
\nonumber\\
&&\leq -\frac{1}{\eps_0}  + \frac{1}{\sqrt{{\cal E}}}
\sqrt{1+K^2 \Pi_{x,ub}^2(\eps_0)} \; ,
\label{eq:3.52}
\end{eqnarray}
whence, by (\ref{eq:3.15}), we can ``shrink'' 
the maximal domain $\D(\eps,\nu)$ to the $\eps$-independent domain 
$W(\eps_0)\times\R$ where
\bea
&&\hspace{-8mm} 
 W(\eps):= \R\times (\chi_{lb}(\eps),\infty) \; ,
\label{eq:3.27} 
\eea
with
\bea
&& \hspace{-8mm} \chi_{lb}(\eps):= -\frac{1}{\eps} + 
\frac{1}{\sqrt{{\cal E}}}\sqrt{
1+K^2 \Pi_{x,ub}^2(\eps)} \; .
\label{eq:3.28} 
\eea
\subsection{Resonant, nonresonant, $\Delta$-nonresonant, 
near-to-resonant}
\label{3.2}
Now that the structure of the $g_i$ have been characterized at the level needed
for the averaging theorems, we discuss the structure of the $f_i$ defined in
(\ref{eq:3.21}),(\ref{eq:3.22}).
Clearly $f_1$ is $2\pi$ periodic in $\zeta$. We write, by 
(\ref{eq:2.67}),(\ref{eq:3.22}),
\begin{eqnarray}
&&  \hspace{-12mm}  f_2(\theta,\zeta;\nu) = -K^2
(\cos\zeta+ \Delta P_{x0})\cos\biggl(\nu\theta-\nu\zeta
-\nu\Upsilon_0\sin\zeta - \nu\Upsilon_1\sin 2\zeta\biggr) 
\nonumber\\
&&=:\check{f}_2(\theta,\zeta,\nu\zeta;\nu)
\; ,
\label{eq:3.31} 
\end{eqnarray}
where $\check{f}_2(\theta,\zeta_1,\zeta_2;\nu):=-K^2
(\cos\zeta_1+ \Delta P_{x0})\\
\times\cos\biggl(\nu\theta-\zeta_2
-\nu\Upsilon_0\sin\zeta_1 - \nu\Upsilon_1\sin 2\zeta_1\biggr)$.
Since $\check{f}_2(\theta,\zeta_1,\zeta_2;\nu)$ is 
of class $C^\infty$ in $(\zeta_1,\zeta_2)$ and
$2\pi$-periodic in $\zeta_1$ and $\zeta_2$
we conclude from (\ref{eq:3.31}) that 
$f_2$ is a quasiperiodic function 
of $\zeta$ with two base frequencies $1$ and $\nu$ (for the
definition of quasiperiodic functions, see, e.g., \cite{LM}).
To make the resonant structure explicit we write $f_2$ as
\begin{eqnarray}
&&  \hspace{-8mm}  
f_2(\theta,\zeta;\nu) 
= -\frac{K^2}{2}\exp(i\nu(\theta-\zeta))jj(\zeta;\nu,\Delta P_{x0})  + cc \; ,
\label{eq:3.32} 
\end{eqnarray}
where
\begin{eqnarray}
&&  \hspace{-15mm}  
jj(\zeta;\nu,\Delta P_{x0}) :=
(\cos\zeta+ \Delta P_{x0})\exp(-i\nu[\Upsilon_0\sin\zeta
+\Upsilon_1\sin 2\zeta]) \; ,
\label{eq:3.33} 
\end{eqnarray}
is $2\pi$-periodic in $\zeta$. The Fourier series of $jj$ is 
\begin{eqnarray}
jj(\zeta;\nu,\Delta P_{x0})\sim\sum_{n\in{\mathbb Z}} 
\widehat{jj}(n;\nu,\Delta P_{x0})
e^{in\zeta} \; ,
\label{eq:3.34} 
\end{eqnarray}
with
\begin{eqnarray}
&&  \widehat{jj}(n;\nu,\Delta P_{x0}):=\frac{1}{2\pi}
\int_{[0,2\pi]} \; d\zeta jj(\zeta;\nu,\Delta P_{x0})
\,{\rm e}^{-in\zeta} \; ,
\label{eq:3.40} 
\end{eqnarray}
and $\Z$ being the set of integers.
Since $jj(\cdot;\nu,\Delta P_{x0})$ is a $2\pi$-periodic
$C^\infty$ function its Fourier series (\ref{eq:3.34}) is absolutely
convergent, i.e., \\$\sum_{n\in{\mathbb Z}} 
|\widehat{jj}(n;\nu,\Delta P_{x0})|<\infty$ whence $\sim$ in
(\ref{eq:3.34}) can replaced by $=$.
The $f_2$ in Eq. (\ref{eq:3.20}) can now be written
\bea
&& \hspace{-5mm} 
f_2(\theta,\zeta;\nu) =-\frac{K^2}{2}e^{i\nu\theta}
\sum_{n\in\Z}\widehat{jj}(n;\nu,\Delta P_{x0})e^{i(n-\nu)\zeta}+ cc 
\; ,
\label{eq:3.41}
\eea
which clearly shows the resonant structure in that the 
$\zeta$ average of $f_2$ is zero
for $\nu\neq$ integer.
In Appendix \ref{A} we find
\begin{eqnarray}
&&  \widehat{jj}(n;\nu,\Delta P_{x0})
=\frac{1}{2}{\mathcal J}(n,1,\nu,\Upsilon_0,\Upsilon_1)
+\frac{1}{2}{\mathcal J}(n,-1,\nu,\Upsilon_0,\Upsilon_1)
\nonumber\\
&&\quad + \Delta P_{x0}{\mathcal J}(n,0,\nu,\Upsilon_0,\Upsilon_1)\; ,
\label{eq:3.42} 
\end{eqnarray}
where
\begin{eqnarray}
&&  \hspace{-0mm} 
{\mathcal J}(n,m,\nu,\Upsilon_0,\Upsilon_1):=
\sum_{l\in{\mathbb Z}} J_{m-n-2l}(\nu\Upsilon_0)J_l(\nu\Upsilon_1)\; ,
\label{eq:3.43} 
\end{eqnarray}
and $J_k$ is the $k$-th-order Bessel function of the first kind.
Note that\\ 
$jj(-\zeta;\nu,\Delta P_{x0})=jj(\zeta;\nu,\Delta P_{x0})^*$
which implies $\widehat{jj}(n;\nu,\Delta P_{x0})$ is real. This is confirmed
in the explicit form of (\ref{eq:3.42}),(\ref{eq:3.43}) since
the $J_k$ are real valued.

The time average of $f_1$ 
in (\ref{eq:3.21}) is clearly
\begin{eqnarray}
&& \hspace{-8mm}
\bar f_1(\chi):= \lim_{T\rightarrow\infty}[\frac{1}{T}\int_0^T \;
 f_1(\chi,\zeta)d\zeta] =
2\chi \; .
\label{eq:3.44}
\end{eqnarray}
Since the series in (\ref{eq:3.41}) converges uniformly in $\zeta$
and since $\overline{\exp(i(n-\nu)\zeta)}=\delta_{n,\nu}$, 
the time average of the quasiperiodic $f_2$ is
\begin{eqnarray}
&& \bar f_2(\theta;\nu):=
\lim_{T\rightarrow\infty}[\frac{1}{T}\int_0^T \;
f_2(\theta,\zeta; \nu)d\zeta]
\nonumber\\
&&
=\left\{ \begin{array}{ll}  
 0  & \;\; {\rm if\;} \nu\not\in\N\\ 
-K^2 \widehat{jj}(k;k,\Delta P_{x0})\cos(k\theta) 
&\;\;{\rm if\;} \nu=k\in\N \; , \end{array} 
                  \right.
 \label{eq:3.45}
\end{eqnarray}
where $\N$ denotes the set of positive integers and where
we have used the fact that $\widehat{jj}$ is real.
This forms the basis
of our definitions of resonant, nonresonant and near-to-resonant
frequencies $\nu$.

\setcounter{definition}{0}
\begin{definition} (Resonant, nonresonant, $\Delta$-nonresonant, 
near-to-resonant)
\label{D1} \\
\noindent 
Let $\nu\geq 1/2$. We say $\nu$ is nonresonant (NR) if $\nu \not\in \N$
and resonant otherwise. 
We also say that $\nu$ is $\Delta$-nonresonant ($\Delta$-NR) 
when $\nu \in [k+\Delta, k+1-\Delta]$ with 
$\Delta\in(0,0.5)$ and $k\in\N$. Note that $\nu$ is NR if it is
$\Delta$-NR. We say that
$\nu$ is near-to-resonant (NtoR) if 
$\nu=k +\eps a$ where $k\in\N,a\in[-1/2,1/2]$.
Recall $0<\eps \leq \eps_0\leq 1$ and 
that we take
$\N$ to denote the set of positive integers.
\hfill $\Box$ 
\end{definition}

\noindent{\bf Remark:}\\
In our various estimates
we need to keep $\nu$ away from zero but want to include $\nu=1$ since
it is the primary resonance. Thus we require $\nu\geq 1/2$ 
and since $\eps\leq 1$ we require $|a|\leq 1/2$.\\

\noindent
It follows from the Fourier form of (\ref{eq:3.41}) that 
it is only possible to have a nontrivial normal form, i.e.,
$\bar{f}_2\neq0$, if $\nu$ is an integer.
Thus $\nu=1$ is the primary resonance as discussed in the Introduction, 
justifying
the choice of $K_r$ in (\ref{eq:1.12}) and
(\ref{eq:2.64}). The resonant normal form at $\nu=k$
is of the pendulum form with 
\begin{eqnarray}
&& \theta'=\eps 2\chi \; , \quad
\chi'=-\eps K^2 \widehat{jj}(k;k,\Delta P_{x0})\cos(k\theta) \; .
\label{eq:3.46}
\end{eqnarray}
From Appendix \ref{A} we have, for $\Delta P_{x0}=0$,
\begin{eqnarray}
&&  \widehat{jj}(k;k,0)
=\left\{ \begin{array}{ll}  
\frac{1}{2}(-1)^n[J_n(x_n) - J_{n+1}(x_n)]
& \;\; {\rm if\;} k=2n+1 
\\ 
0 & \;\; {\rm if\;} k {\rm \;even} \; , \end{array} 
                  \right.
\label{eq:3.47} 
\end{eqnarray}
where $x_n:=(2n+1)\Upsilon_1$ and $n=0,1,...$ with $\Upsilon_1$ defined in
(\ref{eq:2.68}). Thus, for $\Delta P_{x0}=0$, (\ref{eq:3.46})
gives the standard FEL pendulum system (see also 
\cite{KHL},\cite{MP},\cite{SSY},\cite{HK}):
\begin{eqnarray}
&& \theta'=\eps 2\chi \; , \quad
\chi'=-\eps K^2 \widehat{jj}(k;k,0)\cos(k\theta) \; .
\label{eq:3.46a}
\end{eqnarray}
For a general quasiperiodic function with base frequencies $1$ and $\nu$
it is possible to have a nontrivial normal form for every rational
$\nu$ and thus $\nu$ would be defined to be resonant if it were rational.

Since $\bar f_1(\chi)$ is independent of $\nu$ it 
plays no role in Definition \ref{D1}.
Clearly $\bar f_2(\theta;\nu)=0$ if $\nu$ is NR.
We state our NR theorem in Theorem \ref{T1} for the 
$\Delta$-NR case. In fact because of a
small divisor problem the theorem will require $\nu$
to stay away from neighborhoods of resonances in order to get 
an $o(1)$ error bound as $\eps\rightarrow 0+$. 
We will obtain an $O(\eps^{1-\beta})$
bound for $\beta\in(0,1]$ depending on the distance from the resonance 
by letting $\Delta=O(\eps^\beta)$. In the resonant case we will 
explore an $O(\eps)$ neighborhood of the resonance. 
This will allow us to at least partially fill the gap between
the $\Delta$-NR $\nu$s in the NR theorem
and the $\nu$s in the NtoR theorem.
The way this occurs will be seen in the error analysis in the proofs
of Theorems \ref{T1} and \ref{T2}.
\subsection{The nonresonant case and its normal form}
\label{3.3}
The exact ODE's in the NR case are (\ref{eq:3.19}),(\ref{eq:3.20}).
Clearly they are the same in the $\Delta$-NR subcase.
By definition, the NR normal form, i.e., the normal form with 
$\nu$ NR, is obtained from 
(\ref{eq:3.19}),(\ref{eq:3.20})
by dropping the $O(\eps^2)$ terms
and averaging the rhs over $\zeta$
holding $\theta,\chi$ fixed whence, by
(\ref{eq:3.44}),(\ref{eq:3.45}),
\begin{eqnarray}
&&v_1'= \eps \bar f_1(v_2) = \eps 2v_2 \;, 
\label{eq:3.48} \\
&&v_2'=\eps \bar f_2(v_1;\nu) = 0 \; ,
\label{eq:3.49}
\end{eqnarray}
with the same initial conditions as in the exact ODE's, i.e., 
$v_1(0,\eps)=\theta_0,v_2(0,\eps)=\chi_0$ 
and solution
\begin{eqnarray}
&& v_1(\zeta,\eps)
= 2\chi_0\eps \zeta  + \theta_0 \; , \quad
v_2(\zeta,\eps) = \chi_0 \; .
\label{eq:3.60}
\end{eqnarray}
The solutions of (\ref{eq:3.48}),(\ref{eq:3.49})
with $\eps=1$ play an important role in the
statement and proof of Theorem \ref{T1} and we refer to 
\begin{eqnarray}
&& \vb(\cdot,1)=(v_1(\cdot,1),v_2(\cdot,1))
\; , 
\label{eq:3.61}
\end{eqnarray}
as the guiding solution at $(\theta_0,\chi_0)$. Note that the $\vb$ in
(\ref{eq:3.61}) should not be confused with the velocity vector 
$\vb$ in (\ref{eq:2.12}).

Our basic result in the NR case
will be that $|\theta(\zeta)-v_1(\zeta,\eps)|$
and $|\chi(\zeta)-v_2(\zeta,\eps)|$ are $O(\eps/\Delta)$ 
in the $\Delta$-NR subcase.
If $\Delta=O(1)$ then the error is $O(\eps)$. Putting 
$\Delta$ into the order symbol allows one to discuss $\Delta$ small, e.g.,
as a function of $\eps$. 
The precise statement is given in \S\ref{3.5.1} and its
proof is given in \S\ref{4.1}.

\setcounter{proposition}{0}
\begin{proposition} \label{P1}
Let $0<\eps \leq \eps_0\leq 1$ and let $\nu\geq 1/2$. Then
\begin{eqnarray}
&&  W(\eps_0)\times\R \subset W(\eps)\times\R \subset  \D (\eps,\nu) \; .
\label{eq:3.56}
\end{eqnarray}
Moreover
$g_1(\cdot;\eps,\nu),g_2(\cdot;\eps,\nu)$ 
are $C^\infty$ functions on $W(\eps_0)\times\R$. 
Furthermore, for $(\theta,\chi,\zeta)\in W(\eps_0)\times\R$,
\bea
&&\hspace{-8mm}
\lim_{\eps\rightarrow 0+}\;[g_1( \theta,\chi,\zeta;\eps,\nu)] 
=  -\frac{q(\zeta)}{4\bar{q}} (   \frac{3q(\zeta)}{{\cal E}}  +  12\chi^2)
\nonumber\\
&&\quad - \frac{K^2}{2\nu}
\biggl( \sin(\nu[\theta - Q(\zeta)]) -\sin(\nu\theta_0) \biggr)
(\cos\zeta + \Delta P_{x0}) \; ,
\label{eq:3.29}\\
&&\hspace{-10mm}
\lim_{\eps\rightarrow 0+}\;
[g_2( \theta,\chi,\zeta;\eps,\nu)] 
=K^2 \chi \cos( \nu[\theta - Q(\zeta)]) (\cos\zeta + \Delta P_{x0}) \; .
\label{eq:3.30}
\eea
\end{proposition}
\noindent{\bf Remark:}\\
Proposition \ref{P1} entails that the
vector field on the rhs of (\ref{eq:3.19}),(\ref{eq:3.20}) 
is a  $C^\infty$ function on $W(\eps_0)\times\R$ (whence
the vector field on the rhs of (\ref{eq:3.17}),(\ref{eq:3.18})
is a $C^\infty$ function on $W(\eps_0)\times\R$, too).
Proposition \ref{P1} will allow us 
to use, in Theorem \ref{T1},
the domain $W(\eps_0)\times\R$.
Furthermore the domain is large enough to contain the $\chi$ of
physical interest (see Proposition \ref{P3} in 
\S\ref{3.5.3}).
\\

\noindent {\em Proof of Proposition \ref{P1}:} 
Let $(\theta,\chi,\zeta)\in W(\eps)\times\R$. Then, by
(\ref{eq:3.25}),(\ref{eq:3.27}),(\ref{eq:3.28}),
\begin{eqnarray}
&&\hspace{-8mm} 
\chi>  -\frac{1}{\eps} 
+ \frac{1}{\sqrt{{\cal E}}}\sqrt{1+K^2\Pi_x^2(\theta,\zeta,\eps,\nu)} \; ,
\nonumber
\end{eqnarray}
whence, by (\ref{eq:3.15}),
$(\theta,\chi,\zeta)\in \D(\eps,\nu)$ which proves the second
inclusion in (\ref{eq:3.56}). The first inclusion in (\ref{eq:3.56})
follows from (\ref{eq:3.27}) and from the fact that, by (\ref{eq:3.28}),
$\chi_{lb}(\eps)$ is increasing with $\eps$. 
Moreover, by the remarks after (\ref{eq:3.24}),
$g_1(\cdot;\eps,\nu),g_2(\cdot;\eps,\nu)$ 
are $C^\infty$ functions on $\D(\eps,\nu)$ whence, by (\ref{eq:3.56}),
they are $C^\infty$ functions on $W(\eps_0)\times\R$. 
Finally, (\ref{eq:3.29}),(\ref{eq:3.30})
are proven in Appendix \ref{B} (see (\ref{eq:nB.63}),(\ref{eq:nB.632})).
\hfill $\Box$ 
\subsection{The Near-to-Resonant case and its normal form}
\label{3.4}
\subsubsection{The Near-to-Resonant system}
\label{3.4.1}
According to Definition \ref{D1} we have, in the NtoR case,
\begin{eqnarray}
&& \nu=k+\eps a, 
\label{eq:3.62}
\end{eqnarray}
where $k\in\N$ and $a\in[-1/2,1/2]$ is a measure of the distance of 
$\nu$ from $k$. 
The $O(\eps)$ neighborhood of $k$ is natural in first-order averaging.
If $|\nu-k|$ is too small then the normal form will be close to
the resonant normal form
and if $|\nu-k|$ is too big, then 
$\nu$ will be in the NR regime. 
Eq. (\ref{eq:3.62})
clearly includes the resonant case for $a=0$.
We start from (\ref{eq:3.19}),(\ref{eq:3.20}),(\ref{eq:3.22})
use (\ref{eq:3.62}) and obtain
\bea
&&\hspace{-8mm} \theta'= \eps f_1(\chi,\zeta)
+ \eps^2 g_1( \theta,\chi,\zeta;\eps,k+\eps a) \; , \label{eq:3.63}\\
&& \hspace{-8mm} \chi'=  \eps f_2(\theta,\zeta;k+\eps a)
+  \eps^2 g_2( \theta,\chi,\zeta;\eps,k+\eps a)
\; ,
\label{eq:3.64}
\eea
with initial conditions $\theta(0,\eps)=\theta_0,\chi(0,\eps)=\chi_0$.

By the remarks after (\ref{eq:3.24}),
the vector field in 
(\ref{eq:3.63}),(\ref{eq:3.64})
is of class $C^\infty$ on the maximal domain $\D(\eps,k+\eps a)$.
Since $f_1$ in (\ref{eq:3.63}) is independent of $\eps$ 
the normal form associated with it
will be the same as in the NR case. We now need to study the $\eps$
dependence of $f_2$ in (\ref{eq:3.64}).
From (\ref{eq:3.32}),
\begin{eqnarray}
&&  \hspace{-8mm}  
f_2(\theta,\zeta;k+\eps a) 
= -\frac{K^2}{2}\exp(i(k+\eps a)(\theta-\zeta))jj(\zeta;k+\eps a,\Delta P_{x0})
  + cc 
\nonumber\\
&& = -\frac{K^2}{2}\exp(i[k\theta-\eps a\zeta])\exp(-ik\zeta)
jj(\zeta;k,\Delta P_{x0})
\nonumber\\
&&\quad\times\exp(i\eps a[\theta- \Upsilon_0\sin\zeta
-\Upsilon_1\sin 2\zeta])  + cc \; ,
\label{eq:nn2.200} 
\end{eqnarray}
where we have used from (\ref{eq:3.33}) that
\begin{eqnarray}
&&  \hspace{-10mm}  
jj(\zeta;k+\eps a,\Delta P_{x0}) =
(\cos\zeta+ \Delta P_{x0})\exp(-i(k+\eps a)[\Upsilon_0\sin\zeta
+\Upsilon_1\sin 2\zeta]) 
\nonumber\\
&&  = jj(\zeta;k,\Delta P_{x0})\exp(-i\eps a[\Upsilon_0\sin\zeta
+\Upsilon_1\sin 2\zeta]) \; .
\label{eq:nn2.2000} 
\end{eqnarray}
For $a=0$ the resonant normal form of (\ref{eq:3.45}) is obtained
in (\ref{eq:nn2.200}). For $a\neq 0$ (\ref{eq:nn2.200}) displays
two $\eps$ dependencies. The first is the $\eps a\zeta$
one which cannot be expanded since it is $O(1)$ for $\zeta=O(1/\eps)$
the upper range of our averaging theorem.
The second is the $\eps a$ factor in the final exponential which can be expanded
and makes an $O(1)$ contribution to $g_2$ in (\ref{eq:3.64}) for 
all $\zeta$. Therefore we rewrite $f_2$ as
\begin{eqnarray}
&&  \hspace{-8mm}  
f_2(\theta,\zeta;k+\eps a) 
= f_2^R(\theta,\eps\zeta,\zeta;k,a)  + O(\eps) 
\; ,
\label{eq:nn2.200aa} 
\end{eqnarray}
where
\begin{eqnarray}
&&  \hspace{-15mm}  
f_2^R(\theta,\tau,\zeta;k,a):=
-\frac{K^2}{2}\exp(i[k\theta-a\tau])\exp(-ik\zeta)
jj(\zeta;k,\Delta P_{x0}) + cc
\nonumber\\
&& = -\frac{K^2}{2}\exp(i[k\theta-a\tau])
\sum_{n\in{\mathbb Z}} 
\widehat{jj}(n;k,\Delta P_{x0})
e^{i\zeta[n-k]}  + cc
\; .
\label{eq:430010an}
\end{eqnarray}

We can now write the basic system for the MoA, in this NtoR case.
From (\ref{eq:3.63})-(\ref{eq:430010an}) we obtain
\begin{eqnarray}
&&\hspace{-10mm}\theta' = \eps f^R_1(\chi,\zeta)
+\eps^2 g^R_1(\theta,\chi,\zeta,\eps,k,a) \; , 
\label{eq:430011} \\
&&\hspace{-10mm}\chi'=\eps f^R_2(\theta,\eps\zeta,\zeta;k,a)
+\eps^2 g^R_2(\theta,\chi,\zeta,\eps,k,a) \; ,
\label{eq:430010} 
\end{eqnarray}
where 
\bea
&&\hspace{-8mm}
f^R_1(\chi,\zeta):= f_1(\chi,\zeta) = \frac{2q(\zeta)\chi}{\bar{q}} \; ,
\label{eq:430011an} \\
&&\hspace{-8mm}
g^R_1(\theta,\chi,\zeta,\eps,k,a) 
:=g_1(\theta,\chi,\zeta;\eps,k+\eps a) \; ,
\label{eq:n2.101aaca} \\
&&\hspace{-8mm}
g^R_2(\theta,\chi,\zeta,\eps,k,a) 
:= g_2(\theta,\chi,\zeta;\eps,k+\eps a) 
\nonumber\\
&&\quad
+\frac{1}{\eps}[ f_2(\theta,\zeta;k+\eps a) - 
f_2^R(\theta,\eps\zeta,\zeta;k,a) ] \; ,
\label{eq:n2.101aada}
\eea
and where $g^R_2$ can be rewritten as follows. By (\ref{eq:3.31}) we have
\begin{eqnarray}
&&  \hspace{-12mm}  f_2(\theta,\zeta;k+\eps a) =
-K^2(\cos\zeta+ \Delta P_{x0})
\nonumber\\
&&\cos\biggl((k+\eps a)[\theta-\zeta
-\Upsilon_0\sin\zeta - \Upsilon_1\sin 2\zeta]\biggr) \; ,
\nonumber\\
\label{eq:nnn2.200} 
\end{eqnarray}
and, by (\ref{eq:3.33}),(\ref{eq:430010an}),
\begin{eqnarray}
&&  \hspace{-10mm}  
f_2^R(\theta,\eps\zeta,\zeta;k,a)=
-\frac{K^2}{2}\exp(i[k\theta-\eps a\zeta])\exp(-ik\zeta)
(\cos\zeta+ \Delta P_{x0})
\nonumber\\
&&
\times\exp(-ik[\Upsilon_0\sin\zeta
+\Upsilon_1\sin 2\zeta])
+ cc
\nonumber\\
&& = -K^2(\cos\zeta+ \Delta P_{x0})\cos\biggl( k[\theta-\zeta
-\Upsilon_0\sin\zeta -\Upsilon_1\sin 2\zeta] - \eps a\zeta\biggr)
\; .
\nonumber\\
\label{eq:430010aaa} 
\end{eqnarray}
Using (\ref{eq:nnn2.200}),(\ref{eq:430010aaa}) we can write
(\ref{eq:n2.101aada}) as
\bea
&&\hspace{-8mm}
g^R_2(\theta,\chi,\zeta,\eps,k,a) 
= g_2(\theta,\chi,\zeta;\eps,k+\eps a) 
\nonumber\\
&& -\frac{K^2}{\eps}(\cos\zeta+ \Delta P_{x0})
\Biggl(\cos\biggl((k+\eps a)[\theta-\zeta
-\Upsilon_0\sin\zeta - \Upsilon_1\sin 2\zeta]\biggr)
\nonumber\\
&&\quad
-\cos\biggl( k[\theta-\zeta
-\Upsilon_0\sin\zeta -\Upsilon_1\sin 2\zeta] - \eps a\zeta\biggr)\Biggr) \; ,
\label{eq:nnn2.101aada}
\eea
which will be useful in obtaining bounds for $g^R_2$ in Appendix \ref{E}.

The following proposition is the analogue of Proposition \ref{P1}
for the NtoR case.
\setcounter{proposition}{1}
\begin{proposition} \label{P2}
Let $0<\eps \leq \eps_0\leq 1$ and let $a\in[-1/2,1/2],k\in\N$.
Then $g_1^R(\cdot;\eps,k,a),g_2^R(\cdot;\eps,k,a)$ 
are $C^\infty$ functions on $W(\eps_0)\times\R$. 
Furthermore for $(\theta,\chi,\zeta)\in W(\eps_0)\times\R$
\bea
&&\hspace{-8mm}
\lim_{\eps\rightarrow 0+}\;[g^R_1( \theta,\chi,\zeta,\eps,k,a)] 
=  -\frac{q(\zeta)}{4\bar{q}} (   \frac{3}{{\cal E}} q(\zeta)  +  12\chi^2)
\nonumber\\
&&\quad - \frac{K^2}{2k}
\biggl( \sin(k[\theta - Q(\zeta)]) -\sin(k\theta_0) \biggr)
(\cos\zeta + \Delta P_{x0}) \; ,
\label{eq:xnH.20} \\
&&\hspace{-8mm}
\lim_{\eps\rightarrow 0+}\;[g^R_2( \theta,\chi,\zeta,\eps,k,a)] 
= \chi K^2\cos( k[\theta - Q(\zeta)])
(\cos\zeta + \Delta P_{x0}) 
\nonumber\\
&& 
+ K^2  a (\theta-\Upsilon_0\sin\zeta - \Upsilon_1\sin 2\zeta)
\nonumber\\
&&\quad \times\sin(k[\theta-\zeta-\Upsilon_0\sin\zeta - \Upsilon_1\sin 2\zeta])
(\cos\zeta + \Delta P_{x0}) \; .
\label{eq:xnH.60}
\eea
\end{proposition}
\noindent
Remark: Proposition \ref{P2} entails that the
vector field on the rhs of 
(\ref{eq:430011}),(\ref{eq:430010})
is a  $C^\infty$ function on $W(\eps_0)\times\R$.
Proposition \ref{P2} will allow us 
to use, in Theorem \ref{T2},
the domain $W(\eps_0)\times\R$.
\\

\noindent {\em Proof of Proposition \ref{P2}:} 
The $C^\infty$ property of $g_1^R(\cdot;\eps,k,a),g_2^R(\cdot;\eps,k,a)$ 
follows from Proposition \ref{P1} and
(\ref{eq:n2.101aaca}),(\ref{eq:n2.101aada}).   
Moreover (\ref{eq:xnH.20}),(\ref{eq:xnH.60})
are proven in Appendix \ref{D} (see (\ref{eq:nD.20}),(\ref{eq:nD.60})).
\hfill $\Box$ 
\subsubsection{The NtoR normal form}
\label{3.4.2}
The NtoR normal form ODE's 
are obtained from (\ref{eq:430011}),(\ref{eq:430010}) 
by dropping the $O(\eps^2)$ 
terms and averaging the rhs over $\zeta$
holding the slowly varying quantities 
$\theta,\chi,\eps a\zeta$ fixed. We thus obtain from 
(\ref{eq:430010an}),(\ref{eq:430011}),(\ref{eq:430010}),
(\ref{eq:430011an}) that
\begin{eqnarray}
&&\hspace{-10mm} v_1'
= \eps \bar{f}^R_1(v_2) 
= 2\eps v_2 \;  ,   
\label{eq:nn430016} \\
&&\hspace{-10mm} v_2'
= \eps \bar{f}^R_2(v_1,\eps\zeta;k) 
= -\eps K_0(k) \cos(kv_1-\eps a\zeta) \; , 
\label{eq:nn430015} 
\end{eqnarray}
where
\begin{eqnarray}
&&\hspace{-10mm} K_0(k):=K^2 \widehat{jj}(k;k,\Delta P_{x0}) \; ,
\label{eq:nn430015x} 
\end{eqnarray}
and the same initial conditions as in the exact ODE's, i.e.,
$v_1(0,\eps)=\theta_0,v_2(0,\eps)=\chi_0$.
For $a=0$, eq.'s
(\ref{eq:nn430016}),(\ref{eq:nn430015}) become the resonant normal form
(\ref{eq:3.46}).
For $\Delta P_{x0}=a=0$, eq.'s
(\ref{eq:nn430016}),(\ref{eq:nn430015}) are
the standard FEL pendulum equations, given by
(\ref{eq:3.47}),(\ref{eq:3.46a}). In the special case when $K_0(k)=0$ the ODE's
(\ref{eq:nn430016}),(\ref{eq:nn430015}) are the same as 
NR equations (\ref{eq:3.48}),(\ref{eq:3.49}) and so this case
needs no further comment. Note that the special case $K_0(k)=0$ occurs,
e.g., when $\Delta P_{x0}=0$ and $k$ even (see the remark after
(\ref{eq:A.48})).

The ultimate justification for the normal form 
(\ref{eq:nn430016}),(\ref{eq:nn430015}) comes from the
averaging theorem itself. However, if
we replace $\eps\zeta$ in (\ref{eq:430010}) by $\tau$ and add the
equation $\tau'=\eps$ then this, together 
with (\ref{eq:430011}),(\ref{eq:430010}),
is in a standard form for ``periodic averaging'' (=averaging over a periodic
function) and the normal form (\ref{eq:nn430016}),(\ref{eq:nn430015}) is
obtained by averaging over $\zeta$ holding $\theta,\chi,\tau$ fixed.
In this $\theta,\chi,\tau$
formulation standard periodic averaging theorems apply for the
3D system of $\theta,\chi,\tau$, see, e.g., \cite{M1,ESD}
and Section 3.3 in \cite{SVM}.
We will however prove an averaging theorem directly tuned to 
(\ref{eq:430011}),(\ref{eq:430010}) both to show the reader
a proof in a simple context and in the process we
obtain nearly optimal error bounds which are stronger than in those
standard theorems.
\subsubsection{Structure of the NtoR normal form solutions}
\label{3.4.3}
Here we write the solution of the IVP for the normal form system
(\ref{eq:nn430016}),(\ref{eq:nn430015}) in terms of solutions of the
simple pendulum system and discuss their behavior.
Therefore in this Section we exclude the simple subcase where $K_0=0$.
Let $\vb=(v_1,v_2)$, then it is easy to see that
\begin{eqnarray}
&&\hspace{-10mm} \vb(\zeta,\eps)=\vb(\eps\zeta,1) \; .
\label{eq:n3.1} 
\end{eqnarray}
We first make the transformation 
$\vb(\tau,1)\rightarrow\hat{\vb}(\tau)$ via
\begin{eqnarray}
&&\hspace{-10mm} 
\hat{\vb}(\tau)= \left( \begin{array}{c} 
\hat{v}_1(\tau) \\ \hat{v}_2(\tau)
\end{array}\right) :=
\left( \begin{array}{c} 
kv_1(\tau,1) - a \tau \\ v_2(\tau,1)
\end{array}\right) \; ,
\label{eq:n3.2} 
\end{eqnarray}
which gives
\begin{eqnarray}
&&\hspace{-10mm} \frac{d\hat{v}_1}{d\tau}  
=  2k \hat{v}_2 -a  \;  ,  \quad  \hat{v}_1(0)=k\theta_0 \; ,
\label{eq:430016} \\
&&\hspace{-10mm} \frac{d\hat{v}_2}{d\tau}  = -K_0(k)
\cos \hat{v}_1 \; , \quad \hat{v}_2(0)=\chi_0 \; .
\label{eq:430015} 
\end{eqnarray}
Thus we have scaled away the $\eps$ and made the transformed system
autonomous. Solution properties of (\ref{eq:430016}),(\ref{eq:430015})
are easily understood in terms of its phase plane portrait (PPP).
However it is more convenient to transform
it to the simple pendulum system
\begin{eqnarray}
&& \hspace{-10mm} 
X'=Y, \quad Y' = -\sin X \; ,
\label{eq:237} \\
&& \hspace{-10mm} 
X(0;Z_0) =: X_0 \; , \quad Y(0;Z_0)=:Y_0 \; , \quad
Z_0:= \left( \begin{array}{c} X_0 \\ Y_0
\end{array}\right) \; .
\label{eq:237c} 
\end{eqnarray}
The required transformation is
\begin{eqnarray}
&& \hat{v}_1(\tau)=X(\Omega\tau;Z_0)-{\rm sgn}(K_0)\frac{\pi}{2} \; , 
\label{eq:430031} \\
&& \hat{v}_2(\tau)=
\frac{\Omega Y(\Omega\tau;Z_0) +a}{2k} \; , 
\label{eq:430030}
\end{eqnarray}
where 
\begin{eqnarray}
&& \Omega = \Omega(k):=\sqrt{2k|K_0(k)|} \; .
\label{eq:n3.3}
\end{eqnarray}
From (\ref{eq:n3.1}),(\ref{eq:n3.2})
(\ref{eq:430031}) and (\ref{eq:430030}), the solutions of
(\ref{eq:nn430016}),(\ref{eq:nn430015}) are represented by
\begin{eqnarray}
&& v_1(\zeta,\eps) = \frac{ X(\Omega\eps\zeta;Z_0)
-{\rm sgn}(K_0)\frac{\pi}{2} + \eps a\zeta}{k}
\; , 
\label{eq:n3.4} \\
&&  v_2(\zeta,\eps) = \frac{\Omega Y(\Omega\eps\zeta;Z_0) + a}{2k}
\; , 
\label{eq:n3.5}
\end{eqnarray}
where
\begin{eqnarray}
&&  \hspace{-10mm} Z_0(\theta_0,\chi_0,k,a)
=\left( \begin{array}{c} 
 X_0(\theta_0,k) \\ Y_0(\chi_0,k,a)
\end{array}\right) 
= \left( \begin{array}{c} 
k\theta_0 + {\rm sgn}(K_0(k))\frac{\pi}{2} \\ 
 (2k\chi_0 -a)/\Omega(k)
\end{array}\right) \; .
\label{eq:237a} 
\end{eqnarray}

We now discuss the solution properties of 
(\ref{eq:nn430016}),(\ref{eq:nn430015}) in terms of the simple 
pendulum PPP, \cite{SPPPP}, for (\ref{eq:237})
using (\ref{eq:n3.4}) and (\ref{eq:n3.5}). The  equilibria
of (\ref{eq:237}) are at $(X,Y)=(\pi l,0)$ with integer $l$.

The systems obtained by linearizing about these equilibria are centers
for $l$ even and saddle points for $l$ odd. From the theory of
Almost Linear Systems (see, e.g., \cite{BN}), it follows that the
equilibria are centers and saddle points for the nonlinear system.
A conservation law for the simple pendulum system is easily derived by first noting that the direction field is given by
\begin{eqnarray}
&& \hspace{-10mm} 
\frac{dY}{dX} = - \frac{\sin X}{Y} \; .
\label{eq:237ca} 
\end{eqnarray}
This equation is separable and has solutions given implicitly by $\frac{1}{2}Y^2 + 1-\cos X =const$. Thus
\begin{eqnarray}
&& {\cal E}_{Pen}(X,Y):=\frac{1}{2}Y^2 +U(X) \; , \quad U(X)=1-\cos X
\label{eq:430040} 
\end{eqnarray}
is a constant of the motion which is easily checked directly.
Incidentally ${\cal E}_{Pen}$ is also a Hamiltonian for the ODE's 
(\ref{eq:nn430016}),(\ref{eq:nn430015}) 
but this plays no role here. 
The PPP is easily constructed from the so-called potential plane which is simply a plot of the potential $U(X)$ vs. $X$, see \cite{PP}.
The PPP shows that 
the solutions of the simple pendulum system has four types of behavior, the equilibria
mentioned above, libration, rotation and separatrix motion.
These can be characterized in terms of ${\cal E}_{Pen}$.
Clearly, ${\cal E}_{Pen}$ is nonnegative, the centers correspond to
${\cal E}_{Pen}(X,Y)=0$ and the saddle points and separatrices to
${\cal E}_{Pen}(X,Y)=2$. The motion is libration for 
$0<{\cal E}_{Pen}(X,Y)<2$, rotation for ${\cal E}_{Pen}(X,Y)>2$ and 
separatrix motion for ${\cal E}_{Pen}(X,Y)=2$ with $Y\neq 0$.
In the libration case the solutions are periodic, which is easy to show,
and the period as a function of amplitude, \cite{Period}, is given by
\begin{eqnarray}
&& T(A) =2\sqrt{2}\int^A_0\frac{dt}{[\cos t- \cos A]^{1/2}} \; , \quad
(0<A<\pi)
\end{eqnarray}
where $T(A)$ is the period associated with the initial conditions
$X_0=A,Y_0=0$. It is easy to show that 
$\lim_{A\rightarrow 0}\;T(A)=2\pi$.

We denote by ${\cal B}_n$ the $n$-th pendulum bucket which is
defined by
\begin{eqnarray}
&&  \hspace{-15mm}  {\cal B}_n:=\lbrace (X,Y)\in\R^2:{\cal E}_{Pen}(X,Y)<2,
|X -2\pi n|<\pi\rbrace \; ,
\label{eq:430040aa} 
\end{eqnarray}
with $n\in\Z$. Note that, by (\ref{eq:237a}),(\ref{eq:430040}),
\begin{eqnarray}
&& \hspace{-10mm}
{\cal E}_{Pen}(Z_0(\theta_0,\chi_0,k,a))
= {\cal E}_{R}(\theta_0,\chi_0,k,a):=\frac{1}{2}
 [\frac{2k\chi_0 -a}{\Omega(k)}]^2 
\nonumber\\
&&
+ 1 + {\rm sgn}(K_0)\sin(k\theta_0) \; .
\label{eq:430040a} 
\end{eqnarray}
Note also that, by (\ref{eq:n3.4}),(\ref{eq:n3.5}),(\ref{eq:237a}),
\begin{eqnarray}
&&  \hspace{-10mm} |v_1(\zeta,\eps)-\theta_0| = 
\Big{|} \frac{ X(\Omega\eps\zeta;Z_0) - X_0 + \eps a\zeta}{k}\Big{|}
\leq \frac{ |X(\Omega\eps\zeta;Z_0) - X_0| + \eps |a|\zeta}{k}
\; , 
\nonumber\\
\label{eq:n3.4a} \\
&&  \hspace{-10mm}  
|v_2(\zeta,\eps) -\chi_0| = \frac{\Omega}{2k}
|Y(\Omega\eps\zeta;Z_0)-Y_0| \; ,
\label{eq:n3.5a} \\
&&  \hspace{-10mm}  
|v_2(\zeta,\eps)| \leq \frac{\Omega |Y(\Omega\eps\zeta;Z_0)|+|a|}{2k} \; .
\label{eq:n3.6a}
\end{eqnarray}

We can now discuss 
the four cases of
equilibria, libration, rotation and separatrix motion.
In each case, using (\ref{eq:n3.4a}),(\ref{eq:n3.5a}),
(\ref{eq:n3.6a}), we will find $d^{min}_1,d^{min}_2,\chi_\infty  \geq 0$
such that, for all $\zeta\geq 0$,
\begin{eqnarray}
&&  \hspace{-10mm} |v_1(\zeta,\eps)-\theta_0| \leq 
d^{min}_1(\theta_0,\chi_0,\eps\zeta,k,a) \; , \quad
|v_2(\zeta,\eps) -\chi_0| \leq d^{min}_2(\theta_0,\chi_0,k,a)
\; ,
\nonumber\\
\label{eq:n3.5b} \\
&&  \hspace{-10mm} |v_2(\zeta,\eps)| \leq \chi_\infty(\theta_0,\chi_0,k,a) \; ,
\label{eq:n3.5e}
\end{eqnarray}
and we will at the same time observe that
$d^{min}_1(\theta_0,\chi_0,\tau,k,a)$ is increasing w.r.t. $\tau$.

\begin{itemize}
\item[(I)] Equilibria regime:
$Y_0=0$ and either
${\cal E}_{Pen}(X_0,Y_0)=0$ or $2$.

Clearly $X_0=\pi l$ where $l\in\Z$ and, by (\ref{eq:237a}),
\begin{eqnarray}
&&  \hspace{-15mm} 
 \left( \begin{array}{c} 
k\theta_0 + {\rm sgn}(K_0(k))\frac{\pi}{2} \\ 
 (2k\chi_0 -a)/\Omega(k)
\end{array}\right)
= Z_0(\theta_0,\chi_0,k,a) =\left( \begin{array}{c} 
X \\ Y
\end{array}\right)
=\left( \begin{array}{c} 
\pi l \\ 0
\end{array}\right) \; ,
\label{eq:237aa} 
\end{eqnarray}
so that $\theta_0=(\pi l - {\rm sgn}(K_0(k))\frac{\pi}{2})/k$ and $\chi_0=a/2k$.
Thus, by (\ref{eq:n3.4}),(\ref{eq:n3.5}),
\begin{eqnarray}
&& v_1(\zeta,\eps) = \theta_0 + \frac{\eps a\zeta}{k}
\; , 
\label{eq:n3.6} \\
&&  v_2(\zeta,\eps) = \chi_0
\; . 
\label{eq:n3.7}
\end{eqnarray}
Clearly, by direct substitution, these are solutions of
(\ref{eq:nn430016}),(\ref{eq:nn430015}). Incidentally
these solutions are stable for $l$ even and unstable for
$l$ odd. 

Clearly, due to (\ref{eq:n3.5b}),(\ref{eq:n3.5e}),(\ref{eq:n3.6}),
(\ref{eq:n3.7}), we can choose
\begin{eqnarray}
&&  \hspace{-10mm} 
d^{min}_1(\theta_0,\chi_0,\eps\zeta,k,a)
:= \frac{\eps |a| \zeta}{k} \; , \quad d^{min}_2(\theta_0,\chi_0,k,a)
:= 0 \; ,
\label{eq:n3.5c} \\
&&  \hspace{-10mm}  \chi_\infty(\theta_0,\chi_0,k,a):= |\chi_0| \; .
\label{eq:n3.5ca} 
\end{eqnarray}
\item[(II)] Libration regime:
$0<{\cal E}_{Pen}(X_0,Y_0)<2$.

In this case $Z_0(\theta_0,\chi_0,k,a)\in{\cal B}_{n(\theta_0,k)}$
where the integer $n=n(\theta_0,k)$ 
is determined by the condition $|X_0(\theta_0,k)-2\pi n(\theta_0,k)|<\pi$. 
From (\ref{eq:n3.4}),(\ref{eq:n3.5}) we see that
\begin{eqnarray}
&& \vb(\zeta,\eps) = \vb_{per}(\zeta,\eps) + \vb_{lin}(\eps\zeta) \; ,
\label{eq:n3.5v}
\end{eqnarray}
and it is easy to show that 
the periodic part has 
amplitude determined by  the max and min values of $X$ and $Y$ and the linear
growth term is 
\begin{eqnarray}
&&  \hspace{-10mm} 
\vb_{lin}(\eps\zeta) =
 \left( \begin{array}{c} 
 \eps a\zeta/k \\ 0
\end{array}\right) \; .
\label{eq:n3.5va} 
\end{eqnarray}
The maximum values $X_{max}$ and $Y_{max}$ 
of $X$ and $Y$ satisfy, by (\ref{eq:430040}),
\begin{eqnarray}
&& 
\hspace{-15mm} 
{\cal E}_{Pen}(Z_0)=
\frac{1}{2}Y_0^2 + 1-\cos X_0 =\frac{1}{2}Y_{max}^2 = 1-\cos X_{max} \; , 
\label{eq:n3.8} 
\end{eqnarray}
whence 
\begin{eqnarray}
&& 
\hspace{-5mm} 
X_{max}(\theta_0,\chi_0,k,a) = 2\pi n(\theta_0,k) 
+ \arccos(\cos X_0 - \frac{1}{2}Y_0^2) \; ,
\nonumber\\
&&
= 2\pi n(\theta_0,k) 
+ \arccos\biggl( 1 - {\cal E}_{R}(\theta_0,\chi_0,k,a) \biggr) \; ,
\nonumber\\
&& \hspace{-5mm} 
Y_{max}(\theta_0,\chi_0,k,a)
:=\sqrt{2{\cal E}_{Pen}(Z_0(\theta_0,\chi_0,k,a))} 
\nonumber\\
&&= \sqrt{2{\cal E}_{R}(\theta_0,\chi_0,k,a)} 
\; ,
\nonumber\\
\label{eq:n3.8b} 
\end{eqnarray}
and the minimum values $X_{min}$ and $Y_{min}$ 
of $X$ and $Y$ are given by
\begin{eqnarray}
&& 
\hspace{-5mm} X_{min}
:=4\pi n - X_{max} \; , \quad
Y_{min}:= -Y_{max} \;  .
\label{eq:n3.8c} 
\end{eqnarray}
Here $\arccos$ is the principle branch of the inverse $\cos$ mapping $[-1,1]\rightarrow[0,\pi]$. 

We now determine $d_1^{min},d_2^{min}$ and $\chi_\infty$.
It follows from (\ref{eq:n3.4a}),(\ref{eq:n3.5a}),(\ref{eq:n3.6a}),
(\ref{eq:n3.5b}),(\ref{eq:n3.5e}),(\ref{eq:n3.8b}),(\ref{eq:n3.8c}) that
\begin{eqnarray}
&&  \hspace{-10mm} |v_1(\zeta,\eps)-\theta_0| 
\leq \frac{ |X(\Omega\eps\zeta;Z_0) - X_0| + \eps |a|\zeta}{k}
\nonumber\\
&&  \hspace{-5mm}
\leq \frac{ 2X_{max}(\theta_0,\chi_0,k,a) -4\pi n(\theta_0,k) + \eps |a|\zeta}{k}
\nonumber\\
&& = \frac{ 2\arccos\biggl( 1 - {\cal E}_{R}(\theta_0,\chi_0,k,a) \biggr)
+ \eps |a|\zeta}{k}
=:d_1^{min}(\theta_0,\chi_0,\eps\zeta,k,a)
\; , \nonumber\\
\label{eq:n3.4d} \\
&&  \hspace{-10mm}  
|v_2(\zeta,\eps) -\chi_0| = \frac{\Omega}{2k}
|Y(\Omega\eps\zeta;Z_0)-Y_0| 
\leq \frac{\Omega}{k}Y_{max}(\theta_0,\chi_0,k,a)
\nonumber\\
&& = \frac{\Omega(k)}{k}\sqrt{2{\cal E}_{R}(\theta_0,\chi_0,k,a)} 
=:d_2^{min}(\theta_0,\chi_0,k,a)
\; ,
\label{eq:n3.5d} \\
&&  \hspace{-10mm}  
|v_2(\zeta,\eps)| \leq \frac{\Omega |Y(\Omega\eps\zeta;Z_0)|+|a|}{2k}
 \leq \frac{\Omega Y_{max}(\theta_0,\chi_0,k,a) +|a|}{2k}
\nonumber\\
&& = \frac{ \Omega(k)\sqrt{2{\cal E}_{R}(\theta_0,\chi_0,k,a)} +|a|}{2k}
=:\chi_\infty(\theta_0,\chi_0,k,a) \; .
\label{eq:n3.5da} 
\end{eqnarray}
\item[(III)] Separatrix regime: $Y_0\neq 0$ and
${\cal E}_{Pen}(X_0,Y_0)=2$.

In this case $(X,Y)\in\overline{{\cal B}_{n(\theta_0,k)}}$
where the integer $n=n(\theta_0,k)$ is determined such that 
$|X_0(\theta_0,k)-2\pi n(\theta_0,k)|<\pi$. Clearly
\begin{eqnarray}
&& \hspace{-5mm} |X - X_0| \leq 2\pi  \; , \quad 
 |Y - Y_0| \leq \sqrt{2{\cal E}_{Pen}(X_0,Y_0)} = 2 \; , \quad 
 |Y | \leq 2 \; .
\nonumber\\
\label{eq:n3.9b} 
\end{eqnarray}
For $Y_0>0$, $(X(t),Y(t))\rightarrow ((2n+1)\pi,0)$ as 
$t\rightarrow\infty$ and, for 
$Y_0<0$, $(X(t),Y(t))\rightarrow ((2n-1)\pi,0)$ 
as $t\rightarrow\infty$. Thus for large $\zeta$ 
\begin{eqnarray}
&&  \hspace{-10mm} 
v(\eps\zeta) \approx  \frac{1}{k}
 \left( \begin{array}{c} 
 (2n \pm 1)\pi  -{\rm sgn}(K_0(k))\frac{\pi}{2} + \eps a\zeta \\ 
a/2
\end{array}\right) \; ,
\label{eq:n3.5vb} 
\end{eqnarray}
which is the odd $l$ solution in case I.

We now determine $d_1^{min},d_2^{min}$ and $\chi_\infty$.
By (\ref{eq:n3.4a}),(\ref{eq:n3.5a}),(\ref{eq:n3.6a}),
(\ref{eq:n3.5b}),\\
(\ref{eq:n3.5e}),(\ref{eq:n3.9b}) 
\begin{eqnarray}
&&  \hspace{-10mm} |v_1(\zeta,\eps)-\theta_0| 
\leq \frac{ |X(\Omega\eps\zeta;Z_0) - X_0| + \eps |a|\zeta}{k}
\nonumber\\
&&\leq \frac{ 2\pi + \eps |a|\zeta}{k}
=:d_1^{min}(\theta_0,\chi_0,\eps\zeta,k,a) \; , 
\label{eq:n3.9d} \\
&&  \hspace{-10mm}  
|v_2(\zeta,\eps) -\chi_0| = \frac{\Omega}{2k}
|Y(\Omega\eps\zeta;Z_0)-Y_0| 
\leq \frac{\Omega(k)}{k}
\nonumber\\
&&
=:
d_2^{min}(\theta_0,\chi_0,k,a)\; ,
\label{eq:n3.9e} \\
&&  \hspace{-10mm} |v_2(\zeta,\eps)| \leq 
\frac{\Omega |Y(\Omega\eps\zeta;Z_0)|+|a|}{2k} 
\nonumber\\
&&
\leq \frac{2\Omega(k) +|a|}{2k} =:
\chi_\infty(\theta_0,\chi_0,k,a) \; . 
\label{eq:n3.9ea}
\end{eqnarray}
\item[(IV)] 
Rotation regime: ${\cal E}_{Pen}(X_0,Y_0)>2$.

For $Y_0>0$, $X$ is increasing and $Y$ is periodic such that 
\begin{eqnarray}
&& \sqrt{2}\sqrt{{\cal E}_{Pen}(X_0,Y_0)-2}\leq Y \leq 
\sqrt{2}\sqrt{{\cal E}_{Pen}(X_0,Y_0)} \; ,
\label{eq:n3.10a} 
\end{eqnarray}
and for $Y_0<0$, $X$ is decreasing and $Y$ is periodic such that 
\begin{eqnarray}
&& -\sqrt{2}\sqrt{{\cal E}_{Pen}(X_0,Y_0)} \leq Y \leq 
-\sqrt{2}\sqrt{{\cal E}_{Pen}(X_0,Y_0)-2} \; .
\label{eq:n3.10b} 
\end{eqnarray}
Clearly $v_2(\cdot,\eps)$ is periodic.
We now determine $d_1^{min},d_2^{min}$ and $\chi_\infty$.
It follows from (\ref{eq:n3.10a}),(\ref{eq:n3.10b}) that
for any choice of $Y_0$
\begin{eqnarray}
&&  \hspace{-12mm} |Y - Y_0| \leq 
\sqrt{2}\sqrt{ {\cal E}_{R}(\theta_0,\chi_0,k,a)  }-
\sqrt{2}\sqrt{  {\cal E}_{R}(\theta_0,\chi_0,k,a) -2 } 
\; ,
\label{eq:n3.10c} \\
&&  \hspace{-12mm} |Y| \leq 
\sqrt{2{\cal E}_{R}(\theta_0,\chi_0,k,a) } \; .
\label{eq:n3.10ca} 
\end{eqnarray}
It follows from (\ref{eq:n3.5a}),(\ref{eq:n3.6a}),(\ref{eq:n3.10c}),
(\ref{eq:n3.10ca}) that
\begin{eqnarray}
&&  \hspace{-10mm}  
|v_2(\zeta,\eps) -\chi_0| = \frac{\Omega}{2k}
|Y(\Omega\eps\zeta;Z_0)-Y_0| 
\nonumber\\
&&
\leq \frac{\Omega}{2k}\biggl( 
\sqrt{2}\sqrt{ {\cal E}_{R}(\theta_0,\chi_0,k,a) }
-\sqrt{2}\sqrt{  {\cal E}_{R}(\theta_0,\chi_0,k,a) -2} \biggr)
\nonumber\\
&&
=:d_2^{min}(\theta_0,\chi_0,k,a)
\; ,
\label{eq:n3.10d} \\
&&  \hspace{-10mm}  
|v_2(\zeta,\eps)| \leq \frac{\Omega |Y(\Omega\eps\zeta;Z_0)|+|a|}{2k} 
 \leq \frac{\Omega(k) \sqrt{2 {\cal E}_{R}(\theta_0,\chi_0,k,a)} +|a|}{2k}
\nonumber\\
&&
=:\chi_\infty(\theta_0,\chi_0,k,a)
\; .
\label{eq:n3.10da}
\end{eqnarray}
It follows from (\ref{eq:237}),(\ref{eq:n3.10ca}) that
\begin{eqnarray}
&& \hspace{-10mm}   | X(\Omega\eps\zeta;Z_0) -X_0 |  
=|\int_0^{\Omega\eps\zeta}\;X'(s)ds| 
=|\int_0^{\Omega\eps\zeta}\;Y(s)ds| 
\nonumber\\
&&
\leq 
\int_0^{\Omega\eps\zeta}\;|Y(s)|ds 
\leq \sqrt{2} \int_0^{\Omega\eps\zeta}\;\sqrt{ {\cal E}_{Pen}(X(s),Y(s))}ds 
\nonumber\\
&&
=\sqrt{2}\Omega\eps\zeta\sqrt{ {\cal E}_{Pen}(X_0,Y_0)} 
=\sqrt{2}\Omega\eps\zeta\sqrt{ {\cal E}_R(\theta_0,\chi_0,k,a)} \; ,
\label{eq:n3.10e}
\end{eqnarray}
whence, by (\ref{eq:n3.4a}),
\begin{eqnarray}
&&  \hspace{-10mm} |v_1(\zeta,\eps)-\theta_0| 
\leq \frac{ |X(\Omega\eps\zeta;Z_0) - X_0| + \eps |a|\zeta}{k}
\nonumber\\
&&  \hspace{-5mm}
\leq \frac{ 
\sqrt{2}\Omega(k)\eps\zeta\sqrt{ {\cal E}_R(\theta_0,\chi_0,k,a)}
+ \eps |a|\zeta}{k}
\nonumber\\
&&
=:d_1^{min}(\theta_0,\chi_0,\eps\zeta,k,a)
\; .
\label{eq:n3.10f}
\end{eqnarray}
\end{itemize}

Clearly the simple pendulum system is central to our NtoR normal form
approximation.
Every student who has taken a course in ODE's  or Classical Mechanics has studied the pendulum equation at some level.
However, not every reader of this paper may know the general settings of the equation.
So, as an aside, we thought some might be interested in knowing how it fits in a broader context.
First, the pendulum equation is a special case
of the nonlinear oscillator $\ddot{x}+g(x)=0$ and second, the nonlinear oscillator is an important subclass
of the class of second-order autonomous systems $\dot{x}=f(x,y),\dot{y}=g(x,y)$.
The nonlinear oscillator is discussed in many texts, and here we mention
\cite{BN} and \cite{Ar}. Its PPP is easily constructed from the
potential plane as mentioned above and in \cite{PP}.
After the class  of linear systems, the class of second-order
autonomous systems has the most well developed theory  \cite{SOAS}. Here the qualitative behavior is completely captured in the 
PPP's. What's missing from a PPP is the time it takes to go from one point
on an orbit to another, but this is easily determined using a good ODE solver. 
The limiting behavior of all solutions bounded in forward time is given by the celebrated 
${\rm Poincar\acute{e}}$-Bendixson theorem
and as a consequence existence of periodic solutions can be inferred
and the possibility of chaotic behavior is eliminated. It also follows
that a closed orbit in the phase plane corresponds to a periodic solution.
\subsubsection{NR limit far away from the pendulum buckets}
\label{3.4.4}
Even though for small $\eps$ 
there will be gaps in $\nu$ between the $\Delta$-NR and NtoR cases,
as we will discuss in the context of Theorems \ref{T1},\ref{T2},
we show here that far away from the pendulum buckets
the NR normal form emerges.
While not a rigorous argument since we do not quantify ``large''
it is a consistency check. As in Section \ref{3.4.3}
we exclude the simple subcase where $K_0=0$.

For $Z_0$ far away from the pendulum buckets
in the sense that
$|Y_0|=|2k\chi_0-a|/\Omega\gg 2$, we are in the rotation regime.
Letting $X(\tilde{s})=\hat{X}(s),Y(\tilde{s})=Y_0\hat{Y}(s),s=Y_0\tilde{s}$,
(\ref{eq:237}),(\ref{eq:237c}) become
\begin{eqnarray}
&& \hspace{-10mm} 
\frac{d\hat{X}}{ds}=\hat{Y} \; , 
\quad \frac{d\hat{Y}}{ds} = -\epsilon\sin \hat{X} \; , \quad 
\hat{X}(0)=X_0 \; , \quad
\hat{Y}(0)=Y_0=1 \; ,
\label{eq:n3.16}
\end{eqnarray}
where $\epsilon=1/Y_0^2$. 
A regular perturbation expansion
yields $\hat{X}(s)=s+ X_0 + O(\epsilon),\hat{Y}(s)=1+ O(\epsilon)$ 
as we show in Appendix \ref{F} therefore
$X(\tilde{s})= Y_0\tilde{s} + X_0 
+ O(1/Y_0^2),Y(\tilde{s})= Y_0 + O(1/Y_0)$ and thus from
(\ref{eq:n3.4}),(\ref{eq:n3.5}),(\ref{eq:237a})
\begin{eqnarray}
&& \hspace{-10mm}  v_1(\zeta,\eps) = \frac{ Y_0\Omega\eps\zeta + X_0
+ O(1/Y_0^2)
-{\rm sgn}(K_0(k))\pi/2 + \eps a\zeta}{k}
\nonumber\\
&&
= \theta_0 + \frac{ Y_0\Omega + a}{k}\eps\zeta  + O(1/Y_0^2)
= 2\chi_0\eps\zeta + \theta_0  + O(1/Y_0^2)
\; , 
\label{eq:n3.17} \\
&&  \hspace{-10mm}  v_2(\zeta,\eps) = \frac{\Omega Y_0 + a}{2k} 
 + O(1/Y_0) = \chi_0  + O(1/Y_0)
\; , 
\label{eq:n3.18}
\end{eqnarray}
consistent with (\ref{eq:3.60}).
\subsection{Averaging theorems}
\label{3.5}
Recall that we have gone from our basic Lorentz system, (\ref{eq:2.25})-(\ref{eq:2.18}), to  (\ref{eq:3.19}),(\ref{eq:3.20}) with no approximations. 
We have also derived two related normal forms for $\nu\ge 1/2$ in the NR 
(\S\ref{3.3}) and NtoR (\S\ref{3.4}) cases.
Here we state theorems which conclude that the solutions of these normal form systems yield good approximations to the solutions of (\ref{eq:2.25})-(\ref{eq:2.18}) in the appropriate $\nu$ domains.

Our NR theorem in \S\ref{3.5.1} will cover the $\Delta$-NR case, i.e.,
closed subintervals
$[k+\Delta, k+1-\Delta]$ of $(k,k+1)$, where 
$k=0,1,...,0<\Delta<0.5$, and we will obtain error bounds
of $O(\eps/\Delta)$ (Here $\Delta$ can be small as mentioned in 
\S\ref{3.2} and \S\ref{3.3}). Our NtoR theorem in \S\ref{3.5.2} will cover
the case where $\nu=k+\eps a$ which includes the resonant $\nu=k$ 
case and we will obtain error bounds
of $O(\eps)$.
%
\subsubsection{$\Delta$-nonresonant case: 
$\nu \in [k+\Delta, k+1-\Delta]$ (Quasiperiodic Averaging)}
\label{3.5.1}
The exact ODE's to be analyzed are
(\ref{eq:3.19}),(\ref{eq:3.20})
with the initial conditions $\theta(0,\eps)=\theta_0,\chi(0,\eps)=\chi_0$ and 
where $f_1,f_2$ are defined by (\ref{eq:3.21}),(\ref{eq:3.22})
and where $\widehat{jj}(n;\nu,\Delta P_{x0})$ is defined by
(\ref{eq:3.40}) and $g_1,g_2$ by
(\ref{eq:3.23}),(\ref{eq:3.24}).
The normal form ODE's are (\ref{eq:3.48}),(\ref{eq:3.49})
with initial conditions $v_1(0,\eps)=\theta_0,v_2(0,\eps)=\chi_0$ 
and solution (\ref{eq:3.60}).
Note that $v_i(\zeta,\eps)=v_i(\eps\zeta,1)$. 

We are now ready to state the NR theorem which roughly concludes that 
$|\theta(\zeta,\eps) - 2\chi_0\eps\zeta-\theta_0|
=O(\eps/\Delta)$ and $|\chi(\zeta,\eps)- \chi_0| =O(\eps/\Delta)$ for
$0\leq \zeta\leq O(1/\eps)$ with $\eps$ 
sufficiently small.
To make the statement of the theorem concise, we now set up the 
theorem in nine steps.

\begin{itemize}
\item[(1)] (Basic parameters)\\
Let $0<\eps \leq \eps_0\leq 1$, fix $0<\Delta<0.5$ and 
let $\nu\in[k+\Delta, k+1-\Delta]$ where $k$ is a nonnegative integer. 
\item[(2)] (Initial data)\\
Choose $\theta_0,\chi_0$ such that
$(\theta_0,\chi_0)\in (\R\times [-\chi_M,\chi_M])$
where $\chi_M>0$ is chosen such that $-\chi_M>\chi_{lb}(\eps_0)$
where $\chi_{lb}$ is defined by (\ref{eq:3.28}).
Clearly $(\R\times [-\chi_M,\chi_M])\subset W(\eps_0)$ where $W(\eps_0)$
is defined by (\ref{eq:3.27}).
Note also that, by (\ref{eq:3.60}), the 
corresponding guiding
solution $\vb(\zeta,1)=(2\chi_0\zeta + \theta_0,\chi_0)$ belongs to
$(\R\times [-\chi_M,\chi_M])$ for all $\zeta\in[0,\infty)$.
\item[(3)] (Guiding solution) \\
Choose $T>0$ and define the compact (=closed and bounded)  
subset 
\begin{eqnarray}
&& \hspace{-12mm}
S:=\{ \vb(\tau,1): \tau\in[0,T]\}
=\{ (2\chi_0 \tau  + \theta_0,\chi_0):\tau\in[0,T]\}
\label{eq:n2.201an}
\end{eqnarray}
of $(\R\times [-\chi_M,\chi_M])\subset W(\eps_0)$.
Recall that $\vb(\zeta,\eps)=\vb(\eps\zeta,1)$.
\item[(4)] (Rectangle around initial value $(\theta_0,\chi_0)$:
the basic domain for averaging theorem) \\
Let $\hat{W}(\theta_0,\chi_0,d_1,d_2)$ 
be the following open rectangle around $S$ where
\begin{eqnarray}
&&  \hspace{-15mm} 
\hat{W}(\theta_0,\chi_0,d_1,d_2):=
(\theta_0-d_1,\theta_0+d_1)\times (\chi_0-d_2,\chi_0+d_2) \; ,
\label{eq:n2.201aa}
\end{eqnarray}
where 
\begin{eqnarray}
&&  \hspace{-15mm} 
2|\chi_0|T< d_1 \; , \quad  0< d_2 < \chi_0-\chi_{lb}(\eps_0) \; .
\label{eq:n2.201aab}
\end{eqnarray}
Note that the closure, 
$\overline{\hat{W}(\theta_0,\chi_0,d_1,d_2)}=
[\theta_0-d_1,\theta_0+d_1]\times [\chi_0-d_2,\chi_0+d_2]$, of 
$\hat{W}(\theta_0,\chi_0,d_1,d_2)$ is compact and
that, by (\ref{eq:3.27}),(\ref{eq:n2.201an}),
(\ref{eq:n2.201aa}),(\ref{eq:n2.201aab}),
$(\theta_0,\chi_0)\in S\subset\hat{W}(\theta_0,\chi_0,d_1,d_2)\subset
\overline{\hat{W}(\theta_0,\chi_0,d_1,d_2)}\subset W(\eps_0)$. 
Thus, by Proposition \ref{P1} in \S\ref{3.3}, 
the vector field of the ODE's 
(\ref{eq:3.19}),(\ref{eq:3.20})
is $C^\infty$ on $\hat{W}(\theta_0,\chi_0,d_1,d_2)\times\R$.
\item[(5)] (Restriction on $\eps_0$)\\
Choose $\eps_0$ so small that $\chi_{lb}(\eps_0)< -\chi_M -d_2$.
Note that this is made possible since, by (\ref{eq:3.28}), 
\bea
&& \hspace{-8mm} 
\chi_{lb}(\eps_0)\leq -\frac{1}{\eps_0} + 
\frac{1}{\sqrt{{\cal E}}}\sqrt{
1+K^2 \Pi_{x,ub}^2(1)} \; ,
\nonumber
\eea
whence $\chi_{lb}(\eps_0)< -\chi_M -d_2$ if
\bea
&& \hspace{-8mm} 
\eps_0 <  \biggl( \frac{1}{\sqrt{{\cal E}}}\sqrt{ 1+K^2 \Pi_{x,ub}^2(1)}
+\chi_M + d_2\biggr)^{-1} \; .
\label{eq:3.28b} 
\eea
Since the RHS of (\ref{eq:3.28b}) is positive $\eps_0$ can indeed be
chosen sufficiently small.
\item[(6)] (Exact solution in rectangle) \\
Since the vector fields in (\ref{eq:3.19}),(\ref{eq:3.20})
are $C^\infty$, solutions in 
$\hat{W}(\theta_0,\chi_0,d_1,d_2)$
with initial condition $\theta(0,\eps)=\theta_0,\chi(0,\eps)=\chi_0$ exist uniquely in $\hat{W}(\theta_0,\chi_0,d_1,d_2)$
on a maximum forward interval of
existence $[0,\beta(\eps))$.
Here $d_1,d_2$ satisfy (\ref{eq:n2.201aab}).
Either $\beta(\eps)=\infty$ or the solution approaches the boundary of 
$\hat W$ as $\zeta \rightarrow \beta(\eps)-$.
See Chapter 1 of \cite{Hale} for a discussion of existence, uniqueness and
continuation to a maximum forward interval of existence.

For convenience we define $I(\eps,T):=[0,T/\eps]\cap[0,\beta(\eps))$.
\item[(7)] (Lipschitz constants for $f_1,f_2$ on rectangle)\\
Let $L_1,L_2$ be defined by
\begin{eqnarray}
&&\hspace{-5mm} L_1: = \frac{2}{\bar{q}} max_{\zeta\in[0,2\pi]}\; |q(\zeta)| 
= 2 [ 1 + \frac{2K^2}{\bar{q}}|\Delta P_{x0}| +\frac{K^2}{2\bar{q}}] \; ,
\label{eq:2303b} \\
&&\hspace{-5mm} L_2:= \nu K^2 (1 + |\Delta P_{x0}|) \; .
\label{eq:2303d} 
\end{eqnarray}
It follows by (\ref{eq:3.21}),(\ref{eq:3.22}),
(\ref{eq:2303b}),(\ref{eq:2303d}) and for
$\theta_1,\theta_2,\chi_1,\chi_2,\zeta\in\R$, that
\begin{eqnarray}
&&\hspace{-5mm} |f_1(\chi_2,\zeta) - f_1(\chi_1,\zeta)|
\leq  \frac{2|q(\zeta)|}{\bar{q}}  |\chi_2 - \chi_1| \leq  L_1 |\chi_2 - \chi_1| \; ,
\label{eq:2303a} \\
&&\hspace{-5mm} |f_2(\theta_2,\zeta;\nu) - f_2(\theta_1,\zeta;\nu)|
\nonumber\\
&& = K^2 |\cos\zeta + \Delta P_{x0}| \;
|\cos( \nu[\theta_2 - Q(\zeta)])-\cos( \nu[\theta_1 - Q(\zeta)])|
\nonumber\\
&&
\leq 
K^2 (1 + |\Delta P_{x0}|) \;
| \nu[\theta_2 - Q(\zeta)] - \nu[\theta_1 - Q(\zeta)] |
\nonumber\\
&&
= \nu K^2 (1 + |\Delta P_{x0}|) \; |\theta_2 - \theta_1 | 
=  L_2 |\theta_2 - \theta_1| \; ,
\label{eq:2303c} 
\end{eqnarray}
where we have also used the fact that 
$|\cos x-\cos y|\leq |x-y|$.
Thus $L_1,L_2$ are Lipschitz constants for $f_1,f_2$ on 
$\hat{W}(\theta_0,\chi_0,d_1,d_2)$ respectively (in fact
even on $\R^2$).
\item[(8)] (Bounds for $g_1,g_2$ on rectangle) \\
Appendix \ref{C} gives a very detailed derivation of quite explicit minimal bounds for $g_1$ and $g_2$. There we show, for $(\theta,\chi,\zeta)$ in 
$\hat{W}(\theta_0,\chi_0,d_1,d_2)\times\R$,
\begin{eqnarray}
&&\hspace{-5mm}
|g_i(\theta,\chi,\zeta,\eps,\nu)|\le 
C_i(\chi_0,\eps_0,\nu,d_2) \; ,
\label{eq:2297ax} 
\end{eqnarray}
where $i=1,2$ and $d_1,d_2$ satisfy
(\ref{eq:n2.201aab}) and where the 
finite $C_1$ and $C_2$ are defined by (\ref{eq:nC.43n}),(\ref{eq:nC.95n}).
\item[(9)] (Besjes terms) \\
Let $B_1,B_2$ be defined by
\begin{eqnarray}
&&\hspace{-10mm} 
B_1(\zeta):=|\int_0^\zeta \tilde f_1(v_2(s,\eps),s)\,ds| 
=|\int_0^\zeta \tilde f_1(\chi_0,s)\,ds| 
\; , \nonumber\\
&&\hspace{-10mm} 
B_2(\zeta):=|\int_0^\zeta \tilde{f}_2(v_1(s,\eps),s;\nu)ds| 
=|\int_0^\zeta \tilde{f}_2(2\chi_0\eps s+\theta_0,s;\nu)ds| 
\; ,
\nonumber\\
\label{eq:2383}
\end{eqnarray}
where
\begin{eqnarray}
&&\hspace{-10mm} \tilde f_1(v_2,s):=f_1(v_2,s) - \bar{f}_1(v_2) 
= 2(\frac{q(s)}{\bar{q}}-1)v_2
 \; , \nonumber\\
&&\hspace{-10mm} 
\tilde f_2(v_1,s;\nu):=f_2(v_1,s;\nu) - \bar{f}_2(v_1;\nu)  
=f_2(v_1,s;\nu)
\; .
\nonumber\\
\label{eq:2353} 
\end{eqnarray}
In (\ref{eq:2383}) we have used (\ref{eq:3.60}).
We will also need $B_{1,\infty},B_{2,\infty}$  defined by
\begin{eqnarray}
&&\hspace{-10mm} 
B_{i,\infty}(\zeta):=\sup_{s\in[0,\zeta)}\; B_i(s) \; , 
\label{eq:2383n}
\end{eqnarray}
for $i=1,2$. 

We refer to $B_1,B_2$ as ``Besjes terms'' and their
importance will be seen both in the bounds presented in Theorem \ref{T1} and in the proof of the theorem  where they eliminate the need for a near identity transformation (for the latter, see
\cite{M1,LM,SVM,M2,Sanders}).
\end{itemize}

With this setup we can now state the NR approximation theorem.

\setcounter{theorem}{0}
\begin{theorem}(Averaging theorem in $\Delta$-NR case:
$\nu \in [k+\Delta, k+1-\Delta]$,
$k=0,1,...,0<\Delta<0.5$) \label{T1}\\

\noindent
With the setup given by items 1-9 of the above 
preamble we obtain, for $\zeta\in I(\eps,T)$, that 
\begin{eqnarray}
&&\hspace{-5mm}
|\theta(\zeta,\eps) - 2\chi_0\eps\zeta-\theta_0|
=O(\eps/\Delta) \; , \quad
|\chi(\zeta,\eps)- \chi_0| =O(\eps/\Delta) \; .
\label{eq:2383aa}
\end{eqnarray}
More precisely
\begin{eqnarray}
&&\hspace{-5mm}|\theta(\zeta,\eps)- 2\chi_0 \eps\zeta -\theta_0| \le 
\eps \biggl( [B_{1,\infty}(T/\eps)+C_1T] \cosh(T\sqrt{L_1 L_2})
\nonumber\\
&& +[B_{2,\infty}(T/\eps)+C_2T] \sqrt{\frac{L_1}{L_2}}\sinh(T\sqrt{L_1 L_2})\biggr)
\; , \label{eq:204} \\
&& \hspace{-5mm}|\chi(\zeta,\eps)- \chi_0| \le 
\eps \biggl( 
[B_{1,\infty}(T/\eps)+C_1T]\sqrt{\frac{L_2}{L_1}}\sinh(T\sqrt{L_1 L_2})
\nonumber\\
&&
+ [B_{2,\infty}(T/\eps)+C_2T] \cosh(T\sqrt{L_1 L_2})\biggr)\; .
\label{eq:222} 
\end{eqnarray}
Moreover
%
\begin{eqnarray}
&& B_{1,\infty}(T/\eps)\le \check B_1 \; ,  \quad  
 B_{2,\infty}(T/\eps)\le \check B_2(T,\Delta) \; ,
\label{eq:2383a}
\end{eqnarray}
where $i=1,2$ and the $\check{B}_1,\check B_2(T,\Delta)
\in[0,\infty)$ are finite, $\eps$-independent and are defined in terms
of our basic parameters and initial conditions by
\begin{eqnarray}
&&  \hspace{-12mm} \check B_1:=
 \frac{2K^2|\chi_0|}{\bar{q}} ( 2|\Delta P_{x0}| + \frac{1}{4}) \; ,
\label{eq:222a} \\
&& \hspace{-12mm}
\check B_2(T,\Delta):= \frac{1}{\Delta}\check B_{21}(T)+\check B_{22}(T) \; ,
\label{eq:3900dcn} \\
&& \hspace{-12mm} \check B_{21}(T):=
2K^2[ 1 + (k+1) |\chi_0|T]
\biggl( |\widehat{jj}(k;\nu,\Delta P_{x0})| + 
 |\widehat{jj}(k+1;\nu,\Delta P_{x0})| \biggr)
\; ,
\nonumber\\
\label{eq:3900cbn} \\
&& \hspace{-12mm} \check B_{22}(T):= 
2K^2\biggl(1 + (k+1) |\chi_0|T\biggr)\sum_{n\in(\Z\setminus\lbrace k,k+1\rbrace)} 
|\widehat{jj}(n;\nu,\Delta P_{x0})| \; .
\label{eq:3900cn} 
\end{eqnarray}
Furthermore, for $\eps_0$ sufficiently small, $(\theta(\zeta,\eps),
\chi(\zeta,\eps))$ 
stays away from the boundary of the rectangle 
$\hat{W}(\theta_0,\chi_0,d_1,d_2)$ for $\zeta\in I(\eps,T)$. 
Thus the ODE continuation theorem (see \cite[Section 1.2]{Hale}) gives
$\beta(\eps)>T/\eps$, hence $I(\eps,T)=[0,T/\eps]$.
\end{theorem}
The proof of Theorem \ref{T1} is presented in \S\ref{4.1}.
Note that the symbol $O(\eps/\Delta)$ conveys that the error
contains the factor $\frac{1}{\Delta}$.
\subsubsection{NtoR case:  $\nu=k+\eps a$ (Periodic Averaging)}
\label{3.5.2}
The NtoR case was defined in \S\ref{3.2}.
The exact ODE's to be analyzed in this case were
derived in \S\ref{3.4} and are given by
(\ref{eq:430011}),(\ref{eq:430010})
with initial conditions $\theta(0,\eps)=\theta_0,\chi(0,\eps)=\chi_0$ 
and where $g^R_1,g^R_2$ are defined by
(\ref{eq:n2.101aaca}),(\ref{eq:n2.101aada})
and $f^R_1,f^R_2$ by (\ref{eq:430010an}),(\ref{eq:430011an}).
The normal form ODE's are (\ref{eq:nn430016}),(\ref{eq:nn430015})
with initial conditions $v_1(0,\eps)=
\theta_0,v_2(0,\eps)=\chi_0$ solved by
(\ref{eq:n3.4}),(\ref{eq:n3.5}).
where $X,Y$ satisfy
the standard pendulum equations (\ref{eq:237}) with the initial
conditions (\ref{eq:237a}).

The setup for the theorem is as follows.
\begin{itemize}
\item[(1)] (Basic parameters)\\
Let $0<\eps \leq \eps_0\leq 1$, 
$a\in[-1/2,1/2]$ and $k$ be a positive integer.
\item[(2)] (Initial data) \\
Choose $\theta_0,\chi_0$ such that
$(\theta_0,\chi_0)\in (\R\times [-\chi_M,\chi_M])$
where $\chi_M>0$ is chosen such that $-\chi_M>\chi_{lb}(\eps_0)$.
Clearly $(\R\times [-\chi_M,\chi_M])\subset W(\eps_0)$.
\item[(3)] (Guiding solution) \\
Choose $T>0$ and define the compact subset 
$S_R:=\{ \vb(\tau,1): \tau\in[0,T]\}$ of $W(\eps_0)$ where 
$\vb=(v_1,v_2)$ with $v_1,v_2$ given by 
(\ref{eq:n3.4}),(\ref{eq:n3.5}).
Note that $S_R\subset W(\eps_0)$ holds for arbitrary $T>0$ if
\begin{eqnarray}
&& \hspace{-10mm} 
\chi_{lb}(\eps_0) < \chi_0 - d_2^{min}(\theta_0,\chi_0,k,a)
\label{eq:430031aa} 
\end{eqnarray}
since $|v_2(\tau,1)-\chi_0|\leq  d_2^{min}(\theta_0,\chi_0,k,a)$
where $d_2^{min}$ is defined in \S\ref{3.4.3}.
\item[(4)] (Rectangle around initial value $(\theta_0,\chi_0)$:
the basic domain for averaging theorem) \\
Define an open 
rectangle $\hat{W}_{R}(\theta_0,\chi_0,d_1,d_2)$ around $S_R$ by
\begin{eqnarray}
&&  \hspace{-15mm} 
\hat{W}_{R}(\theta_0,\chi_0,d_1,d_2):=
(\theta_0-d_1,\theta_0+d_1)\times (\chi_0-d_2,\chi_0+d_2) \; ,
\label{eq:n2.201aaa}
\end{eqnarray}
where $d_1,d_2$ satisfy
\begin{eqnarray}
&&  \hspace{-18mm} 
0 \leq d_1^{min}(\theta_0,\chi_0,T,k,a) < d_1 \; ,
\label{eq:n2.201aaan} \\
&&  \hspace{-18mm} 
0 \leq  d_2^{min}(\theta_0,\chi_0,k,a) < d_2 
< \chi_0 - \chi_{lb}(\eps_0) \; ,
\label{eq:n2.201aan} 
\end{eqnarray}
with $d_1^{min},d_2^{min}$ defined in \S\ref{3.4.3}.
Note that (\ref{eq:n2.201aan}) entails (\ref{eq:430031aa}).
Note also that, by (\ref{eq:n3.5b}),(\ref{eq:n2.201aaan}),
(\ref{eq:n2.201aan}),
\begin{eqnarray}
&&  \hspace{-10mm} |v_1(\tau,1)-\theta_0| 
\leq d^{min}_1(\theta_0,\chi_0,\tau,k,a) 
\leq d^{min}_1(\theta_0,\chi_0,T,k,a)
< d_1 \; , 
\nonumber\\
&& \hspace{-10mm} 
|v_2(\tau,1) -\chi_0| \leq d^{min}_2(\theta_0,\chi_0,k,a) < d_2
\; ,
\nonumber\\
\label{eq:n3.5en}
\end{eqnarray}
where we also used that
$d^{min}_1(\theta_0,\chi_0,\tau,k,a)$ is increasing w.r.t. $\tau$.
It follows from (\ref{eq:n2.201aaa}),(\ref{eq:n3.5en}) that
$(\theta_0,\chi_0)\in S_R\subset\hat{W}_{R}(\theta_0,\chi_0,d_1,d_2)$ and, by
(\ref{eq:3.27}),(\ref{eq:430031aa})
that $\overline{\hat{W}_{R}(\theta_0,\chi_0,d_1,d_2)}
\subset W(\eps_0)$. 
Thus, by Proposition \ref{P2} in \S\ref{3.4}, 
the vector field of the ODE's 
(\ref{eq:430011}),(\ref{eq:430010})
is of class $C^\infty$ on $\hat{W}_R(\theta_0,\chi_0,d_1,d_2)\times\R$. Note that
the closure, $\overline{\hat{W}_{R}(\theta_0,\chi_0,d_1,d_2)}
=[\theta_0-d_1,\theta_0+d_1]\times [\chi_0-d_2,\chi_0+d_2]$, of 
$\hat{W}_{R}(\theta_0,\chi_0,d_1,d_2)$ is compact.
\item[(5)] (Restriction on $\eps_0$)\\
Choose $\eps_0$ so small that 
$\chi_{lb}(\eps_0)< -\chi_M -d_2$. Recall from item 5 of the preamble
to Theorem \ref{T1} that such a choice is always possible.
\item[(6)] (Exact solution in rectangle) \\
Since the vector fields in 
(\ref{eq:430011}),(\ref{eq:430010})
are $C^\infty$, solutions in 
$\hat{W}(\theta_0,\chi_0,d_1,d_2)$
with initial condition $\theta(0,\eps)=\theta_0,\chi(0,\eps)=\chi_0$ exist uniquely on a maximum forward interval of
existence $[0,\beta(\eps))$.
Here $d_1,d_2$ satisfy (\ref{eq:n2.201aaan}),(\ref{eq:n2.201aan}).
Either $\beta(\eps)=\infty$ or the solution approaches the boundary of $\hat W$ as $\zeta\rightarrow \beta(\eps)-$.
See Chapter 1 of \cite{Hale} for a discussion of existence, uniqueness and
continuation to a maximum forward interval of existence.

It is convenient to introduce $I(\eps,T):=[0,T/\eps]\cap[0,\beta(\eps))$.

\item[(7)] (Lipschitz constants for $f_1^R,f_2^R$ on rectangle)\\
Let $L^R_1,L^R_2$ be defined by
\begin{eqnarray}
&&\hspace{-5mm} L^R_1: = L_1 
= 2 [ 1 + \frac{2K^2}{\bar{q}}|\Delta P_{x0}| +\frac{K^2}{2\bar{q}}] \; ,
\label{eq:2303bn} \\
&&\hspace{-5mm} L^R_2:= K^2 k (1 + |\Delta P_{x0}|) \; ,
\label{eq:2303dn} 
\end{eqnarray}
where we have also used (\ref{eq:2303b})
and where $d_1,d_2$ satisfy (\ref{eq:n2.201aaan}),(\ref{eq:n2.201aan}).
It follows by (\ref{eq:430011an}),(\ref{eq:430010aaa}),
(\ref{eq:2303a}),(\ref{eq:2303bn}),(\ref{eq:2303dn}) and, for
$\theta_1,\theta_2,\chi_1,\chi_2,\zeta\in\R$,
\begin{eqnarray}
&&\hspace{-5mm} 
|f^R_1(\chi_2,\zeta) - f^R_1(\chi_1,\zeta)| 
= |f_1(\chi_2,\zeta) - f_1(\chi_1,\zeta)|
\nonumber\\
&&\leq   L_1 |\chi_2 - \chi_1| =  L^R_1 |\chi_2 - \chi_1| \; ,
\label{eq:2303an} \\
&&\hspace{-5mm} 
|f^R_2(\theta_2,\eps\zeta,\zeta;k,a) - f^R_2(\theta_1,\eps\zeta,\zeta;k,a)| 
\nonumber\\
&& = K^2 |\cos\zeta+ \Delta P_{x0}|
\Big{|} \cos\biggl( k[\theta_2-\zeta
-\Upsilon_0\sin\zeta -\Upsilon_1\sin 2\zeta] - \eps a\zeta\biggr)
\nonumber\\
&&\quad
- \cos\biggl( k[\theta_1-\zeta
-\Upsilon_0\sin\zeta -\Upsilon_1\sin 2\zeta] - \eps a\zeta\biggr)\Big{|}
\nonumber\\
&& \leq  k K^2 (1 + |\Delta P_{x0}|) |\theta_2-\theta_1|
= L^R_2 |\theta_2 - \theta_1|
\; ,
\label{eq:2303cn} 
\end{eqnarray}
where we have also used the fact that 
$|\cos x-\cos y|\leq |x-y|$.
Thus $L^R_1,L^R_2$ are Lipschitz constants for 
$f^R_1,f^R_2$ on $\hat{W}_{R}(\theta_0,\chi_0,d_1,d_2)$ 
(in fact even on $\R^2$).
\item[(8)] (Bounds for $g_1^R,g_2^R$ on rectangle)\\ 
Appendix \ref{E} gives a very detailed derivation of quite explicit minimal bounds for $g_1^R$ and $g_2^R$. 
There we show  that, for $(\theta,\chi,\zeta)\in 
\hat{W}_{R}(\theta_0,\chi_0,d_1,d_2)\times \R$,
\begin{eqnarray}
&&\hspace{-5mm} 
|g^R_1(\theta,\chi,\zeta,\eps,k,a)|\le 
C_1^R(\chi_0,\eps_0,d_2)\; , 
\nonumber\\
&& \hspace{-5mm}
|g^R_2(\theta,\chi,\zeta,\eps,k,a)|\le 
C_2^R(\theta_0,\chi_0,\eps_0,a,d_1,d_2)
\; , 
\nonumber\\
\label{eq:2297axn} 
\end{eqnarray}
where $i=1,2$ and $d_1,d_2$ satisfy
(\ref{eq:n2.201aaan}),(\ref{eq:n2.201aan}) and where the 
finite $C_1^R$ and $C_2^R$ are defined by (\ref{eq:nE.10b}),(\ref{eq:nE.80}).
\item[(9)] (Besjes terms) \\
Let $B^R_1,B^R_2$ be defined by
\begin{eqnarray}
&&\hspace{-10mm} 
B^R_1(\zeta):=|\int_0^\zeta \tilde{f}_1^R(v_2(s,\eps),s)\,ds| \; , 
\nonumber\\
&& \hspace{-10mm}
B^R_2(\zeta):=|\int_0^\zeta \tilde{f}_2^R(v_1(s,\eps),\eps s,s;k,a)ds| \; ,
\nonumber\\
\label{eq:x2383}
\end{eqnarray}
where
\begin{eqnarray}
&&\hspace{-10mm} \tilde{f}_1^R(\chi,s):=f^R_1(\chi,s) 
- \bar{f}^R_1(\chi)  \; , 
\nonumber\\
&& \hspace{-10mm} \tilde{f}_2^R(\theta,\eps s,s;k,a):=
f^R_2(\theta,\eps s,s;k,a) 
- \bar{f}^R_2(\theta,\eps s;k) 
 \; .
\nonumber\\
\label{eq:x2353} 
\end{eqnarray}
We will also need $B^R_{1,\infty},B^R_{2,\infty}$ defined by
\begin{eqnarray}
&&\hspace{-10mm} 
B^R_{i,\infty}(\zeta):=\sup_{s\in[0,\zeta)}\; B^R_i(s) \; , 
\label{eq:x2383n}
\end{eqnarray}
where $i=1,2$. 

We refer to $B_1^R,B_2^R$ as ``Besjes terms'' and their
importance will be seen both in the bounds presented in Theorem \ref{T2} and in the proof of the theorem  where they eliminate the need for a near identity transformation.
\end{itemize}

With this setup we can now state the NtoR approximation theorem.

\setcounter{theorem}{1}
\begin{theorem}(Averaging theorem in NtoR case: $\nu=k+\eps a,
0<\eps \leq \eps_0,\\k\in\N,|a|\leq 0.5$) \label{T2}

\noindent
With the setup given by items 1-9 of the above 
preamble we obtain, for $\zeta\in I(\eps,T)$, that 
\begin{eqnarray}
&&\hspace{-5mm}
|\theta(\zeta,\eps) - v_1(\zeta,\eps)|
=O(\eps) \; , \qquad
|\chi(\zeta,\eps)- v_2(\zeta,\eps)| =O(\eps) \; .
\nonumber
\end{eqnarray}
More precisely
\begin{eqnarray}
&&\hspace{-5mm}|\theta(\zeta)-v_1(\zeta,\eps)| \le 
\eps \biggl( [B_{1,\infty}^R(T/\eps)+C_1^RT] \cosh(T\sqrt{L_1^R L_2^R})
\nonumber\\
&& +[B_{2,\infty}^R(T/\eps)+C_2^RT] 
\sqrt{\frac{L_1^R}{L_2^R}}\sinh(T\sqrt{L_1^R L_2^R})\biggr)
\; , \label{eq:x204} \\
&& \hspace{-5mm}|\chi(\zeta)-v_2(\zeta,\eps)| \le 
\eps \biggl( 
[B_{1,\infty}^R(T/\eps)+C_1^RT]\sqrt{\frac{L_2^R}{L_1^R}}\sinh(T\sqrt{L_1^R L_2^R})
\nonumber\\
&&
+ [B_{2,\infty}^R(T/\eps)+C_2^R T] \cosh(T\sqrt{L_1^R L_2^R})\biggr)\; .
\label{eq:x222} 
\end{eqnarray}
Moreover
\begin{eqnarray}
&& B_{i,\infty}^R(T/\eps)\le \check B_i^R(T) \; ,
\label{eq:x2383a}
\end{eqnarray}
where $i=1,2$ and $\check{B}_i^R(T)\in[0,\infty)$ 
are independent of $\eps$ and defined by
\begin{eqnarray}
&&  \hspace{-12mm} \check{B}_1^R(T):=
\frac{2K^2}{\bar{q}} [2|\Delta P_{x0}| + \frac{1}{4}]
\biggl(  \chi_\infty(\theta_0,\chi_0,k,a) \nonumber\\
&& 
+  K^2 T |\widehat{jj}(k;k,\Delta P_{x0})|\biggr)\; ,
\label{eq:x222a} \\
&& \hspace{-12mm}
\check{B}_2^R(T):= K^2 \biggl( 2 +  
T\; [ |a| + 2k \chi_\infty(\theta_0,\chi_0,k,a) ]\biggr)
\nonumber\\
&&\times \sum_{n\in\Z\setminus\lbrace k\rbrace}
\frac{|\widehat{jj}(n;k,\Delta P_{x0})|}{|n-k|}\; .
\label{eq:x3900dcn} 
\end{eqnarray}
Furthermore, there exists an $0<\eps_0\leq 1$ such that for 
$0<\eps\le \eps_0$, $(\theta(\zeta,\eps),\chi(\zeta,\eps))$ 
stays away from the boundary of the rectangle 
$\hat{W}_{R}(\theta_0,\chi_0,d_1,d_2)$ for $\zeta\in I(\eps,T)$.
Thus the ODE continuation theorem (see \cite[Section 1.2]{Hale}) 
gives $\beta(\eps)>T/\eps$, hence $I(\eps,T)=[0,T/\eps]$.
\end{theorem}
The proof of Theorem \ref{T2} is presented in \S\ref{4.2}.
\subsubsection{Remarks on the averaging theorems}
\label{3.5.3}
\begin{itemize}
\item[(1)] We have now explored the $\theta,\chi$ dynamics as a function
of $\nu$ in the $\Delta$-NR case and $\nu=k+\eps a$
in the NtoR case. However asymptotically there are gaps for
$\nu\in(k+\eps a,k+\Delta)$ when $\eps$ is small. 
For $\Delta=O(\eps)$ the NR normal form
breaks down because the error is $O(1)$, however we can come close
to the NtoR neighborhood by letting $\Delta=O(\eps^\beta)$ with
$\beta$ near $1$ however the error in the NR normal form does
deteriorate to $O(\eps^{1-\beta})$. It could be
interesting to explore the dynamics in these gaps.
\item[(2)] Important for the functioning of the FEL is knowledge
of the fraction of the bunch that occupies a bucket. From the analysis
in \S\ref{3.4.3} this occurs for ICs in the libration case, i.e.,
$0<{\cal E}_{Pen}(Z_0)<2$ where $Z_0$ is given in (\ref{eq:237c}). 
One can thus determine the set of $(\theta_0,\chi_0)$ for
which $Z_0$ occupies the pendulum buckets. 
For more details on the pendulum motion and its impact on
the low gain theory 
see \S\ref{3.7}.
\item[(3)] Mathematically we want to make sure the buckets are covered by our
domain $W(\eps_0)\times\R$ for physically reasonable $\chi_0$.
From (\ref{eq:n3.5}) the range of the $v_2$-values
in the buckets for the NtoR normal form is the interval 
$(-\frac{\Omega}{k}+\frac{a}{2k},\frac{\Omega}{k}+\frac{a}{2k})$.
Now $a\geq -1/2$ so, for every $k$, the smallest $v_2$ in a bucket is
$-\frac{\Omega}{k}-\frac{1}{4k}$ whence, since
$k\geq 1$, the very smallest $v_2$ in a bucket is $-\Omega-1/4$. Thus requiring 
\begin{eqnarray}
&& \chi_b:=-\Omega-\frac{1}{4} < 0 \; ,
\label{eq:234b}
\end{eqnarray}
entails that $\chi_b$ is smaller than any $\chi$-value 
inside the buckets and smaller than any $\chi$-value on the separatrix.
It is plausible to restrict the physically interesting $\chi$-values
to be greater than, say $3\chi_b$. 
The condition that $(\theta,3\chi_b)\in W(\eps_0)$
entails that the buckets are covered by $W(\eps_0)$ and 
that $\eps_0$ satisfies the constraint $3\chi_b>\chi_{lb}(\eps_0)$.
The following proposition is a corollary to Propositions \ref{P1},\ref{P2}.

\setcounter{proposition}{2}
\begin{proposition} \label{P3}
Let $0<\eps \leq \eps_0$ where $0<\eps_0\leq 1$ and $\nu\in[1/2,\infty)$.
Let also $\Delta\gamma$ be a positive constant 
and let
\begin{eqnarray}
&& \eps_0 < \sqrt{\cal E}\biggl( \Delta\gamma 
+ \sqrt{1+K^2 \Pi_{x,ub}^2(1)}\biggr)^{-1}
\; .
\label{eq:x3222cixa} 
\end{eqnarray}
If $\chi\in\R$ satisfies the condition:
\bea
&&\hspace{-8mm} 
1\leq \gamma_c-\Delta\gamma\leq \gamma_c(1+\eps\chi) \leq
\gamma_c+\Delta\gamma \; ,
\label{eq:222cbx} 
\eea
then  
\bea
&&\hspace{-8mm} 
\chi> \chi_{lb}(\eps_0) \; .
\label{eq:xx222cbx} 
\eea
In other words if $\eps_0$ satisfies (\ref{eq:x3222cixa}) then
the $\gamma$ values in $[\gamma_c-\Delta\gamma,\gamma_c+\Delta\gamma]$
are covered by $W(\eps_0)$.
\end{proposition}

\noindent

The proposition guarantees, by choosing a sufficiently small
$\eps_0$, that the domain $W(\eps_0)\times\R$ 
is large enough to contain 
the physical relevant values of $\theta,\chi,\zeta$. 
\\

\noindent {\em Proof of Proposition \ref{P3}:} 
Let $\chi\in\R$ satisfy (\ref{eq:222cbx}). Then, by 
(\ref{eq:1.13}),
$\chi\in[-\frac{1}{\sqrt{\cal E}}\Delta\gamma,
\frac{1}{\sqrt{\cal E}}\Delta\gamma]$ whence, by
(\ref{eq:3.26}),(\ref{eq:3.28}),(\ref{eq:x3222cixa}), 
\bea
&&\hspace{-8mm} 
\chi_{lb}(\eps_0) 
=-\frac{1}{\eps_0} + \frac{1}{\sqrt{{\cal E}}}\sqrt{1+K^2 \Pi_{x,ub}^2(\eps_0)} 
\nonumber\\
&& \leq -\frac{1}{\eps_0} + \frac{1}{\sqrt{{\cal E}}}
\sqrt{1+K^2 \Pi_{x,ub}^2(1)} 
< -\frac{1}{\sqrt{\cal E}}\Delta\gamma\leq \chi
 \; ,
\nonumber
\eea
which entails (\ref{eq:xx222cbx}).
\hfill $\Box$ 

Note that the condition: $1\leq \gamma_c-\Delta\gamma$ in
(\ref{eq:222cbx}) is not used in the proof of Proposition \ref{P3}
but serves to guarantee that $\chi$ satisfies the physical condition:
$\gamma\geq 1$, i.e.,
$1\leq \gamma_c(1+\eps\chi)$.
\item[(4)]
In applications of Theorems \ref{T1},\ref{T2}, 
$T$ should be chosen so that $z\in [0,T/\eps k_u]$ is the 
domain of interest, e.g., so that $T/(\eps k_u)$ is the length of the 
undulator.
\item[(5)] 
In many discussions of this nature, researchers often just assert the existence of bounds, for example by using the well known fact that a continuous function on a compact set is bounded, or bounds are obtained which are crude. Here we wanted to do more. By using, in the proofs of Theorems \ref{T1} and \ref{T2},
a system of differential inequalities instead of the Gronwall inequality we have been able to use two Lipschitz constants in each proof instead of their maximum and in a similar manner can treat the two Besjes' terms independently as well as the components of $g$ and $g^R$. Furthermore, we believe the Besjes bounds and the bounds on $g_1,g_2,g_1^R,g_2^R$ are nearly optimal. 

We also note that there are only 3 restrictions on the size of $\eps_0$ and thus $\eps$. The first is that we require $\eps_0\le 1$. But this is only a matter of convenience and is really no restriction at all since the averaging theorems are only useful for $\eps$ small. The second restriction is in item 5 of the preambles to the two theorems, however as indicated there this is not a significant restriction. Thus the only real restriction is keeping the solution away from the boundary of $\hat W,\hat W_R$ in order to obtain $I(\eps,T)=[0,T/\eps]$. 
This is an optimization problem; by making 
$\hat W,\hat W_R$ larger, $\eps$ can be larger, however this is compensated to some extent in the Lipschitz constants as well as the bounds on 
$g_1,g_2,g_1^R,g_2^R$ which 
would become larger. Nonetheless, the situation is quite good in comparison to say KAM or Nekhoroshev theorems (see e.g., \cite{HPT}), where the restrictions on $\eps$ are quite severe and it is with great effort that the restrictions on $\eps$ have been improved in some applications, e.g., solar system problems.

\item[(6)] 
We here clarify the contributions of $\widehat{jj}$ to the
error bounds of Theorems \ref{T1} and \ref{T2} by finding simple
upper bounds for $\check B_{21}(T),\check{B}_1^R(T),\check B_{22}(T)$ and
$\check{B}_2^R(T)$. First of all we note from (\ref{eq:3.33}) and 
(\ref{eq:3.40}) that 
\begin{eqnarray}
&& \hspace{-12mm} 
|\widehat{jj}(n;\nu,\Delta P_{x0})| \leq 1+|\Delta P_{x0}| \; ,
\label{eq:3.99} 
\end{eqnarray}
where $\nu\geq 1/2$. Clearly (\ref{eq:3.99}) gives upper bounds for
$\check B_{21}(T),\check{B}_1^R(T)$ in (\ref{eq:3900cbn}),(\ref{eq:x222a}).
Secondly, we obtain from the Cauchy-Schwarz inequality that
\bea
&& \hspace{-10mm} 
\sum_{0\neq n\in\Z}\;|\widehat{jj}(n;\nu,\Delta P_{x0})|
=\sum_{0\neq n\in\Z}\;\frac{1}{|n|}|n|\;
|\widehat{jj}(n;\nu,\Delta P_{x0})|
\nonumber\\
&& \hspace{-5mm}  \leq 
\biggl( \sum_{0\neq n\in\Z}\;n^2|\widehat{jj}(n;\nu,\Delta P_{x0})|^2
\biggr)^{1/2} \biggl( \sum_{0\neq n\in\Z}\;\frac{1}{n^2}\biggr)^{1/2}
\nonumber\\
&& 
=\frac{\pi}{\sqrt{3}}
\biggl( \sum_{0\neq n\in\Z}\;n^2|\widehat{jj}(n;\nu,\Delta P_{x0})|^2
\biggr)^{1/2} 
\; ,
\label{eq:3.41a}
\eea
where the finiteness of the rhs follows from the fact that the function
$jj(\cdot;\nu,\Delta P_{x0})$ is of class $C^\infty$. Since
$jj(\cdot;\nu,\Delta P_{x0})$ is also $2\pi$-periodic we can apply Parseval's
theorem to get
\bea
&& \hspace{-5mm} 
\frac{1}{2\pi}
\int_{[0,2\pi]} \; d\zeta 
|\frac{d}{d\zeta} jj(\zeta;\nu,\Delta P_{x0})|^2
= \sum_{0\neq n\in\Z}\;n^2|\widehat{jj}(n;\nu,\Delta P_{x0})|^2 \; .
\nonumber\\
\label{eq:3.41b}
\eea
It also follows from (\ref{eq:3.33}) that
\begin{eqnarray}
&&  \hspace{-10mm}  
\frac{d}{d\zeta}jj(\zeta;\nu,\Delta P_{x0})
=  -\exp(-i\nu[\Upsilon_0\sin\zeta+\Upsilon_1\sin 2\zeta]) 
\biggl( \sin\zeta 
\nonumber\\
&&  +i\nu(\cos\zeta+ \Delta P_{x0})
[\Upsilon_0\cos\zeta+2\Upsilon_1\cos 2\zeta]\biggr) \; ,
\nonumber
\end{eqnarray}
whence
\begin{eqnarray}
&&  \hspace{-15mm}  
|\frac{d}{d\zeta}jj(\zeta;\nu,\Delta P_{x0})|^2
\leq 1 + \nu^2 (1 + |\Delta P_{x0}|)^2[|\Upsilon_0|+2\Upsilon_1]^2 \; ,
\nonumber
\end{eqnarray}
so that, by (\ref{eq:3.41a}),(\ref{eq:3.41b}),
\bea
&& \hspace{-10mm} 
\sum_{0\neq n\in\Z}\;|\widehat{jj}(n;\nu,\Delta P_{x0})|
\leq \frac{\pi}{\sqrt{3}}
\biggl( 1 + \nu^2 (1 + |\Delta P_{x0}|)^2[|\Upsilon_0|+2\Upsilon_1]^2
\biggr)^{1/2} 
\; ,
\nonumber\\
\label{eq:3.41c}
\eea
which entails, by (\ref{eq:3.99}),
\begin{eqnarray}
&& \hspace{-12mm} 
\sum_{n\in(\Z\setminus\lbrace k,k+1\rbrace)} 
|\widehat{jj}(n;\nu,\Delta P_{x0})| 
\leq 1 + |\Delta P_{x0}| +
\sum_{0\neq n\in\Z}\;|\widehat{jj}(n;\nu,\Delta P_{x0})|
\nonumber\\
&& \leq  1 + |\Delta P_{x0}| 
+ \frac{\pi}{\sqrt{3}}
\biggl( 1 + \nu^2 (1 + |\Delta P_{x0}|)^2[|\Upsilon_0|+2\Upsilon_1]^2
\biggr)^{1/2} \Biggr) \; .
\nonumber\\
\label{eq:3.41d} 
\end{eqnarray}
Clearly (\ref{eq:3.41d}) gives an upper bound for
$\check B_{22}(T)$ in (\ref{eq:3900cn}).
Moreover, by (\ref{eq:3.99}),(\ref{eq:3.41c}),
\begin{eqnarray}
&&  \hspace{-12mm} 
\sum_{n\in\Z\setminus\lbrace k\rbrace}
\frac{|\widehat{jj}(n;k,\Delta P_{x0})|}{|n-k|}
\leq |\widehat{jj}(0;k,\Delta P_{x0})|
+ \sum_{0\neq n\in\Z}\;|\widehat{jj}(n;k,\Delta P_{x0})|
\nonumber\\
&&\leq
1 + |\Delta P_{x0}| +
 \frac{\pi}{\sqrt{3}}
\biggl( 1 + \nu^2 (1 + |\Delta P_{x0}|)^2[|\Upsilon_0|+2\Upsilon_1]^2
\biggr)^{1/2} 
\; ,
\nonumber
\end{eqnarray}
which gives an upper bound for $\check{B}_2^R(T)$ in 
(\ref{eq:x3900dcn}).
\end{itemize}
\subsection{Approximation for the phase space variables in 
(\ref{eq:2.25})-(\ref{eq:2.18})}
\label{3.6}
Here we discuss the approximate solutions of
(\ref{eq:2.25})-(\ref{eq:2.18}) and
(\ref{eq:2.27}) in terms of the normal form approximations
given in (\ref{eq:3.60}),(\ref{eq:n3.4}),(\ref{eq:n3.5}), namely
\begin{eqnarray}
&& \hspace{-12mm} \theta_{NF}(\tau) 
:=\left\{ \begin{array}{ll}  
2\chi_0\tau + \theta_0 & \;\; {\rm NR\;case} \\ 
\biggl( X(\Omega\tau;Z_0) - {\rm sgn}(K_0(k))\pi/2 + a\tau \biggr)/k
&\;\;{\rm NtoR\;case} \; ,  \end{array} 
                  \right.
\nonumber\\
\label{eq:430112ab}
\end{eqnarray}
and
\begin{eqnarray}
&& \chi_{NF}(\tau) 
:=\left\{ \begin{array}{ll}  
 \chi_0  & \;\; {\rm NR\;case} \\ 
\biggl( \Omega Y(\Omega\tau;Z_0)+a\biggr)/2k 
&\;\;{\rm NtoR\;case} \; , \end{array} 
                  \right.
\label{eq:430112a}
\end{eqnarray}
where $K_0$ is given in (\ref{eq:nn430015x}) and $\Omega$ in 
(\ref{eq:n3.3}). Recall from Theorems \ref{T1} and \ref{T2} that
\begin{eqnarray}
&& \theta(\zeta,\eps) = \theta_{NF}(\eps\zeta)  + O(\eps) \; ,
\label{eq:3.200} \\
&& \chi(\zeta,\eps) = \chi_{NF}(\eps\zeta)  + O(\eps) \; ,
\label{eq:3.201} 
\end{eqnarray}
for $\zeta\in I(\eps,T)$. 
From (\ref{eq:1.11}),(\ref{eq:2.67}),(\ref{eq:2.64}),(\ref{eq:2.80})
\begin{eqnarray}
&& \hspace{-10mm}
\theta(\zeta,\eps) = \frac{2{\cal E}}{\eps^2\bar{q}}
\biggl( \zeta - k_u ct(\zeta/k_u)\biggr) + Q(\zeta) \; ,
\label{eq:3.205}
\end{eqnarray}
and from (\ref{eq:2.39})
\begin{eqnarray}
\gamma(\zeta/k_u)=\gamma_c(1+\eps\chi(\zeta,\eps)) \; .
\label{eq:3.206}
\end{eqnarray}

Now we can determine the approximate solution of
(\ref{eq:2.25})-(\ref{eq:2.18}) and (\ref{eq:2.27}).
From (\ref{eq:3.200}),(\ref{eq:3.205}) the arrival time,
$t(z)$, of a particle at $z$ is given by
\begin{eqnarray}
&& \hspace{-10mm}
t(z) = \frac{z}{c} -\frac{\eps^2\bar{q}}{2{\cal E}k_u c}
\biggl( \theta_{NF}(\eps k_u z) - Q(k_u z) + O(\eps)\biggr) \; .
\label{eq:3.210}
\end{eqnarray}
Furthermore from (\ref{eq:1.13}),(\ref{eq:3.201}),(\ref{eq:3.206}) 
the energy in (\ref{eq:2.27}) is given by
\begin{eqnarray}
\gamma(z) = \sqrt{\cal E}(\frac{1}{\eps}+\chi_{NF}(\eps k_u z) +O(\eps)) \; ,
\label{eq:3.212}
\end{eqnarray}
and is clearly slowly varying.
From (\ref{eq:2.36}),(\ref{eq:3.10}),(\ref{eq:3.11}) we have 
\begin{eqnarray}
&& p_x(z) = mcK [\cos(k_u z) +  \Delta P_{x0} + O(\eps^2)]   \; .
\label{eq:3.215}
\end{eqnarray}
It is tedious but straightforward to derive from
(\ref{eq:1.13}),(\ref{eq:2.36}),(\ref{eq:3.10}),(\ref{eq:3.11}),
(\ref{eq:3.212})
\begin{eqnarray}
&& \hspace{-8mm}
p_z(z) =  mc\sqrt{{\cal E}}\biggl(\frac{1}{\eps} +\chi_{NF}(\eps k_uz)
 + O(\eps) \biggr) 
\; .
\label{eq:3.220} 
\end{eqnarray}
Finally we can now determine $x(z)$. From (\ref{eq:2.25}),(\ref{eq:3.215}) and
(\ref{eq:3.220})
\begin{eqnarray}
&& \hspace{-8mm} \frac{d}{dz}x(z)
= \frac{p_x(z)}{p_z(z)}
\nonumber\\
&& = \Biggl(  mcK [\cos(k_u z) +  \Delta P_{x0} + O(\eps^2)]   \Biggr)/
\Biggl( mc\sqrt{\cal E}\biggl(\frac{1}{\eps} +\chi_{NF}(\eps k_uz)
 + O(\eps) \biggr) \Biggr)
\nonumber\\
&& = \eps \frac{(K/\sqrt{\cal E})[\cos(k_u z) +  \Delta P_{x0} + O(\eps^2)]}
            { 1 +\eps\chi_{NF}(\eps k_uz)+ O(\eps^2)}
\nonumber\\
&&
= \frac{\eps K}{\sqrt{\cal E}} 
\biggl( \cos(k_u z) +  \Delta P_{x0} + O(\eps^2)\biggr)
            \biggl( 1 - \eps\chi_{NF}(\eps k_uz)+ O(\eps^2)\biggr)\nonumber\\
&&
= \frac{\eps K}{\sqrt{\cal E}} 
[\cos(k_u z) +  \Delta P_{x0}] [1 - \eps\chi_{NF}(\eps k_uz)]
+ O(\eps^3)
 \; .
\label{eq:x430130}
\end{eqnarray}
Integrating (\ref{eq:x430130}) gives
\begin{eqnarray}
&& \hspace{-10mm} 
x(z) = x(0) 
\nonumber\\
&&
+ \frac{\eps K}{\sqrt{\cal E}}
\biggl( \frac{\sin(k_u z)}{k_u} +  z\Delta P_{x0} 
-\eps\int_0^z\;[\cos(k_u s) +  \Delta P_{x0}]\chi_{NF}(\eps k_us)ds\biggr) 
\nonumber\\
&& + O(\eps^3z) \; .
\label{eq:430145} 
\end{eqnarray}

For $\eps$ sufficiently small, $I(\eps,T)=[0,T/\eps]$ and then
(\ref{eq:3.210})-(\ref{eq:3.220}) and (\ref{eq:430145}) hold for
$0\leq k_u z\leq T/\eps$.
\subsection{Low Gain Calculation in the NtoR regime}
\label{3.7}
Low gain theories in \cite{KHL,SDR,MP} are done in the context of the
pendulum equations, i.e., (\ref{eq:nn430016}),(\ref{eq:nn430015}) with
$a=0,\Delta P_{x0}=0$, and $k=1$. Here we will not make those
assumptions and we define the gain by
\begin{eqnarray}
&&\hspace{-15mm} G(\zeta,\eps)
:=\eps\overline{(v_2(\zeta,\eps)-\chi_0)}_{\theta_0} 
=\eps\overline{(v_2(\eps\zeta,1)-\chi_0)}_{\theta_0} 
\; ,
\label{eq:nn430015ra} 
\end{eqnarray}
where $v_2$ is given in (\ref{eq:n3.5}) and $\overline{(\;)}_{\theta_0}$ 
denotes the average over $\theta_0$. 
This is consistent with \cite{KHL,SDR,MP}.

The gain $G$ could be calculated numerically using a quadrature formula and an ODE solver, however standard treatments calculate it perturbatively using a regular (and thus short time) perturbation
expansion. We could do a regular perturbation expansion in
(\ref{eq:nn430016}),(\ref{eq:nn430015}) by letting
$v_i=\sum_{k=0}^4\;\eps^k A_{ik}+O(\eps^5)$ and using Grownwall techniques to make 
the $O(\eps^5)$ error rigorous (see \cite[p.594]{SSC} for an example
of a regular perturbation theorem at first order and its proof).
However at the fourth order needed here
this would be quite cumbersome. Because of the special scaling structure 
in (\ref{eq:nn430016}),(\ref{eq:nn430015}) as given in (\ref{eq:n3.1})
we can use a Taylor expansion.
For $\eps=1$ we get from (\ref{eq:nn430016}),(\ref{eq:nn430015})
\begin{eqnarray}
&&\hspace{-10mm} v_1'(\cdot,1) = 2v_2(\cdot,1) \; , 
\quad v_1(0,1)=\theta_0 \; , 
\nonumber\\
&&\hspace{-10mm} 
v_2'(\cdot,1) = -K_0(k) \cos(kv_1(\cdot,1)-a\tau) \; , \quad v_2(0,1)=\chi_0 \; ,
\nonumber\\
\label{eq:nn430015v} 
\end{eqnarray}
and we expand $v_2(\cdot,1)$ about $\tau=0$ so that
\begin{eqnarray}
&&\hspace{-10mm} 
v_2(\tau,1) = \chi_0 + \sum_{k=1}^4\;\frac{1}{k!} v_2^{(k)}(0,1)\tau^k
 + \frac{\tau^5}{4!}\;\int_0^1\;(1-t)^4v_2^{(5)}(t\tau,1)dt \; .
\label{eq:nn430015w} 
\end{eqnarray}
From (\ref{eq:G.30}) in Appendix \ref{G} we have
\begin{eqnarray}
&&\hspace{-10mm} v_2'(0,1) =-K_0(k) \cos(k\theta_0) \; ,
\nonumber\\
&&\hspace{-10mm} 
v_2''(0,1) =K_0(k)(2k\chi_0-a)   \sin(k\theta_0) \; , 
\nonumber\\
&&\hspace{-10mm} 
v_2'''(0,1) = K_0(k)\biggl( -kK_0(k) \sin(2k\theta_0) 
+ [ 2k\chi_0 -a]^2  \cos(k\theta_0)\biggr)
\; ,
\nonumber\\
&&\hspace{-10mm} 
v_2''''(0,1) = K_0(k)\biggl( 2kK_0(k)(2k\chi_0-a)[\sin^2(k\theta_0) 
-3\cos^2(k\theta_0)]
\nonumber\\
&& -[ 2k\chi_0 -a]^3\sin(k\theta_0)\biggr) \; .
\label{eq:nn430015t} 
\end{eqnarray}
It follows from (\ref{eq:nn430015w}),(\ref{eq:nn430015t}) that
the average over $\theta_0$ leads to
\begin{eqnarray}
&&\hspace{-10mm} \overline{(v_2(\tau,1)-\chi_0)}_{\theta_0} 
= \frac{\tau^4}{4!} \overline{v_2''''(0,1)}_{\theta_0} + O(\tau^5) 
= -\frac{\tau^4}{12} kK_0^2(k) [ 2k\chi_0 -a] + O(\tau^5) 
\; ,
\nonumber\\
\label{eq:nn430015s} 
\end{eqnarray}
which gives, by (\ref{eq:nn430015ra}),
\begin{eqnarray}
&&\hspace{-15mm} G(\zeta,\eps)=\eps\overline{(v_2(\eps\zeta,1)-\chi_0)}_{\theta_0} 
= -\frac{\eps^5\zeta^4}{12} kK_0^2(k) [ 2k\chi_0 -a] + O(\eps^6) \; .
\label{eq:nn430015r} 
\end{eqnarray}
This shows the effect of $a$ and $k$ on the gain.

We now compare our gain formula in (\ref{eq:nn430015r}) with the
corresponding calculation in \cite{KHL}, where
$a=0,\Delta P_{x0}=0$, and $k=1$.
From our NtoR normal form system 
(\ref{eq:nn430016}),(\ref{eq:nn430015})
and letting $\theta=v_1$ and $\eta=\eps v_2$ we obtain the IVP
\begin{eqnarray}
&&\theta'=2\eta\;, \quad \theta(0)=\theta_0\; , \label{eq:RP10} \\
&&\eta'=-\epsilon\cos \theta\;, \quad \eta(0)=\eps\chi_0=:\eta_0 \;,
\label{eq:RP11}
\end{eqnarray}
where $\epsilon=\eps^2K_0(1)$.
The procedure in \cite{KHL} is a regular perturbation expansion in $\epsilon$ that does not assume that $\eta_0$ is small.
Proceeding as they do, we write
\begin{eqnarray}
&&\theta(\zeta,\epsilon)=\theta^0(\zeta) +\epsilon\theta^1(\zeta) 
+\epsilon^2\theta^2(\zeta) +O(\epsilon^3) \;, \label{eq:RP14}\\
&&\eta(\zeta,\epsilon)=\eta^0(\zeta) +\epsilon\eta^1(\zeta) 
+\epsilon^2\eta^2(\zeta) +O(\epsilon^3) \;. \label{eq:RP15}
\end{eqnarray}
We find
\begin{eqnarray}
&&\eta^0(\zeta)=\eta_0 \;, \label{eq:RP20a}\\
&&\theta^0(\zeta)=2\eta_0\zeta+\theta_0 \;, \label{eq:RP20b}\\
&&\eta^1(\zeta)=\frac{1}{2\eta_0}[\sin \theta_0-\sin(2\eta_0\zeta +\theta_0)] \;, \label{eq:RP20c}\\
&&\theta^1(\zeta)=\frac{1}{\eta_0}\{\zeta\sin \theta_0+\frac{1}{2\eta_0}[\cos(2\eta_0\zeta+\theta_0)-\cos \theta_0]\} \;, \label{eq:RP20d}\\
&&\eta^2(\zeta)=\frac{1}{\eta_0}\int^\zeta_0 dt \sin(2\eta_0t +\theta_0) \{t\sin \theta_0 \nonumber \\
&&+\frac{1}{2\eta_0}[\cos(2\eta_0t+\theta_0)-\cos \theta_0] \} \;. \label{eq:RP20e}
\end{eqnarray}
It follows that $\overline{\eta^1(\zeta)}_{\theta_0}=0$ and 
\begin{eqnarray}
&&\overline{\eta^2(\zeta)}_{\theta_0} =\frac{1}{2\eta_0}\int^\zeta_0(t\cos 2\eta_0t - \frac{1}{2\eta_0} \sin 2\eta_0t) dt\;. 
\label{eq:RP24}
\end{eqnarray}
We can rewrite (\ref{eq:RP24}) as
\begin{eqnarray}
&&\overline{\eta^2(\zeta)}_{\theta_0} = \frac{\zeta^3}{4} \frac{d}{d\tau}\big(\frac{\sin \tau}{\tau}\big)^2 \;, \quad \tau:=\eta_0\zeta\; ,
\label{eq:RP25}
\end{eqnarray}
and the gain becomes
\begin{eqnarray}
&& G(\zeta,\eps)=\epsilon^2 \overline{\eta^2(\zeta)}_{\theta_0}=\eps^4K_0^2(1)
\frac{1}{4} \zeta^3 \frac{d}{d\tau}\big(\frac{\sin \tau}{\tau}\big)^2 \; ,
\label{eq:RP26}
\end{eqnarray}
consistent with \cite{KHL}.
For $\eta_0$ small, which is required by our averaging approximation
(since $\eta_0=\eps\chi_0$ and $\chi_0=O(1)$),
we obtain from (\ref{eq:RP24}) that 
\begin{eqnarray}
&&\overline{\eta^2(\zeta)}_{\theta_0} =\frac{1}{2\eta_0}\int^\zeta_0[-\frac{4}{3}\eta_0^2t^3+O(\eta_0t)^4] dt   \approx -\frac{1}{6}\eta_0\zeta^4 \;. 
\label{eq:RP27}
\end{eqnarray}
It follows from (\ref{eq:RP26}),(\ref{eq:RP27}) that 
\begin{eqnarray}
&& G(\zeta,\eps)\approx -\epsilon^2\frac{1}{6}\eta_0\zeta^4 = -\frac{\eps^5\zeta^4}{6} K_0^2(1) \chi_0 \; ,
\label{eq:RP30}
\end{eqnarray}
as in (\ref{eq:nn430015r}) with $a=0$ and $k=1$.

Thus we see that (\ref{eq:nn430015r}) is consistent with the standard gain formula for $\tau=\eta_0\zeta$ small.
The $O(\eps^6)$ error in  (\ref{eq:nn430015r}) can be made precise by estimating the remainder term in (\ref{eq:nn430015w}).
However, we cannot justify the gain formula either in (\ref{eq:nn430015r}) or in (\ref{eq:RP26}) in the context of our Lorentz system in (\ref{eq:2.25}) - (\ref{eq:2.18}), because our NtoR normal form approximation only gives an approximation to $O(\eps)$.
Thus a justification of the gain formulas, based on our Lorentz system, would need to come from elsewhere, e.g., 
a numerical calculation based on (\ref{eq:3.17}) and (\ref{eq:3.18}).
\setcounter{equation}{0}
\section{Proof of averaging theorems}
\label{4}
In \S\ref{4.1} we prove the NR theorem, Theorem 1 of \S\ref{3.5.1}, and in
\S\ref{4.2} we prove the NtoR theorem, Theorem 2 of \S\ref{3.5.2}.
\subsection{Proof of Theorem \ref{T1} 
(Averaging theorem in $\Delta$-NR case)} 
\label{4.1}
Here we compare solutions of the exact IVP (\ref{eq:3.19}),(\ref{eq:3.20}):
\bea
&&\hspace{-8mm} \theta'= \eps f_1(\chi,\zeta)
+ \eps^2 g_1( \theta,\chi,\zeta;\eps,\nu) \; , 
\quad \theta(0,\eps)=\theta_0 \; ,
\label{eq:n4.10}\\
&& \hspace{-8mm} \chi'=  \eps f_2(\theta,\zeta;\nu)
+  \eps^2 g_2( \theta,\chi,\zeta;\eps,\nu)
\; , \quad \chi(0,\eps)=\chi_0 \; ,
\label{eq:n4.11}
\eea
where
\bea
&&\hspace{-8mm}  f_1(\chi,\zeta) = 
\frac{2q(\zeta)\chi}{\bar{q}} \; , 
\label{eq:n4.15}\\
&& \hspace{-8mm}  f_2(\theta,\zeta;\nu) = - K^2
(\cos\zeta + \Delta P_{x0}) \cos( \nu[\theta - Q(\zeta)])
\nonumber\\
&&=-\frac{K^2}{2}e^{i\nu\theta}
\sum_{n\in\Z}\widehat{jj}(n;\nu,\Delta P_{x0})e^{i(n-\nu)\zeta}+ cc \; ,
\label{eq:n4.16}
\eea
with the normal form IVP of 
(\ref{eq:3.48}),(\ref{eq:3.49}):
\begin{eqnarray}
&&  v_1'=\eps \bar{f}_1(v_2) \; , \quad  v_1(0,\eps)
=\theta_0 \;,\label{eq:n4.20} \\
&&  v_2'= \eps \bar{f}_2(v_1;\nu) \; , \quad v_2(0,\eps)=\chi_0 \; , 
\label{eq:n4.21}
\end{eqnarray}
where 
\begin{eqnarray}
&& \bar{f}_1(v_2) = 2v_2 \; , \quad \bar{f}_2(v_1;\nu) = 0 \; ,
\label{eq:n4.25}
\end{eqnarray}
for $\nu \in [k+\Delta, k+1-\Delta]$.

Subtracting and integrating, we obtain from (\ref{eq:2353}),
(\ref{eq:n4.10}),(\ref{eq:n4.11}),(\ref{eq:n4.20}),(\ref{eq:n4.21}) that
\begin{eqnarray}
&&\hspace{-10mm} \theta(\zeta,\eps)-v_1(\zeta,\eps) = 
\eps \int_0^\zeta\big[ f_1(\chi(s,\eps),s)- f_1(v_2(s,\eps),s) 
\nonumber\\
&& \hspace{-5mm} 
+ f_1(v_2(s,\eps),s) - \bar{f}_1(v_2(s,\eps)) 
+\eps g_1(\theta(s,\eps),\chi(s,\eps),s;\eps,\nu)  \big] \,ds 
\nonumber\\
&& \hspace{-5mm}  =\eps \int_0^\zeta\big[ f_1(\chi(s,\eps),s)
- f_1(v_2(s,\eps),s) 
\nonumber\\
&&
+ \tilde{f}_1(\chi_0,s) +\eps g_1(\theta(s,\eps),
\chi(s,\eps),s;\eps,\nu)  \big] \,ds 
\; ,
\label{eq:2193} 
\end{eqnarray}
and
\begin{eqnarray}
&&\hspace{-10mm} \chi(\zeta,\eps)-v_2(\zeta,\eps) = 
\eps \int_0^\zeta\big[ f_2(\theta(s,\eps),s;\nu)- f_2(v_1(s,\eps),s;\nu) 
\nonumber\\
&& \hspace{-5mm} 
+ f_2(v_1(s,\eps),s;\nu) +\eps g_2(\theta(s,\eps),\chi(s,\eps),
s;\eps,\nu)  \big] \,ds 
\nonumber\\
&& \hspace{-5mm} = \eps \int_0^\zeta\big[ f_2(\theta(s,\eps),s;\nu)
- f_2(v_1(s,\eps),s;\nu) 
\nonumber\\
&&
+ \tilde{f}_2(v_1(s,\eps),s;\nu) +\eps g_2(\theta(s,\eps),\chi(s,\eps),
s;\eps,\nu)  \big] \,ds 
\; ,
\label{eq:2223} 
\end{eqnarray}
for $\zeta\in I(\eps,T)=[0,T/\eps]\cap[0,\beta(\eps))$. 
Important for our analysis below is that the points $(\theta(\zeta,\eps),\chi(s,\eps))$ and
$(v_1(s,\eps),v_2(s,\eps))$ belong to the rectangle  
$\hat{W}(\theta_0,\chi_0,d_1,d_2)$ for $\zeta\in I(\eps,T)$.
Note that we have added and subtracted
$f_1(v_2(s,\eps),s)$ in (\ref{eq:2193}) and $f_2(v_1(s,\eps),s;\nu)$ in 
(\ref{eq:2223}), an idea introduced by Besjes \cite{Besjes}
(see also \cite{ESD}).

Taking absolute values, applying the Lipschitz condition 
on $\hat{W}(\theta_0,\chi_0,d_1,d_2)$ and defining
\begin{eqnarray}
&&\hspace{-5mm} e_1(s):=|\theta(s,\eps)-v_1(s,\eps)|
\; , \label{eq:204a} \\
&&\hspace{-5mm}
e_2(s):=|\chi(s,\eps)-v_2(s,\eps)|
\; ,
\label{eq:204b} 
\end{eqnarray}
gives, by (\ref{eq:2303b}),(\ref{eq:2303d}),(\ref{eq:2297ax}),
(\ref{eq:2383}),(\ref{eq:2383n}),
(\ref{eq:2193}),(\ref{eq:2223})
for $\zeta \in I(\eps,T)$,
\begin{eqnarray}
&& \hspace{-10mm} 0\leq e_1(\zeta)\leq \eps [L_1
\int_0^\zeta e_2(s)ds + |\int_0^\zeta \tilde f_1(\chi_0,s)ds|
 \nonumber\\
&&
+ \eps \int_0^\zeta\;|g_1(\theta(s,\eps),\chi(s,\eps),s;\eps,\nu)|] 
\leq \eps [L_1
\int_0^\zeta e_2(s)ds + B_1(\zeta) + T C_1] 
 \nonumber\\
&& 
\leq \eps [L_1
\int_0^\zeta e_2(s)ds + B_{1,\infty}(T/\eps) + T C_1] 
=: R_1(\zeta) \; , 
\label{eq:2243} \\
&& \hspace{-10mm} 0\leq e_2(\zeta)\leq \eps [L_2
\int_0^\zeta e_1(s)ds + |\int_0^\zeta \tilde{f}_2(2\chi_0\eps s+\theta_0,s;\nu)ds|
 \nonumber\\
&&
+ \eps \int_0^\zeta\;|g_2(\theta(s,\eps),\chi(s,\eps),s;\eps,\nu)|] 
\leq \eps [L_2
\int_0^\zeta e_1(s)ds + B_2(\zeta) + T C_2] 
 \nonumber\\
&& 
\leq \eps [L_2
\int_0^\zeta e_1(s)ds + B_{2,\infty}(T/\eps) + T C_2] 
=: R_2(\zeta) \; , 
\label{eq:2293} 
\end{eqnarray}
where we also used that $I(\eps,T)\subset[0,T/\eps]$ and where
we have introduced the $R_i$ as in the proof of the Gronwall 
inequality for a single integral inequality (the Gronwall inequality
is discussed in many ODE books, see, e.g., \cite[p.36]{Hale} and 
\cite[p.310 and 317]{Walter}). $\zeta\in I(\eps,T)$.

Recall that $L_1,L_2,C_1,C_2,B_1,B_2$ are defined in items 7,8 and 9 of 
the preamble to the theorem. 
For convenience we have suppressed the $\eps$ dependence of 
$e_1$ and $e_2$.

Before we proceed with the proof,
several comments are in order.
\begin{enumerate}
\item
We refer to the terms $B_1(\zeta),B_2(\zeta)$ in (\ref{eq:2383}) 
as Besjes terms since they were introduced by him
in order to prove an averaging theorem without a near identity transformation; a simplification. Standard proofs use the near identity transformation 
(see e.g., \cite{M1,LM,SVM}).

One may fear that
the Besjes terms could grow as large as $O(1/\eps)$ 
for $\zeta \in [0,T/\eps]$, i.e., that $B_{i,\infty}(T/\eps)=O(1/\eps)$. 
However this doesn't happen here since, by (\ref{eq:2383a}), 
$\check{B}_1,\check{B}_2(T,\Delta)$ are upper bounds for 
$B_{i,\infty}(T/\eps)$ and are $\eps$ independent. Two facts are mainly responsible for this: (a) the fact that for fixed $v_1$ and $v_2$ the integrands have zero mean, i.e., the quantities in (\ref{eq:2353}) have zero mean in $s$, and (b) the fact that  $v_1(s,\eps)$ and $v_2(s,\eps)$ are slowly varying.

\item We maintain the system form in (\ref{eq:2243}),(\ref{eq:2293}). 
We could add these two inequalities and obtain an error estimate 
using a Gronwall inequality. That is, let $L_\infty=max(L_1,L_2)$,
$B_\infty=B_{1,\infty}+B_{2,\infty}$, $C_\infty=C_1+C_2$, then
adding gives
\begin{eqnarray}
&& \hspace{-10mm} 0\leq e_\infty(\zeta)\leq \eps [L_\infty
\int_0^\zeta e_\infty(s)ds + B_\infty(T/\eps) + C_\infty T] \; , 
\end{eqnarray}
where $e_\infty=e_1+e_2$. The Gronwall inequality 
gives \\
$e_\infty(\zeta)\leq \eps [B_\infty(T/\eps) + C_\infty T] \exp(\eps L_\infty\zeta)$.
However our system approach gives better bounds.
\item
We have a draft of a general paper on quasiperiodic averaging which uses the Besjes idea and deals with the small divisor problem (See \cite{DEH}). However the proof we are presenting here is simple, the small divisor problem is trivial and the error bounds are quite explicit.  Thus we feel it is good to give complete proofs here rather than appealing to a more general theory. Also it serves the pedagogical purpose of showing how an averaging theorem is proved in a simple context; here the context of (\ref{eq:3.19}), (\ref{eq:3.20}) and (\ref{eq:430011}), (\ref{eq:430010}). We have incorporated the Besjes idea in much of our previous averaging work, see \cite{ESD,SSC,Annals,DEG,DEV}. 
\end{enumerate}

We now proceed with the proof.
It follows from (\ref{eq:2243}),(\ref{eq:2293}) that 
\begin{eqnarray}
&& \hspace{-5mm} R_1' = \eps L_1 e_2(\zeta) \leq \eps L_1 R_2(\zeta) \; , \quad
R_1(0) = \eps  [ B_{1,\infty}(T/\eps)+C_1T] \; , \label{eq:2433} \\
&& \hspace{-5mm} R_2' = \eps L_2 e_1(\zeta) \leq \eps L_2 R_1(\zeta) \; , \quad
R_2(0) =  \eps  [ B_{2,\infty}(T/\eps)+C_2T] \; ,
\label{eq:2443} 
\end{eqnarray}
whence, by Appendix \ref{I} for $\zeta \in I(\eps,T)$,
\begin{eqnarray}
&& \hspace{-5mm} R_1(\zeta)\leq \eps w_1(\eps\zeta) \; , \quad
 R_2(\zeta)\leq \eps w_2(\eps\zeta) \; , 
\label{eq:2453} 
\end{eqnarray}
where 
\begin{eqnarray}
&& \hspace{-5mm} w_1' =  L_1 w_2  \; , \quad
w_1(0) =  B_{1,\infty}(T/\eps)+C_1T \; , \label{eq:2463} \\
&& \hspace{-5mm} w_2' =  L_2 w_1  \; , \quad
w_2(0) =  B_{2,\infty}(T/\eps)+C_2T \; .
\label{eq:2473} 
\end{eqnarray}
Note that in Appendix \ref{I} we use the fact that $R_1,R_2$ are of
class $C^1$.

Solving (\ref{eq:2463}),(\ref{eq:2473}) we find
\begin{eqnarray}
&& \hspace{-5mm}\left( \begin{array}{c}  w_1(s) \\
  w_2(s) \end{array}\right) 
\nonumber\\
&& =
\left( \begin{array}{cc} 
\cosh(s\sqrt{L_1 L_2}) & \sqrt{\frac{L_1}{L_2}}\sinh(s\sqrt{L_1 L_2}) \\
\sqrt{\frac{L_2}{L_1}}\sinh(s\sqrt{L_1 L_2}) & \cosh(s\sqrt{L_1 L_2})
\end{array}\right) 
 \left( \begin{array}{c}  B_{1,\infty}(T/\eps)+C_1T \\
 B_{2,\infty}(T/\eps)+C_2T \end{array}\right)
\; ,
\nonumber\\
\label{eq:2483n} 
\end{eqnarray}
whence, by (\ref{eq:2243}),(\ref{eq:2293}),(\ref{eq:2453}),
\begin{eqnarray}
&& \hspace{-5mm} e_1(\zeta)\leq \eps w_1(\eps\zeta)\le \eps w_1(T) 
=\eps \biggl( [B_{1,\infty}(T/\eps)+C_1T] \cosh(T\sqrt{L_1 L_2})
\nonumber\\
&& +[B_{2,\infty}(T/\eps)+C_2T] \sqrt{\frac{L_1}{L_2}}\sinh(T\sqrt{L_1 L_2})\biggr)
\; , 
\label{eq:2453a} \\
&&  \hspace{-5mm}e_2(\zeta)\leq \eps w_2(\eps\zeta)\le \eps w_2(T) 
= \eps \biggl( 
[B_{1,\infty}(T/\eps)+C_1T]\sqrt{\frac{L_2}{L_1}}\sinh(T\sqrt{L_1 L_2})
\nonumber\\
&& + [B_{2,\infty}(T/\eps)+C_2T] \cosh(T\sqrt{L_1 L_2})\biggr)\; ,
\label{eq:2453aa} 
\end{eqnarray}
for $\zeta\in I(\eps,T)$, where, at  the second inequalities, we have used the fact that $w_1$ and $w_2$ are increasing (the latter follows 
from (\ref{eq:2463}),(\ref{eq:2473}),(\ref{eq:2483n})).
We thus have proven (\ref{eq:204}),(\ref{eq:222}) in Theorem 1.

We note that $\check{B}_1$ and $\check{B}_{2,1}(T)$ are finite.
Also, since the Fourier series of
$jj(\cdot;\nu,\Delta P_{x0})$ is absolutely
convergent, we conclude from (\ref{eq:3900cn}) that 
$\check B_{22}(T)$ is finite whence, by (\ref{eq:3900dcn}), 
$\check{B}_2(T,\Delta)$ is finite.

By restricting $\eps_0$, and thus $\eps$ in (\ref{eq:2453a}),(\ref{eq:2453aa}), we can keep $(\theta(\zeta,\eps),\chi(\zeta,\eps))$ 
away from the boundary of $\hat{W}(\theta_0,\chi_0,d_1,d_2)$ for $\zeta \in I(\eps,T)$.
In this case $T/\eps$ must be less than $\beta(\eps)$
thus $I(\eps,T)=[0,T/\eps]$. 

To complete the proof we have to show (\ref{eq:2383a}) which is the
heart of the proof. Thus we have to estimate $B_1,B_2$.
From (\ref{eq:2.59}),(\ref{eq:3.60}),(\ref{eq:2353})
we obtain
\begin{eqnarray}
&& \hspace{-8mm} 
 \tilde{f}_1(v_2(s,\eps),s) =  2\frac{q(s)-\bar{q}}{\bar{q}}v_2(s,\eps)
=\frac{2K^2}{\bar{q}}[2\Delta P_{x0} \cos s
+ \frac{1}{2}\cos(2s)]\chi_0 \; ,
\nonumber
\end{eqnarray}
and thus, by (\ref{eq:2383}),(\ref{eq:222a}),
%
\begin{eqnarray}
&& \hspace{-8mm} 
B_1(\zeta) =\frac{2K^2}{\bar{q}}|\int_0^\zeta \;  [2\Delta P_{x0} \cos s
+ \frac{1}{2}\cos(2s)]\chi_0 \; ds| 
\nonumber\\
&&
=\frac{2K^2|\chi_0|}{\bar{q}} | 2\Delta P_{x0} \sin\zeta
+ \frac{1}{4}\sin(2\zeta)| 
\leq \frac{2K^2|\chi_0|}{\bar{q}} ( 2|\Delta P_{x0}| + \frac{1}{4}) 
\nonumber\\
&& = \check{B} \; ,
\label{eq:43910ad} 
\end{eqnarray}
so that, by (\ref{eq:2383n}), $B_{1,\infty}(T/\eps)\le \check B_1$.
From (\ref{eq:3.60}),(\ref{eq:2353}),(\ref{eq:n4.16})
we obtain
\begin{eqnarray}
&& \hspace{-8mm} 
 \tilde{f}_2(v_1(s,\eps),s;\nu) 
=-\frac{K^2}{2} e^{i\nu[2\eps\chi_0 s+\theta_0]}
\sum_{n\in\Z}\widehat{jj}(n;\nu,\Delta P_{x0})e^{i(n-\nu)s} + cc 
\; ,
\nonumber
\end{eqnarray}
whence, by (\ref{eq:2383}) and for $\zeta\in\R$,
\begin{eqnarray}
&&\hspace{-18mm} 
B_2(\zeta)=\frac{K^2}{2} | \int_0^\zeta \; e^{i\nu[2\eps\chi_0 s+\theta_0]}
\sum_{n\in\Z}\widehat{jj}(n;\nu,\Delta P_{x0})e^{i(n-\nu)s}ds + cc |
\nonumber\\
&&
= \frac{K^2}{2} |\sum_{n\in\Z}\widehat{jj}(n;\nu,\Delta P_{x0})\;
 \int_0^\zeta \; e^{i\nu[2\eps\chi_0 s+\theta_0]} e^{i(n-\nu)s}ds + cc |
\nonumber\\
&&
\leq K^2 \sum_{n\in\Z}|\widehat{jj}(n;\nu,\Delta P_{x0})|\;
 | \int_0^\zeta \; e^{i2\eps\nu\chi_0 s} e^{i(n-\nu)s}ds| \; ,
\label{eq:254n} 
\end{eqnarray}
where in the second equality we used the fact that the Fourier series of
$jj(\cdot;\nu,\Delta P_{x0})$ is uniformly
convergent. Integrating by parts gives, for $0\leq\zeta\leq T/\eps$,
\begin{eqnarray}
&&\hspace{-5mm} 
|\int_0^\zeta\; e^{i2\eps\nu\chi_0 s}e^{i(n-\nu)s}ds| 
= |\frac{e^{i(n-\nu+2\eps\nu\chi_0)\zeta} - 1
-i2\eps\nu \chi_0 \int_0^\zeta\; e^{i(n-\nu+2\eps\nu\chi_0)s}ds}{i(n-\nu)}| 
\nonumber\\
&&
\leq \frac{2 + 2\eps\nu |\chi_0|\zeta}{|n-\nu|}
\leq \frac{2 + 2(k+1) |\chi_0|T}{|n-\nu|} \; ,
\nonumber
\end{eqnarray}
whence, by (\ref{eq:254n}),
for $0\leq\zeta\leq T/\eps$,
\begin{eqnarray}
&& \hspace{-5mm} 
B_2(\zeta) \leq 2K^2[1 + (k+1) |\chi_0|T] 
\sum_{n\in\Z} |\frac{\widehat{jj}(n;\nu,\Delta P_{x0})}{n-\nu}|  \; .
\label{eq:256an} 
\end{eqnarray}
The $n-\nu$ in the denominator is the so-called small divisor problem in this context. It is easily resolved in this $\Delta$-NR case.
In fact, for $\nu$ $\Delta$-NR, i.e.,
$k+\Delta\leq\nu\leq k+1-\Delta$, we have
\begin{eqnarray}
&& \hspace{-5mm} 
\sum_{n\in\Z} |\frac{\widehat{jj}(n;\nu,\Delta P_{x0})}{n-\nu}| 
= \frac{|\widehat{jj}(k;\nu,\Delta P_{x0})|}{|k-\nu|} 
\nonumber\\
&&
 + \frac{|\widehat{jj}(k+1;\nu,\Delta P_{x0})|}{|k+1-\nu|} 
+ \sum_{n\in(\Z\setminus\lbrace k,k+1\rbrace)} 
\frac{|\widehat{jj}(n;\nu,\Delta P_{x0})|}{|n-\nu|} 
\leq  \frac{|\widehat{jj}(k;\nu,\Delta P_{x0})|}{\Delta} 
\nonumber\\
&& + \frac{|\widehat{jj}(k+1;\nu,\Delta P_{x0})|}{\Delta} 
+ \sum_{n\in(\Z\setminus\lbrace k,k+1\rbrace)} 
\; |\widehat{jj}(n;\nu,\Delta P_{x0})| \; ,
\nonumber
\end{eqnarray}
whence, by (\ref{eq:3900dcn}),
(\ref{eq:3900cbn}),(\ref{eq:3900cn}),(\ref{eq:256an}),
\begin{eqnarray}
&& \hspace{-5mm} 
B_2(\zeta) \leq 2K^2\{1 + (k+1) |\chi_0|T\}\{\frac{
|\widehat{jj}(k;\nu,\Delta P_{x0})| 
+ |\widehat{jj}(k+1;\nu,\Delta P_{x0})|}{\Delta}
\nonumber\\
&&
+ \sum_{n\in(\Z\setminus\lbrace k,k+1\rbrace)} 
|\widehat{jj}(n;\nu,\Delta P_{x0})|\} 
=\frac{1}{\Delta}\check B_{21}(T)+\check B_{22}(T) = 
\check B_{2}(T,\Delta) \; ,
\nonumber\\
\label{eq:256n} 
\end{eqnarray}
so that, by (\ref{eq:2383n}),
$B_{2,\infty}(T/\eps)\le \check B_2(T,\Delta)$.

This completes the proof.
\subsection{Proof of Theorem \ref{T2} 
(Averaging theorem in NtoR case where $\nu=k+\eps a$)}
\label{4.2}
The proof goes analogously to the proof of Theorem \ref{T1} in
\S\ref{4.1} and so we omit some details.

Thus we begin by comparing solutions of the exact IVP 
(\ref{eq:430011}),(\ref{eq:430010})
\begin{eqnarray}
&&\hspace{-10mm}\theta' = \eps f^R_1(\chi,\zeta)
+\eps^2 g^R_1(\theta,\chi,\zeta,\eps,k,a) \; , \quad \theta(0,\eps)=\theta_0 \; ,
\label{eq:xn4.10} \\
&&\hspace{-10mm}\chi'=\eps f^R_2(\theta,\eps\zeta,\zeta;k,a)
+\eps^2 g^R_2(\theta,\chi,\zeta,\eps,k,a) \; ,  \quad \chi(0,\eps)=\chi_0 \; ,
\label{eq:xn4.11} 
\end{eqnarray}
where, by (\ref{eq:430010an}),(\ref{eq:430011an}),(\ref{eq:430010aaa}),
\begin{eqnarray}
&&\hspace{-10mm} 
f^R_1(\chi,\zeta) =  \frac{2q(\zeta)\chi}{\bar{q}} \; ,
\label{eq:xn4.15} \\
&&\hspace{-10mm} 
f^R_2(\theta,\eps\zeta,\zeta;k,a)
= -\frac{K^2}{2}\exp(i[k\theta-a\eps\zeta])
\sum_{n\in{\mathbb Z}} 
\widehat{jj}(n;k,\Delta P_{x0})
e^{i\zeta[n-k]}  + cc \; ,
\nonumber\\
\label{eq:xn4.16} 
\end{eqnarray}
with the normal form IVP of 
(\ref{eq:nn430016}),(\ref{eq:nn430015})
\begin{eqnarray}
&&\hspace{-10mm} 
v_1' = \eps\bar{f}^R_1(v_2)
\; , 
\quad  v_1(0,\eps) =\theta_0 \; , 
\label{eq:xn4.20} \\
&&\hspace{-10mm} v_2' 
= \eps \bar{f}^R_2(v_1,\eps\zeta;k) \; ,  \quad  
v_2(0,\eps) =\chi_0 \; , 
\label{eq:xn4.21} 
\end{eqnarray}
where
\bea
&&\hspace{-8mm}
\bar{f}^R_1(v_2) = 2v_2 \; ,
\label{eq:xn4.25} 
\\
&&\hspace{-8mm}
\bar{f}^R_2(v_1,\eps\zeta;k) 
= -\frac{K^2}{2}\exp(i[k v_1-a\eps\zeta])
\widehat{jj}(k;k,\Delta P_{x0}) + cc \; .
\label{eq:xn4.25a} 
\eea

Subtracting and integrating, we obtain from (\ref{eq:x2353}),
(\ref{eq:xn4.10}),(\ref{eq:xn4.11}),(\ref{eq:xn4.20}),(\ref{eq:xn4.21}) that
\begin{eqnarray}
&&\hspace{-8mm} \theta(\zeta)-v_1(\zeta,\eps) = 
\eps \int_0^\zeta\big[ f^R_1(\chi(s),s)
- f^R_1(v_2(s,\eps),s)
\nonumber\\
&& \hspace{-5mm}
+ f^R_1(v_2(s,\eps),s)
- \bar{f}^R_1(v_2(s,\eps)) 
+\eps g^R_1(\theta(s),\chi(s),s,\eps,k,a)  \big] \,ds 
\nonumber\\
&& = 
\eps \int_0^\zeta\big[ f^R_1(\chi(s),s)
- f^R_1(v_2(s,\eps),s)
\nonumber\\
&& \hspace{-5mm}
+ \tilde{f}^R_1(v_2(s,\eps),s)
+\eps g^R_1(\theta(s),\chi(s),s,\eps,k,a)  \big] \,ds 
\; ,
\nonumber\\
\label{eq:x2193} 
\end{eqnarray}
and
\begin{eqnarray}
&&\hspace{-8mm} \chi(\zeta)-v_2(\zeta,\eps) = 
\eps \int_0^\zeta\big[ f^R_2(\theta(s),\eps s,s;k,a)
- f^R_2(v_1(s,\eps),\eps s,s;k,a)
\nonumber\\
&& \hspace{-5mm}
+ f^R_2(v_1(s,\eps),\eps s,s;k,a)
- \bar{f}^R_2(v_1(s,\eps),\eps s;k) 
+\eps g^R_2(\theta(s),\chi(s),s,\eps,k,a)  \big] \,ds 
\nonumber\\
&& = 
\eps \int_0^\zeta\big[ f^R_2(\theta(s),\eps s,s;k,a)
- f^R_2(v_1(s,\eps),\eps s,s;k,a)
\nonumber\\
&& \hspace{-5mm}
+ \tilde{f}^R_2(v_1(s,\eps),\eps s,s;k,a)
+\eps g^R_2(\theta(s),\chi(s),s,\eps,k,a)  \big] \,ds 
\; ,
\nonumber\\
\label{eq:x2223} 
\end{eqnarray}
for $\zeta\in I(\eps,T)=[0,T/\eps]\cap[0,\beta(\eps))$. 
Taking absolute values, applying the Lipschitz condition and defining
\begin{eqnarray}
&&\hspace{-5mm} e_1(s):=|\theta(s)-v_1(s,\eps)|
\; , 
\label{eq:x204a} \\
&&\hspace{-5mm}
e_2(s):=|\chi(s)-v_2(s,\eps)|
\; ,
\label{eq:x204b} 
\end{eqnarray}
gives, by (\ref{eq:2303an}),(\ref{eq:2303cn}),(\ref{eq:2297axn}),
(\ref{eq:x2383}),(\ref{eq:x2383n}),(\ref{eq:x2193}),(\ref{eq:x2223})
for $\zeta \in I(\eps,T)$,
\begin{eqnarray}
&& \hspace{-10mm} 0\leq e_1(\zeta)\leq \eps [L^R_1
\int_0^\zeta e_2(s)ds + |\int_0^\zeta \tilde{f}_1^R(v_2(s,\eps),s)ds|
\nonumber\\
&&
+ \eps\int_0^\zeta  |g^R_1(\theta(s),\chi(s),s,\eps,k,a)|ds ]
\leq \eps [L^R_1
\int_0^\zeta e_2(s)ds + B^R_1(\zeta) + T C^R_1] 
 \nonumber\\
&& 
\leq \eps [L^R_1
\int_0^\zeta e_2(s)ds + B^R_{1,\infty}(T/\eps) + T 
C^R_1] 
\; , 
\label{eq:x2243} \\
&& \hspace{-10mm} 0\leq e_2(\zeta)\leq \eps [L^R_2
\int_0^\zeta e_1(s)ds + |\int_0^\zeta \tilde{f}_2^R(v_1(s,\eps),\eps s,s;k,a)ds|
 \nonumber\\
&& 
+ \eps\int_0^\zeta  |g^R_2(\theta(s),\chi(s),s,\eps,k,a)|ds ]
\leq \eps [L^R_2
\int_0^\zeta e_1(s)ds + B^R_2(\zeta) + T 
C^R_2] 
 \nonumber\\
&& 
\leq \eps [L^R_2
\int_0^\zeta e_1(s)ds + B^R_{2,\infty}(T/\eps) 
+ T C^R_2] 
\; ,
\label{eq:x2293} 
\end{eqnarray}
where we also used that $I(\eps,T)\subset[0,T/\eps]$.
Recall that $L^R_i,C^R_i,B^R_i$ 
are defined in items 7,8 and 9 of 
the preamble to the theorem. 

We are now in the same situation as in the proof
of Theorem \ref{T1} since replacing $L_i,C_i,B_i$ in 
(\ref{eq:2243}),(\ref{eq:2293}) by
$L^R_i,C^R_i,B^R_i$ results in (\ref{eq:x2243}),(\ref{eq:x2293}).
Since, as shown in the proof of Theorem \ref{T1},
(\ref{eq:2243}),(\ref{eq:2293}) entail
(\ref{eq:2453a}),(\ref{eq:2453aa}) we thus conclude here 
that (\ref{eq:x2243}),(\ref{eq:x2293}) entail:
\begin{eqnarray}
&& \hspace{-5mm} e_1(\zeta)\leq 
\eps \biggl( [B^R_{1,\infty}(T/\eps)+C_1T] \cosh(T\sqrt{L^R_1 L^R_2})
\nonumber\\
&& +[B^R_{2,\infty}(T/\eps)+C_2T] \sqrt{\frac{L^R_1}{L^R_2}}\sinh(T\sqrt{L^R_1 L^R_2})\biggr)
\; , 
\label{eq:x2453a} \\
&&  \hspace{-5mm}e_2(\zeta)\leq \eps \biggl( 
[B^R_{1,\infty}(T/\eps)+C_1T]\sqrt{\frac{L^R_2}{L^R_1}}\sinh(T\sqrt{L^R_1 L^R_2})
\nonumber\\
&& + [B^R_{2,\infty}(T/\eps)+C_2T] \cosh(T\sqrt{L^R_1 L^R_2})\biggr)\; ,
\label{eq:x2453aa} 
\end{eqnarray}
for $\zeta\in I(\eps,T)$.
We thus have proven (\ref{eq:x204}),(\ref{eq:x222}).

Clearly, by (\ref{eq:x222a}), $\check{B}^R_1(T)$ is finite.
Also, since $jj(\cdot;\nu,\Delta P_{x0})$ is a $C^\infty$ function, the
series on the rhs of (\ref{eq:x3900dcn}) converges whence 
$\check{B}^R_2(T)$ is also finite.

By restricting $\eps_0$, and thus $\eps$ in (\ref{eq:x2453a}),(\ref{eq:x2453aa}), we can keep $(\theta(\zeta,\eps),\chi(\zeta,\eps))$ 
away from the boundary of $\hat{W}(\theta_0,\chi_0,d_1,d_2)$ for $\zeta \in I(\eps,T)$.
In this case $T/\eps$ must be less than $\beta(\eps)$
thus $I(\eps,T)=[0,T/\eps]$.

To complete the proof we have to show (\ref{eq:x2383a}).
Thus we have to estimate $B_1^R,B_2^R$ and beginning with $B_1^R$ we conclude 
from (\ref{eq:2.59}),(\ref{eq:x2353}),(\ref{eq:xn4.15}),
(\ref{eq:xn4.25}) that, for $\zeta\in\R$,
\begin{eqnarray}
&& \hspace{-8mm} 
 \tilde{f}_1^R(v_2(s,\eps),s) =  2\frac{q(s)-\bar{q}}{\bar{q}}
v_2(s,\eps)
\nonumber\\
&&=\frac{2K^2}{\bar{q}}[2\Delta P_{x0} \cos s
+ \frac{1}{2}\cos(2s)]v_2(s,\eps) \; ,
\nonumber
\end{eqnarray}
whence, by (\ref{eq:n3.5e}),
(\ref{eq:x2383}),(\ref{eq:x222a}),
(\ref{eq:xn4.21}),(\ref{eq:xn4.25a}) for $0\leq\zeta\leq T/\eps$,
\begin{eqnarray}
&& \hspace{-8mm} 
B_1^R(\zeta) =\frac{2K^2}{\bar{q}}\Big{|}\int_0^\zeta \;  [2\Delta P_{x0} \cos s
+ \frac{1}{2}\cos(2s)]v_2(s,\eps)\; ds\Big{|} 
\nonumber\\
&&
=\frac{2K^2}{\bar{q}} \Big{|} [2\Delta P_{x0} \sin\zeta
+ \frac{1}{4}\sin(2\zeta)] v_2(\zeta,\eps)
\nonumber\\
&&
\quad -\int_0^\zeta\;  [2\Delta P_{x0} \sin s
+ \frac{1}{4}\sin(2s)]\frac{dv_2}{ds}(s,\eps) ds\Big{|} 
\nonumber\\
&& = \frac{2K^2}{\bar{q}} \Big{|} [2\Delta P_{x0} \sin\zeta
+ \frac{1}{4}\sin(2\zeta)] v_2(\zeta,\eps)
\nonumber\\
&&\hspace{-5mm} 
+\eps K^2 \widehat{jj}(k;k,\Delta P_{x0})
 \int_0^\zeta \; [2\Delta P_{x0} \sin s
+ \frac{1}{4}\sin(2s)] \cos\biggl(kv_1(s,\eps)-\eps a s\biggr)ds \Big{|}
\nonumber\\
&&
\leq\frac{2K^2}{\bar{q}} \biggl( [2|\Delta P_{x0}| + \frac{1}{4}]
|v_2(\zeta,\eps)|
\nonumber\\
&&\quad +\eps K^2 \big{|}\widehat{jj}(k;k,\Delta P_{x0})\big{|}
[2|\Delta P_{x0}| + \frac{1}{4}]\zeta\biggr)
\nonumber\\
&&
\leq\frac{2K^2}{\bar{q}} [2|\Delta P_{x0}| + \frac{1}{4}]
\biggl(  |v_2(\zeta,\eps)|
+  K^2 \eps\zeta |\widehat{jj}(k;k,\Delta P_{x0})|\biggr)
\nonumber\\
&&
\leq\frac{2K^2}{\bar{q}} [2|\Delta P_{x0}| + \frac{1}{4}]
\biggl(  \chi_\infty(\theta_0,\chi_0,k,a) 
\nonumber\\
&& 
\quad +  K^2 T \big{|}\widehat{jj}(k;k,\Delta P_{x0})\big{|}\biggr)
= \check{B}_1^R(T) \; ,
\label{eq:x43910ad} 
\end{eqnarray}
so that, by (\ref{eq:x2383n}), $B_{1,\infty}^R(T/\eps)\le \check{B}_1^R(T)$
which proves (\ref{eq:x2383a}) for $i=1$.
The key step here is the integration by parts at the second equality which makes explicit the slowly varying nature of $v_2$ by pulling out the explicit $\eps$ after the third equality.

To prove (\ref{eq:x2383a}) for $i=2$ we conclude
from (\ref{eq:x2353}),(\ref{eq:xn4.16}),
(\ref{eq:xn4.25a}) that, for $\zeta\in\R$,
\begin{eqnarray}
&& \hspace{-8mm} 
 \tilde{f}_2^R(v_1(s,\eps),\eps s,s;k,a) 
=-\frac{K^2}{2} e^{i[kv_1(s,\eps)-\eps a s]}
\sum_{n\in\Z\setminus\lbrace k\rbrace}
\widehat{jj}(n;k,\Delta P_{x0})e^{i(n-k)s} + cc 
\; ,
\nonumber
\end{eqnarray}
whence, by (\ref{eq:x2383}) for $\zeta\in\R$,
\begin{eqnarray}
&&\hspace{-18mm} 
B_2^R(\zeta)=\frac{K^2}{2} \Big{|} \int_0^\zeta \; e^{i[kv_1(s,\eps)-\eps a s]}
\sum_{n\in\Z\setminus\lbrace k\rbrace}
\widehat{jj}(n;k,\Delta P_{x0})e^{i(n-k)s}ds + cc \Big{|}
\nonumber\\
&&
\leq K^2 \sum_{n\in\Z\setminus\lbrace k\rbrace}
\big{|}\widehat{jj}(n;k,\Delta P_{x0})\big{|}\;
 \big{|} \int_0^\zeta \;  e^{i[kv_1(s,\eps)-\eps a s]} e^{i(n-k)s}ds\big{|} \; ,
\label{eq:x254n} 
\end{eqnarray}
where in the inequality we used the fact that the Fourier series of
$jj(\cdot;k,\Delta P_{x0})$ is uniformly
convergent. Integrating by parts gives, by (\ref{eq:n3.5e}),
(\ref{eq:xn4.20}),(\ref{eq:xn4.25}) for $0\leq\zeta\leq T/\eps$,
\begin{eqnarray}
&&\hspace{-5mm} 
|\int_0^\zeta\;  e^{i[kv_1(s,\eps)-\eps a s]}   e^{i(n-k)s}ds| 
= \Big{|}\frac{1}{i(n-k)}\biggl\lbrack   e^{i[kv_1(\zeta,\eps)-\eps a \zeta]} 
e^{i(n-k)\zeta}-e^{ik\theta_0} 
\nonumber\\
&&\hspace{-5mm} 
-\int_0^\zeta\; i(k\frac{dv_1}{ds}(s,\eps)-\eps a)
 e^{i[kv_1(s,\eps)-\eps a s]} 
e^{i(n-k)s}ds\biggr\rbrack\Big{|} 
\nonumber\\
&&\hspace{-5mm} 
\leq \frac{1}{|n-k|}\biggl\lbrack 2 +  
\int_0^\zeta\; (k|\frac{dv_1}{ds}(s,\eps)|+\eps|a|) ds\biggr\rbrack 
\nonumber\\
&&\hspace{-5mm} 
\leq \frac{1}{|n-k|}\biggl\lbrack 2 +  
\eps\int_0^\zeta\; (2k|v_2(s,\eps)|+|a|) ds\biggr\rbrack 
\nonumber\\
&&\hspace{-5mm} 
\leq \frac{1}{|n-k|}\biggl( 2 +  
\eps\zeta\; \biggl\lbrack |a| + 2k \chi_\infty(\theta_0,\chi_0,k,a) \biggr\rbrack
\biggr) 
\nonumber\\
&&\hspace{-5mm} 
\leq \frac{1}{|n-k|}\biggl( 2 +  
T\; \biggl\lbrack |a| + 2k \chi_\infty(\theta_0,\chi_0,k,a) \biggr\rbrack
\biggr) \; ,
\nonumber
\end{eqnarray}
whence, by (\ref{eq:x3900dcn}),(\ref{eq:x254n}) for $0\leq\zeta\leq T/\eps$,
\begin{eqnarray}
&&\hspace{-10mm} 
B_2^R(\zeta)\leq K^2 \biggl( 2 +  
T\; [ |a| + 2k \chi_\infty(\theta_0,\chi_0,k,a) ]\biggr)
\nonumber\\
&&\times \sum_{n\in\Z\setminus\lbrace k\rbrace}
\frac{|\widehat{jj}(n;k,\Delta P_{x0})|}{|n-k|}
=\check B_{2}^R(T) \; ,
\label{eq:x256n} 
\end{eqnarray}
so that, by (\ref{eq:x2383n}),
$B_{2,\infty}^R(T/\eps)\le \check B_2^R(T)$.
This completes the proof.
\setcounter{equation}{0}
\section{Summary and future work}
\label{n5}
%
%
%
We started with the 6D Lorentz equations for a planar undulator in
(\ref{eq:2.20}),(\ref{eq:2.50})-(\ref{eq:2.52}) with time as the
independent variable. In \S\ref{2.2} we introduced $z$ as the
independent variable and considered the IVP at $z=0$ with 
$y_0=p_{y0}=0$. 
Solutions of this system are completely determined by the solutions of our basic 2D system (\ref{eq:2.32}),(\ref{eq:2.33}) for $\alpha$ and $\gamma$.
This basic 2D system is the starting point for the rest of the paper and the first step is to transform it into a form for first-order averaging; the subject of \S\ref{2.3}.
We introduce
$\zeta=k_uz$ as the new independent variable, and $\chi$ as a new dependent variable by $\gamma=\gamma_c(1+\eps\chi)$.
Here we are thinking of electrons as part of an electron bunch with $\gamma_c$ as a characteristic value of $\gamma$ and $\eps$ as a measure of the energy spread so that $\chi$ is an $O(1)$ variable. We thus arrive at the
system for $(\theta_{aux}, \chi)$ given in (\ref{eq:2.41}),(\ref{eq:2.42}) and
we are interested, in this FEL application, in an asymptotic analysis for $\eps$ and $1/\gamma_c$ small.
Expanding the vector field for (\ref{eq:2.41}),(\ref{eq:2.42}) gives (\ref{eq:2.48}),(\ref{eq:2.48a}). Here $\theta_{aux}$ is not slowly varying and we thus
introduce the generalized ponderomotive phase, $\theta$, in (\ref{eq:2.66}) which
leads to the slowly varying form of (\ref{eq:2.48n}),(\ref{eq:2.48an}).
Most importantly, we discover that in order for $\theta$ and $\chi$ to interact at first order we must have $\eps=O(1/\gamma_c)$ and without loss of generality we take (\ref{eq:1.13}) as a result of  (\ref{eq:2.49}). 
Finally we obtain (\ref{eq:2.72}),(\ref{eq:2.73}) which is in a
standard form for the MoA.
Consequently this will lead to a pendulum type behavior which is central to the operation of an FEL. 

The MoA can be applied to 
(\ref{eq:2.72}),(\ref{eq:2.73}) after an appropriate $h$ is
defined and the rest of the paper, in Sections \ref{3},\ref{4}, focuses
on the monochromatic case of (\ref{eq:2.170}).

Before continuing with the summary we note that
in the collective case there is a continuous range of frequencies and so
it is natural to ask, ``what happens in the noncollective case considered in this paper if there
is a continuous range of frequencies?''. Here $h$ can be modeled as in (\ref{eq:2.84}), i.e.,
\begin{eqnarray}
h(\alpha)=\int^\infty_{-\infty}\; \tilde{h}(\xi)\exp(-i\xi\alpha) d\xi \; .
\label{eq:n2.45ba}
\end{eqnarray}
In the nonsmooth monochromatic case 
$\tilde{h}(\xi)=[\delta(\xi-\nu)+\delta(\xi+\nu)]/2$ and 
(\ref{eq:n2.45ba}) gives $h(\alpha)=\cos(\nu\alpha)$ as in
the monochromatic case of
(\ref{eq:2.170}), and, as we have discussed in \S\ref{3}, there are
resonances for integer $\nu$. However we have found that in the smooth case
the average of $(\cos\zeta+ \Delta P_{x0})h(\theta-Q(\zeta))$ is zero
and so the averaging normal form for
(\ref{eq:2.72}),(\ref{eq:2.73}) is just the NR normal form
of \S\ref{3.3}. Thus a smooth $\tilde{h}(\xi)$, localized
near the $\nu=1$ monochromatic resonance, washes out the effect
of that resonance in the first-order averaging normal form.
This does not mean that there is no resonant behavior near 
$\nu=1$ because it may not be possible to prove an
averaging theorem. We are pursuing this. Furthermore even if an 
averaging theorem can be proven there might still be an effect in second-order
averaging.

In \S\ref{3} we begin by determining the $O(\eps^2)$ terms of
(\ref{eq:2.72}),(\ref{eq:2.73}) using
(\ref{eq:2.74}),(\ref{eq:2.75}). Thus we obtain
(\ref{eq:3.19})-(\ref{eq:3.24}) as our basic system for
$\theta,\chi$. 
Proposition \ref{P1} gives a domain, $W(\eps_0)\times\R$, 
on which $g_1,g_2$ are well defined
as well as their limits as $\eps\rightarrow 0+$. In particular the
vector field in (\ref{eq:3.19}),(\ref{eq:3.20})
is well defined on $W(\eps_0)\times\R$.

Eq.'s (\ref{eq:3.19}),(\ref{eq:3.20}) are in a standard form
for the MoA and for each $\nu$ the normal form is obtained by dropping
the $O(\eps^2)$ terms and averaging $f_1,f_2$ over $\zeta$. 
However the average of $f_2$ is not clear from
(\ref{eq:3.22}) and it is convenient to expand 
it in a Fourier series which is given in 
(\ref{eq:3.41})-(\ref{eq:3.43}). The average is then easily obtained
in (\ref{eq:3.45}) and leads to the definition of NR, 
$\Delta$-NR, resonant and
NtoR $\nu$. The NR normal form equations are
$\theta'= \eps 2\chi$ and $\chi'=0$ and the resonant normal form
equations are given by (\ref{eq:3.46}).
The NR case is stated precisely in \S\ref{3.3}. Instead of focusing
on the resonant case of (\ref{eq:3.46})  we consider in \S\ref{3.4}
the more general NtoR case where we study the dynamics in neighborhoods
of the $\nu=k$ resonances. If the neighborhood is too small then the
resonant normal form of (\ref{eq:3.46})  will be
dominant thus the natural neighborhood to study with first-order
averaging is $O(\eps)$ and this is the content of \S\ref{3.4}. 
Replacing $\nu$ by $k+\eps a$, our basic equations
(\ref{eq:3.19}),(\ref{eq:3.20}) are rewritten in
(\ref{eq:3.63}),(\ref{eq:3.64}). The function 
$f_2$ in (\ref{eq:3.64}) has two $\eps$ dependencies
one of which contributes to the $O(\eps^2)$ term and we are led
to the basic NtoR system (\ref{eq:430011})-(\ref{eq:n2.101aada}).
Proposition \ref{P2} is analogous to Proposition \ref{P1}
by giving us the domain $W(\eps_0)\times\R$
on which $g_1^R,g_2^R$ are well behaved 
as well as their limits as $\eps\rightarrow 0+$. In particular the
vector field in (\ref{eq:430011}),(\ref{eq:430010})
is well defined on $W(\eps_0)\times\R$.
In \S\ref{3.4.2} the NtoR normal form is presented in
(\ref{eq:nn430016}),(\ref{eq:nn430015}). The solution structure is
conveniently illuminated, in terms of the simple pendulum system, in
\S\ref{3.4.3}. The simple pendulum exhibits four types of
behavior and these are exploited to discuss the structure of solutions
of (\ref{eq:nn430016}),(\ref{eq:nn430015}) in these four cases.

At this stage we have normal forms for 
$\nu\in[k+\Delta,k+1-\Delta]$ and $\nu=k+\eps a$. However there may be gaps between the dynamics covered by the $\Delta$-NR normal form and that of the NtoR normal form. So it is comforting to note that there is a link between the two dynamical behaviors in that the NtoR normal form
is approximated by the NR normal form far away from the pendulum buckets as discussed in
\S\ref{3.4.4}.

In \S\ref{3.5} we state the two averaging theorems which relate
the $\Delta$-NR and NtoR normal form approximations to the corresponding exact systems. Each theorem
has a detailed preamble which sets up a compact statement of the
theorem. The theorems establish the main results of the paper. Namely that the normal form solutions give an $O(\eps)$ approximation to the exact solutions on long time, $O(1/\eps)$, intervals.  In the $\Delta$-NR case, the $\nu$ interval can be made larger by making $\Delta$ smaller but this is at the expense of increasing the error as discussed in Remark (1) of \S3.5.3.

The results of the theorems are applied in
\S\ref{3.6}, where the normal form approximations are used to derive the approximate solutions of the Lorentz equations with $z$ as the independent variable. In \S\ref{3.7} we discuss the small gain theory
for $\nu=k+\eps a$ based on our NtoR normal form and compare it with the
standard theory for $k=1, \;a=0$. We do point out however, that we have not justified the low gain theory in the context of our NtoR averaging theorem as we 
mention at the end of \S 3.7. 

Finally the proofs are given in \S\ref{4}. It can be seen that the proofs themselves are quite simple. 
The proofs are somewhat novel in that they do not use a near identity transformation, due to the Besjes approach, and they use a system of differential inequalities in the calculation of the error bounds, rather than a Gronwall type inequality, which leads to better error bounds. Therefore a solution 
of the system of differential inequalities is presented and verified in Appendix \ref{I}. 
The first theorem, which is stated for the $\Delta$-NR case, 
is an example of a quasiperiodic averaging theorem with its concomitant small divisor problem. It's inherently interesting in that the small divisor problem arises in what must be the simplest possible way.
We develop the general theory of quasiperiodic averaging in \cite{DEH}. 
The second theorem, which is stated for the NtoR case, 
is an example of periodic averaging which has a vast literature, however as mentioned above our approach here is novel.
While the proofs of Theorems 1 and 2  are simple the whole application of the MoA is not. There was considerable work to put the problem into the standard form and considerable effort to calculate the bounds on $g_1,g_2$ in Appendix \ref{C} and  $g_1^R,g_2^R$ in Appendix \ref{E} as well as their $\eps=0$ limits in Appendixes \ref{B} and \ref{D}.

We now comment on future work. First of all it would be interesting to include the $y$ dynamics using  (\ref{eq:2.35}) as we do, but not assuming the zero initial conditions in $y$, thus treating the full 3D dynamics. 

Secondly, it would be interesting to study the helical undulator as we have done here for the planar undulator, i.e., via first-order averaging.

Thirdly, the
work here sets the stage for a second-order averaging study of the
NR case in (\ref{eq:3.19}),(\ref{eq:3.20}) using (\ref{eq:3.29}),(\ref{eq:3.30})
and the NtoR case in (\ref{eq:430011}),(\ref{eq:430010}) using
(\ref{eq:xnH.20}),(\ref{eq:xnH.60}). In both cases we have systems of the
form
\bea
&&\hspace{-8mm} 
\frac{dU}{dt} = \eps F(U,t) + \eps^2 G(U,t) + O(\eps^3) \; ,
\label{eq:n4.11a}
\eea
with approximating normal form given by
\bea
&&\hspace{-8mm} 
\frac{dV}{dt} = \eps \bar{F}(V) + \eps^2 \hat{G}(V) \; ,
\label{eq:n4.11b}
\eea
where $\bar{F}$ is the $t$-average of $F$ and $\hat{G}$ is a linear combination of the $t$-average of $G$ and terms depending on $F$ (See \cite[Section 5, p.610]{SSC} for a construction of the normal form, i.e., $\hat G$, 
and an associated theorem and proof). Such a study
would include a computation of the averages from
(\ref{eq:3.29}),(\ref{eq:3.30}) and
(\ref{eq:xnH.20}),(\ref{eq:xnH.60}) and then a phase plane analysis
of this second order normal form system including a comparison 
with our first-order
normal form system. 
In addition averaging theorems could be proven
which we anticipate will give an $O(\eps^2)$ error on $[0,T/\eps]$ as in \cite{SSC}. Furthermore, it would be interesting to see what happens in the NR case, e.g., is the energy deviation $\chi$ still conserved.
We note that generically second-order averaging gives a better error estimate but the interval of validity remains the same (See \cite{SSC} for situations where the time interval can be extended).
Finally it would be interesting to know if, in the NtoR case, there is a breakdown in the integrability of the NtoR normal form due to separatrix splitting, \cite{SS}, with the concomitant chaotic behavior. 
This is a delicate issue, which cannot be studied with second-order
averaging, since (\ref{eq:n4.11b}) is a second order autonomous system and as such it cannot exhibit chaos as pointed out at the end of \S\ref{3.4.3}.
This work could be a possible future project, however it does not appear to be interesting from the application point of view since collective effects are surely more important than noncollective effects at second order. 

Fourthly, we are therefore eager to move on to the collective case based in part on our understanding here.
As a first step we are studying the consequence of
(\ref{eq:H.10})-(\ref{eq:H.30}). We have not seen this form of the solution
of the 1D wave equation in the FEL literature although the first equality
in (\ref{eq:H.20}) is derived in many elementary PDE books.
In addition, we are pursuing the issue raised in the paragraph containing Eq. (\ref{eq:n2.45ba}), concerning a smooth $\tilde{h}$.
\section*{Acknowledgments}
\addcontentsline{toc}{section}{Acknowledgments}
\setcounter{equation}{0}
The work of JAE and KH was supported by DOE under DE-FG-99ER41104. The work of MV was supported by DESY. Matt Gooden played a significant role in the early stages of this work and was supported by a Teng summer fellowship at ANL and by an NSF EMSW21-MCTP grant, DMS 0739417, at UNM. Discussions with H.S. Dumas, Z. Huang, K-J Kim, R. Lindberg, B.F. Roberts and R. Warnock are gratefully acknowledged. A special thanks to R. Lindberg for several very helpful comments during the formulation of our approach and a special thanks to Z. Huang and K-J Kim for allowing us to sit in on their USPAS FEL course.
\setcounter{equation}{0}
\section*{Table of notation}
\label{n6}
\addcontentsline{toc}{section}{Table of notation}
\vspace{-5mm} 
\bea
&&\hspace{-8mm} 
\begin{array}{ll}  
a & \qquad  (\ref{eq:3.62})\\
B_1,B_2 & \qquad  (\ref{eq:2383})\\
B_1^R,B_2^R & \qquad  (\ref{eq:x2383}) \\
\D(\eps,\nu) & \qquad  (\ref{eq:3.15})\\ 
{\cal E} & \qquad (\ref{eq:1.14}) \\
f_1,f_2 & \qquad  (\ref{eq:3.21}),(\ref{eq:3.22})\\
f_1^R,f_2^R & \qquad  (\ref{eq:430010an}),(\ref{eq:430011an})\\
g_1,g_2 & \qquad  (\ref{eq:3.23}),(\ref{eq:3.24})\\
g_1^R,g_2^R & \qquad  (\ref{eq:n2.101aaca}),(\ref{eq:n2.101aada})\\
h, H & \qquad (\ref{eq:2.170})\\
jj,\hat{jj} & \qquad  (\ref{eq:3.33}),(\ref{eq:3.40})\\
K & \qquad (\ref{eq:1.10}) \\
K_r & \qquad  (\ref{eq:1.12})\\
K_0  & \qquad (\ref{eq:nn430015x}) \\
{\rm MoA} & \qquad {\rm Method\;of\;Averaging} \\
{\rm NR} \;({\rm nonresonant}) & \qquad  {\rm Definition}\;\ref{D1}\; 
(\S\ref{3.2})\\
{\rm NtoR} \;({\rm near-to-resonant}) & \qquad {\rm Definition}\;\ref{D1}\;
(\S\ref{3.2})  \\
\N & \qquad {\rm Set\;of\;positive\;integers} \\
P_x, P_z  & \qquad  (\ref{eq:2.36})\\ 
q,\bar{q},Q & \qquad  (\ref{eq:2.59}),(\ref{eq:2.61}),
(\ref{eq:2.67})\\
W(\eps),\hat{W},\hat{W}_R & \qquad (\ref{eq:3.27}),(\ref{eq:n2.201aa}),
(\ref{eq:n2.201aaa})\\
\Z & \qquad {\rm Set\;of\;integers} \\
\check{\alpha},\alpha
 & \qquad (\ref{eq:2.45}),(\ref{eq:2.19})\\
\gamma_c & \qquad  (\ref{eq:2.39})\\
\Delta  & \qquad  {\rm Definition}\;\ref{D1}\; (\S\ref{3.2})\\
\Delta-{\rm NR} \;(\Delta-{\rm nonresonant}) & \qquad  
{\rm Definition}\;\ref{D1}\; (\S\ref{3.2})\\
\Delta P_{x0} & \qquad  (\ref{eq:2.46})\\
\eps & \qquad (\ref{eq:1.13}) \\
\zeta  & \qquad  (\ref{eq:2.38})\\
\eta  & \qquad  (\ref{eq:2.39})\\
\theta_{aux},\theta 
& \qquad  (\ref{eq:2.85}),(\ref{eq:2.66})\\
\Pi_x,\Pi_z,\Pi_{x,ub},\Pi_{z,lb}  & \qquad  (\ref{eq:3.11}),(\ref{eq:3.12}),
(\ref{eq:3.26}),(\ref{eq:nC.222ig}) \\ 
\Upsilon_0,\Upsilon_1 & \qquad  (\ref{eq:2.68})\\
\chi, \chi_{lb}(\eps)  
& \qquad  (\ref{eq:2.39}),(\ref{eq:3.28})\\
\Omega & \qquad (\ref{eq:n3.3})
\end{array} 
\nonumber
\eea
\renewcommand{\thesection}{\Alph{section}}
\setcounter{subsection}{0}
\setcounter{section}{0}
\setcounter{equation}{0}
\section*{Appendix}
\addcontentsline{toc}{section}{Appendix}
\setcounter{equation}{0}
\section{The Bessel expansion} 
\label{A}
Here we derive the Bessel expansion (\ref{eq:3.42}) of 
$jj(\cdot;\nu,\Delta P_{x0})$. In fact by (\ref{eq:3.33})
\begin{eqnarray}
&&  \hspace{-10mm}
jj(\zeta;\nu,\Delta P_{x0}) 
= (\cos\zeta+ \Delta P_{x0})\exp(-i\nu\Upsilon_0\sin\zeta)
\exp(-i\nu\Upsilon_1\sin 2\zeta)
\nonumber\\
&&=
\frac{1}{2}jj_1(\zeta)
+ \frac{1}{2}jj_{-1}(\zeta)
+ \Delta P_{x0} jj_0(\zeta) \; ,
\label{eq:A.10} 
\end{eqnarray}
where
\begin{eqnarray}
&&  \hspace{-15mm}  
jj_m(\zeta) :=\exp(im\zeta)
\exp(-i\nu[\Upsilon_0\sin\zeta+\Upsilon_1\sin 2\zeta]) \; .
\label{eq:A.15} 
\end{eqnarray}
Now
\begin{eqnarray}
&& \hspace{-5mm}
\exp(ix\sin\theta) = \sum_{n\in{\mathbb Z}} J_n(x)\exp(in\theta) \; , 
\quad  J_{-n}(x)=(-1)^nJ_n(x) \; ,
\label{eq:A.20}
\end{eqnarray}
whence, by (\ref{eq:A.15}),
\begin{eqnarray}
&&  \hspace{-8mm} 
jj_m(\zeta) 
= e^{im\zeta} e^{-i\nu\Upsilon_0\sin\zeta}e^{-i\nu\Upsilon_1\sin2\zeta}
\nonumber\\
&&= e^{im\zeta} [\sum_{k\in\Z} J_k(\nu\Upsilon_1)e^{-i 2k\zeta}]\;
            [\sum_{l\in\Z} J_l(\nu\Upsilon_0)e^{-il\zeta}]
\nonumber\\
&& = \sum_{k,l\in\Z} \; J_l(\nu\Upsilon_0)J_k(\nu\Upsilon_1)
e^{i(m-l-2k)\zeta}
\nonumber\\
&&  = \sum_{n\in{\mathbb Z}} \biggl
(\sum_{k\in{\mathbb Z}} J_{m-n-2k}(\nu\Upsilon_0)J_k(\nu\Upsilon_1)\biggr)
e^{in\zeta} \; .
\label{eq:A.25} 
\end{eqnarray}
Let 
\begin{eqnarray}
&&  \hspace{-0mm} 
{\cal J}(n,m,\nu,\Upsilon_0,\Upsilon_1):=
\sum_{k\in{\mathbb Z}} J_{m-n-2k}(\nu\Upsilon_0)J_k(\nu\Upsilon_1)\; ,
\label{eq:A.30} 
\end{eqnarray}
then, by (\ref{eq:A.25}),
\begin{eqnarray}
&&  \hspace{-8mm}
jj_m(\zeta) =
\sum_{n\in{\mathbb Z}} {\cal J}(n,m,\nu,\Upsilon_0,\Upsilon_1)e^{in\zeta} \; ,
\label{eq:A.35} 
\end{eqnarray}
and thus, by (\ref{eq:A.10}),
\begin{eqnarray}
&&  jj(\zeta;\nu,\Delta P_{x0}) 
= \sum_{n\in{\mathbb Z}} \biggl(  
\frac{1}{2}{\cal J}(n,1,\nu,\Upsilon_0,\Upsilon_1)
+\frac{1}{2}{\cal J}(n,-1,\nu,\Upsilon_0,\Upsilon_1)
\nonumber\\
&& \hspace{-5mm}
+ \Delta P_{x0}{\cal J}(n,0,\nu,\Upsilon_0,\Upsilon_1)\biggr) e^{in\zeta} \; ,
\label{eq:A.40} 
\end{eqnarray}
whence, by (\ref{eq:3.40}),
\begin{eqnarray}
&&  \hspace{-5mm} \widehat{jj}(n;\nu,\Delta P_{x0}) 
= \frac{1}{2}{\cal J}(n,1,\nu,\Upsilon_0,\Upsilon_1)
+\frac{1}{2}{\cal J}(n,-1,\nu,\Upsilon_0,\Upsilon_1)
\nonumber\\
&& 
+ \Delta P_{x0}{\cal J}(n,0,\nu,\Upsilon_0,\Upsilon_1) \; ,
\label{eq:A.45} 
\end{eqnarray}
so that indeed (\ref{eq:3.42}) holds.

It is useful for the discussion after Definition \ref{D1}
to have the following special case. We have, by (\ref{eq:A.45}),
\begin{eqnarray}
&&  \hspace{-5mm} \widehat{jj}(k;k,0) 
= \frac{1}{2}[{\cal J}(k,1,k,0,\Upsilon_1)
+{\cal J}(k,-1,k,0,\Upsilon_1)] \; ,
\label{eq:A.46} 
\end{eqnarray}
where
\begin{eqnarray}
&&  \hspace{-10mm} 
{\cal J}(k,1,k,0,\Upsilon_1) =
\sum_{k'\in{\mathbb Z}} J_{1-k-2k'}(0)J_{k'}(k\Upsilon_1)
\nonumber\\
&&= \left\{ \begin{array}{ll}  
J_{(1-k)/2}(k\Upsilon_1)  & \quad {\rm if\;}k{\rm\;odd}  \\ 
                     0 & \quad {\rm if\;}k{\rm\;even}  
 \; , \end{array} 
                  \right.
\label{eq:A.47} \\
&&  \hspace{-10mm} 
{\cal J}(k,-1,k,0,\Upsilon_1) =
\sum_{k'\in{\mathbb Z}} J_{-1-k-2k'}(0)J_{k'}(k\Upsilon_1)
\nonumber\\
&&
= \left\{ \begin{array}{ll}  
J_{-(1+k)/2}(k\Upsilon_1) & \quad {\rm if\;}k{\rm\;odd}  \\ 
                     0 & \quad {\rm if\;}k{\rm\;even}  
 \; . \end{array} 
                  \right.
\label{eq:A.48} 
\end{eqnarray}
Thus from (\ref{eq:A.46}) $\widehat{jj}(k;k,0)=0$ for $k$ even and,
for $k=2n+1$ with $n\in\Z$,
\begin{eqnarray}
&&  \hspace{-5mm} \widehat{jj}(2n+1;2n+1,0) 
= \frac{1}{2}[ J_{-n}((2n+1)\Upsilon_1) + J_{-(n+1)}((2n+1)\Upsilon_1)]
\nonumber\\
&&= \frac{1}{2}(-1)^n[ J_{n}((2n+1)\Upsilon_1) - J_{n+1}((2n+1)\Upsilon_1)] \; .
\label{eq:A.49} 
\end{eqnarray}
\section{Limit of $g_1,g_2$}
\label{B}
\setcounter{equation}{0}
Let $\eps\in(0,\eps_0]$ with $\eps_0\in(0,1]$, let $\nu\in[1/2,\infty)$
and let $(\theta,\chi,\zeta)\in W(\eps_0)\times\R$. 
In this appendix we will prove the properties (\ref{eq:xnB.10n}),
(\ref{eq:nB.63}),(\ref{eq:xnB.10n2}),(\ref{eq:nB.632}) of $g_1$ and $g_2$.
The properties (\ref{eq:nB.63}),(\ref{eq:nB.632})
are used in the proof of Proposition \ref{P1}. Furthermore
the properties (\ref{eq:xnB.10n}),(\ref{eq:xnB.10n2})
will be used in Appendix \ref{C}. Since all assumptions of this
appendix are also satisfied in Appendix \ref{B}, we can apply the results
of Appendix \ref{B}.

We first consider $g_1$.
Note that, by (\ref{eq:2.59}),(\ref{eq:3.11}),
\begin{eqnarray}
&&\hspace{-8mm} 
1 + K^2\Pi_x^2(\theta,\zeta,\eps,\nu)=
q(\zeta) 
\nonumber\\
&&+ \frac{\eps^2K^2 \bar{q}}{2\nu}
\biggl( \sin(\nu[\theta - Q(\zeta)]) -\sin(\nu\theta_0) \biggr)
\biggl( 2(\cos\zeta + \Delta P_{x0})
\nonumber\\
&&\quad +\frac{\eps^2 \bar{q}}{2\nu} 
( \sin(\nu[\theta - Q(\zeta)]) -\sin(\nu\theta_0) )\biggr) \; .
\label{eq:xn2.101aabaa} 
\end{eqnarray}
We obtain from (\ref{eq:3.23}) that
\bea
&&\hspace{-8mm}
\eps^2 g_1( \theta,\chi,\zeta;\eps,\nu) 
=\frac{2{\cal E}}{\eps^2\bar{q}}(1-\frac{1}{\Pi_z(\theta,\chi,\zeta,\eps,\nu)}) 
+\frac{q(\zeta)}{\bar{q}} (1 - 2\eps\chi) \; , 
\nonumber
\eea
whence
\bea
&&\hspace{-8mm}
\frac{1}{2{\cal E}}\bar{q} \Pi_z(\Pi_z + 1) \eps^4 g_1
=  \Pi_z^2 - 1 + \frac{1}{2{\cal E}}q  \Pi_z(\Pi_z + 1) \eps^2(1 - 2\eps\chi)
\nonumber\\
&&
=\frac{1}{(1 + \eps\chi)^2}
\biggl( -\frac{\eps^2}{{\cal E}}( q + \eps^2\kappa_1 )
 + \frac{1}{2{\cal E}}q  \Pi_z(\Pi_z + 1) \eps^2  (1 + \eps\chi)^2
(1 - 2\eps\chi) \biggr) \; ,
\nonumber\\
\label{eq:xnB.10}
\eea
where we used from (\ref{eq:3.12}),(\ref{eq:xn2.101aabaa})
the fact that
\bea
&&\hspace{-8mm}
\Pi_z^2(\theta,\chi,\zeta,\eps,\nu) - 1 =  
-\frac{\eps^2}{{\cal E}(1 + \eps\chi)^2}
\biggl( q(\zeta) + \eps^2\kappa_1(\theta,\zeta,\eps,\nu) \biggr) \; ,
\label{eq:nB.61a} 
\eea
with
\bea
&&\hspace{-8mm}
\kappa_1(\theta,\zeta,\eps,\nu):=
\frac{K^2 \bar{q}}{2\nu}
\biggl( \sin(\nu[\theta - Q(\zeta)]) -\sin(\nu\theta_0) \biggr)
\biggl( 2(\cos\zeta + \Delta P_{x0})
\nonumber\\
&&\quad +\frac{\eps^2 \bar{q}}{2\nu} 
( \sin(\nu[\theta - Q(\zeta)]) -\sin(\nu\theta_0) )\biggr) \; .
\label{eq:nB.61} 
\eea
Clearly, by (\ref{eq:xnB.10}),(\ref{eq:nB.61a}),
\bea
&&\hspace{-8mm}
\frac{1}{2{\cal E}}\bar{q} \Pi_z(\Pi_z + 1) \eps^4 g_1
\nonumber\\
&&=  -\frac{\eps^2 q}{{\cal E} (1 + \eps\chi)^2}
\biggl(  1 - \frac{1}{2}\Pi_z(\Pi_z + 1)(1-3\eps^2\chi^2 - 2\eps^3\chi^3)
\biggr)
 -\frac{\eps^4\kappa_1}{{\cal E}(1 + \eps\chi)^2}
\nonumber\\
&&=  -\frac{\eps^2q}{{\cal E} (1 + \eps\chi)^2}
\biggl(  -\frac{1}{2}(\Pi_z-1)(\Pi_z + 2) + \frac{1}{2}\Pi_z(\Pi_z + 1)
(3\eps^2\chi^2 + 2\eps^3\chi^3)\biggr)
\nonumber\\
&&\quad
 -\frac{\eps^4\kappa_1}{{\cal E}(1 + \eps\chi)^2} \; ,
\nonumber
\eea
whence 
\bea
&&\hspace{-8mm}
\frac{1}{2{\cal E}}\bar{q} \Pi_z(\Pi_z + 1)^2 \eps^4 g_1
\nonumber\\
&& =  -\frac{\eps^2q}{ 2{\cal E}(1 + \eps\chi)^2}
\biggl(  -(\Pi_z^2-1)(\Pi_z + 2) + \eps^2 \Pi_z(\Pi_z + 1)^2
(3\chi^2 + 2\eps\chi^3)\biggr)
\nonumber\\
&&\quad
 -\frac{\eps^4\kappa_1}{{\cal E}(1 + \eps\chi)^2}
=  -\frac{\eps^2q}{ 2{\cal E}(1 + \eps\chi)^4}
\biggl(   \frac{\eps^2}{{\cal E}}( q + \eps^2\kappa_1 )
(\Pi_z + 2) 
\nonumber\\
&&\quad
+ \eps^2 \Pi_z(\Pi_z + 1)^2
(3\chi^2 + 2\eps\chi^3)(1 + \eps\chi)^2  \biggr)
-\frac{\eps^4\kappa_1}{{\cal E}(1 + \eps\chi)^2} 
\nonumber\\
&&=  -\frac{\eps^2q}{ 2{\cal E}(1 + \eps\chi)^4}
\biggl(   \frac{\eps^2}{{\cal E}} q (\Pi_z + 2) 
+ \eps^2 \Pi_z(\Pi_z + 1)^2
(3\chi^2 + 2\eps\chi^3)(1 + \eps\chi)^2  \biggr)
\nonumber\\
&&\quad -\frac{\eps^6 q(\Pi_z + 2)\kappa_1}{2{\cal E}^2(1 + \eps\chi)^4} 
-\frac{\eps^4\kappa_1}{{\cal E}(1 + \eps\chi)^2} 
\nonumber\\
&&=  -\frac{\eps^2q}{ 2{\cal E}(1 + \eps\chi)^4}
\biggl(   \frac{\eps^2}{{\cal E}} q (\Pi_z + 2) 
+ \eps^2 \Pi_z(\Pi_z + 1)^2
(3\chi^2 + 2\eps\chi^3)(1 + \eps\chi)^2  \biggr)
\nonumber\\
&&\quad -\frac{\eps^4\kappa_1}{ 2{\cal E}(1 + \eps\chi)^4}
\biggl( 2(1 + \eps\chi)^2  + \frac{\eps^2}{{\cal E}}q(\Pi_z + 2)\biggr) \; , 
\nonumber
\eea
so that
\bea
&&\hspace{-8mm}
\bar{q} \Pi_z(\Pi_z + 1)^2 g_1
\nonumber\\
&&=  -\frac{q}{(1 + \eps\chi)^4}
\biggl(   \frac{q}{{\cal E}}  (\Pi_z + 2) 
+  \Pi_z(\Pi_z + 1)^2
(3\chi^2 + 2\eps\chi^3)(1 + \eps\chi)^2  \biggr)
\nonumber\\
&&\quad -\frac{\kappa_1}{(1 + \eps\chi)^4}
\biggl( 2(1 + \eps\chi)^2  + \frac{\eps^2 q}{{\cal E}}(\Pi_z + 2)\biggr) \; ,
\nonumber
\eea
i.e.,
\bea
&&\hspace{-8mm}
g_1( \theta,\chi,\zeta;\eps,\nu) 
=  -\frac{q}{\bar{q} \Pi_z(\Pi_z + 1)^2 (1 + \eps\chi)^4}
\biggl(   \frac{q}{{\cal E}} (\Pi_z + 2) 
\nonumber\\
&& 
+  \Pi_z(\Pi_z + 1)^2
(3\chi^2 + 2\eps\chi^3)(1 + \eps\chi)^2  \biggr)
\nonumber\\
&&\quad -\frac{\kappa_1}{\bar{q} \Pi_z(\Pi_z + 1)^2 (1 + \eps\chi)^4}
\biggl( 2(1 + \eps\chi)^2  + \frac{\eps^2q}{{\cal E}}
(\Pi_z + 2)\biggr) \; .
\label{eq:xnB.10n}
\eea
Clearly, by (\ref{eq:3.12}),(\ref{eq:nB.61}),
\bea
&&\hspace{-12mm}
\lim_{\eps\rightarrow 0+}\;[\Pi_z(\theta,\chi,\zeta,\eps,\nu)] = 1 \; , 
\label{eq:xnB.10na} \\
&& \hspace{-12mm}
\lim_{\eps\rightarrow 0+}\;[\kappa_1(\chi,\zeta,\eps,\nu)]
= \frac{K^2 \bar{q}}{\nu}
\biggl( \sin(\nu[\theta - Q(\zeta)]) -\sin(\nu\theta_0) \biggr)
(\cos\zeta + \Delta P_{x0}) \; ,
\nonumber\\
\label{eq:xnB.10nb}
\eea
whence, by (\ref{eq:xnB.10n}),
\bea
&&\hspace{-8mm}
\lim_{\eps\rightarrow 0+}\;[g_1( \theta,\chi,\zeta;\eps,\nu)] 
=  -\frac{q(\zeta)}{4\bar{q}} (   \frac{3}{{\cal E}} q(\zeta)  +  12\chi^2)
\nonumber\\
&&\quad - \frac{K^2}{2\nu}
\biggl( \sin(\nu[\theta - Q(\zeta)]) -\sin(\nu\theta_0) \biggr)
(\cos\zeta + \Delta P_{x0}) \; .
\label{eq:nB.63}
\eea
We now consider $g_2$ and we obtain from
(\ref{eq:3.24}) that
\bea
&&\hspace{-8mm}
\eps^2 g_2( \theta,\chi,\zeta;\eps,\nu) 
= \eps K^2 
\cos( \nu[\theta - Q(\zeta)])
\biggl( \cos\zeta + \Delta P_{x0} 
\nonumber\\
&&
-\frac{1}{1+\eps\chi}
\frac{\Pi_x(\theta,\zeta,\eps,\nu) }
{\Pi_z(\theta,\chi,\zeta,\eps,\nu)}\biggr) \; ,
\nonumber
\eea
whence
\bea
&&\hspace{-8mm}
\Pi_z(1+\eps\chi) \eps g_2
= K^2 
\cos( \nu[\theta - Q(\zeta)])\biggl( (1+\eps\chi)\Pi_z
(\cos\zeta + \Delta P_{x0}) -\Pi_x\biggr)
\nonumber\\
&& = K^2 
\cos( \nu[\theta - Q(\zeta)])\biggl( (\cos\zeta + \Delta P_{x0}) 
[(1+\eps\chi) \Pi_z - 1] - \eps^2\kappa_2 \biggr) \; ,
\label{eq:xnB.102}
\eea
where we used from (\ref{eq:3.11}) the fact that
\bea
&&\hspace{-8mm}
 \Pi_x(\theta,\zeta,\eps,\nu) =\cos\zeta + \Delta P_{x0} + \eps^2 
 \kappa_2(\theta,\zeta,\nu) \; ,
\label{eq:nB.61a2} 
\eea
with
\bea
&&\hspace{-8mm}
\kappa_2(\theta,\zeta,\nu):=
\frac{\bar{q}}{2\nu}[\sin(\nu[ \theta - Q(\zeta)]) 
-\sin(\nu\theta_0)] \; .
\label{eq:nB.612} 
\eea
Clearly, by (\ref{eq:xnB.102}),
\bea
&&\hspace{-8mm}
\Pi_z(1+\eps\chi) \eps g_2
= K^2 
\cos( \nu[\theta - Q(\zeta)])\biggl( 
(\cos\zeta + \Delta P_{x0})[\Pi_z-1 +\eps\chi\Pi_z]
- \eps^2\kappa_2 \biggr) \; ,
\nonumber
\eea
whence, by (\ref{eq:nB.61a}),
\bea
&&\hspace{-8mm}
  (\Pi_z + 1)\Pi_z(1+\eps\chi) \eps g_2 
\nonumber\\
&&
= K^2 
\cos( \nu[\theta - Q(\zeta)])\biggl( 
(\cos\zeta + \Delta P_{x0})[\Pi_z^2-1 +\eps\chi\Pi_z(\Pi_z + 1)]
\nonumber\\
&&
- \eps^2\kappa_2 (\Pi_z + 1)\biggr) 
\nonumber\\
&& = K^2 
\cos( \nu[\theta - Q(\zeta)])\biggl( 
(\cos\zeta + \Delta P_{x0})[
-\frac{\eps^2}{{\cal E}(1+\eps\chi)^2}
( q + \eps^2\kappa_1 )
\nonumber\\
&&\quad
+\eps\chi\Pi_z(\Pi_z + 1)]
- \eps^2\kappa_2(\Pi_z + 1) \biggr) \; ,
\nonumber
\eea
so that
\bea
&&\hspace{-8mm}
  \Pi_z(\Pi_z + 1)(1+\eps\chi)^3 \eps g_2 
\nonumber\\
&& = K^2 
\cos( \nu[\theta - Q(\zeta)])\biggl( 
(\cos\zeta + \Delta P_{x0})[
-\frac{\eps^2}{{\cal E}}
( q + \eps^2\kappa_1 )
\nonumber\\
&&\quad
+\eps\chi\Pi_z(\Pi_z + 1)(1+\eps\chi)^2]
- \eps^2\kappa_2(\Pi_z + 1)(1+\eps\chi)^2 \biggr) \; ,
\nonumber
\eea
which entails that
\bea
&&\hspace{-8mm}
  \Pi_z(\Pi_z + 1)(1+\eps\chi)^3  g_2 
\nonumber\\
&& = K^2 
\cos( \nu[\theta - Q(\zeta)])\biggl( 
(\cos\zeta + \Delta P_{x0})[
-\frac{\eps}{{\cal E}}
( q + \eps^2\kappa_1 )
\nonumber\\
&&\quad
+\chi\Pi_z(\Pi_z + 1)(1+\eps\chi)^2]
- \eps\kappa_2(\Pi_z + 1)(1+\eps\chi)^2 \biggr) \; ,
\nonumber
\eea
i.e.,
\bea
&&\hspace{-8mm}
g_2( \theta,\chi,\zeta;\eps,\nu)  
= \frac{K^2\cos( \nu[\theta - Q(\zeta)])}{ \Pi_z(\Pi_z + 1)(1+\eps\chi)^3}
\biggl( (\cos\zeta + \Delta P_{x0})[
-\frac{\eps}{{\cal E}}
( q(\zeta) + \eps^2\kappa_1 )
\nonumber\\
&&\quad
+\chi\Pi_z(\Pi_z + 1)(1+\eps\chi)^2]
- \eps\kappa_2(\Pi_z + 1)(1+\eps\chi)^2 \biggr) \; .
\label{eq:xnB.10n2}
\eea
Clearly, by (\ref{eq:xnB.10na}),(\ref{eq:xnB.10n2}),
\bea
&&\hspace{-8mm}
\lim_{\eps\rightarrow 0+}\;[g_2( \theta,\chi,\zeta;\eps,\nu)] 
= \chi K^2\cos( \nu[\theta - Q(\zeta)])
(\cos\zeta + \Delta P_{x0}) \; .
\label{eq:nB.632}
\eea
\section{Bounds on $g_1,g_2$}
\label{C}
\setcounter{equation}{0}
Let $\eps\in(0,\eps_0]$ with $\eps_0\in(0,1]$, let $\nu\in[1/2,\infty)$
and let $(\theta_0,\chi_0)\in W(\eps_0)$. Let also
\bea
&&\hspace{-8mm}
\chi_{lb}(\eps_0) < - \chi_M \; ,
\label{eq:nC.8aa} 
\eea
where $\chi_M$ is the positive constant from Theorem \ref{T1} (see item 2
of the setup list for Theorem \ref{T1}).
We also assume that 
\bea
&&  (\theta,\chi,\zeta)\in\R\times(\chi_0-d_2,\chi_0+d_2)\times\R \; ,
\label{eq:nC.8b}
\eea
where
\bea
&&\hspace{-8mm}
0 < d_2 < \chi_0 - \chi_{lb}(\eps_0) \; .
\label{eq:nC.8a} 
\eea
Note that, by 
(\ref{eq:3.27}),(\ref{eq:3.56}),(\ref{eq:nC.8b}),(\ref{eq:nC.8a}),
\bea
&&\hspace{-15mm}
(\theta,\chi,\zeta)\in\biggl( \R\times(\chi_0-d_2,\chi_0+d_2)\times\R\biggr)
\subset \biggl( W(\eps_0)\times\R \biggr)\subset  \D (\eps,\nu) 
\; .
\label{eq:nC.8c} 
\eea
In this appendix we will prove the properties
(\ref{eq:nC.43n}),(\ref{eq:nC.95n}) of $g_1$ and $g_2$.
We thus show in this appendix that the
properties (\ref{eq:nC.43n}),(\ref{eq:nC.95n})
hold in the situation of Theorem \ref{T1}
(see item 8 of the setup of Theorem \ref{T1}). Moreover
the properties (\ref{eq:nC.43n}),(\ref{eq:nC.95n})
will be used in Appendix \ref{E}. 

We first consider $g_1$ and we obtain from (\ref{eq:xnB.10n}) 
\bea
&&\hspace{-8mm}
|g_1|
=  \Big{|} -\frac{q}{\bar{q} \Pi_z(\Pi_z + 1)^2 (1 + \eps\chi)^4}
\biggl(   \frac{q}{{\cal E}} (\Pi_z + 2) 
\nonumber\\
&& 
+  \Pi_z(\Pi_z + 1)^2
(3\chi^2 + 2\eps\chi^3)(1 + \eps\chi)^2  \biggr)
\nonumber\\
&&\quad -\frac{\kappa_1}{\bar{q} \Pi_z(\Pi_z + 1)^2 (1 + \eps\chi)^4}
\biggl( 2(1 + \eps\chi)^2  + \frac{\eps^2 q}{{\cal E}}(\Pi_z + 2)\biggr) \Big{|}\; .
\label{eq:xnC.10n}
\eea
It follows from (\ref{eq:2.59}),(\ref{eq:2.61}),(\ref{eq:3.15}),
(\ref{eq:3.16}),(\ref{eq:nC.8c}) that
\bea
&&\hspace{-8mm}
q > 0 \; , \quad \bar{q} > 0 \; , \quad 1 + \eps\chi > 0 \; , \quad
0< \Pi_z < 1 \; ,
\nonumber\\
&&\hspace{-8mm} 3\chi^2 + 2\eps\chi^3 
= \chi^2 + 2\chi^2(1 + \eps\chi) \geq 0 \; ,
\nonumber\\
\label{eq:xnC.17n}
\eea
whence, by (\ref{eq:xnC.10n}),
\bea
&&\hspace{-8mm}
|g_1|
\leq  \frac{q}{\bar{q} \Pi_z(\Pi_z + 1)^2 (1 + \eps\chi)^4}
\biggl(  \frac{q}{{\cal E}} (\Pi_z + 2) 
\nonumber\\
&& 
+  \Pi_z(\Pi_z + 1)^2
(3\chi^2 + 2\eps\chi^3)(1 + \eps\chi)^2  \biggr)
\nonumber\\
&&\quad +\frac{|\kappa_1|}{\bar{q} \Pi_z(\Pi_z + 1)^2 (1 + \eps\chi)^4}
\biggl( 2(1 + \eps\chi)^2  + \frac{\eps^2q}{{\cal E}}(\Pi_z + 2)\biggr) 
\nonumber\\
&&
= \frac{q}{\bar{q}(1 + \eps\chi)^2}
\biggl(  \frac{q (\Pi_z + 2)}{{\cal E}\Pi_z(\Pi_z + 1)^2(1 + \eps\chi)^2}
+  3\chi^2 + 2\eps\chi^3  \biggr)
\nonumber\\
&&\quad +\frac{|\kappa_1|}{\bar{q} \Pi_z(\Pi_z + 1)^2 (1 + \eps\chi)^2}
\biggl( 2  + \frac{ \eps^2 q(\Pi_z + 2)}{{\cal E}(1 + \eps\chi)^2}
\biggr) \; .
\label{eq:xnC.10nn}
\eea
Note also that, by (\ref{eq:3.12}),
(\ref{eq:3.25}),
\begin{eqnarray}
&& \hspace{-8mm} 
\Pi_z^2(\theta,\chi,\zeta,\eps,\nu) 
= 1- \frac{\eps^2}{{\cal E}}\frac{1 + K^2 \Pi_x^2(\theta,\zeta,\eps,\nu)}
{(1+\eps\chi)^2} 
\nonumber\\
&&
\geq 1- \frac{\eps^2}{{\cal E}}\frac{1 + K^2  \Pi_{x,ub}^2(\eps)}{(1+\eps\chi)^2} \; .
\label{eq:nC.222ib} 
\end{eqnarray}
Moreover $\eps^2/(1+\eps\chi)^2$ and 
$1 + K^2  \Pi_{x,ub}^2(\eps,\nu)$ are increasing w.r.t. $\eps$
whence, by (\ref{eq:nC.222ib}),
\begin{eqnarray}
&& \hspace{-8mm} 
\Pi_z^2(\theta,\chi,\zeta,\eps,\nu) 
\geq 1- \frac{\eps_0^2}{{\cal E}}
\frac{1 + K^2  \Pi_{x,ub}^2(\eps_0)}{(1+\eps_0\chi)^2}
 \; .
\label{eq:nC.222ic} 
\end{eqnarray}
Since $0<\eps\leq\eps_0$ we have, by (\ref{eq:nC.8b}),
\bea
&&\hspace{-8mm}
1 + \eps\chi > 1 + \eps(\chi_0-d_2)\geq 1 + \inf_{\eps\in(0,\eps_0]}\;
(\eps(\chi_0-d_2)) 
= 1 + min(0,\eps_0(\chi_0-d_2)) 
\nonumber\\
&&
=:\kappa_3(\chi_0,\eps_0,d_2)\; .
\label{eq:nC.80}
\eea
Note that, by (\ref{eq:3.28}),
(\ref{eq:nC.8a}),
\bea
&&\hspace{-8mm}
1+\eps_0(\chi_0-d_2) >  1+\eps_0\chi_{lb}(\eps_0) > 0 \; ,
\label{eq:nC.90}
\eea
whence, by (\ref{eq:nC.80}),
\bea
&&\hspace{-8mm}
\kappa_3(\chi_0,\eps_0,d_2) > 0 \; ,
\label{eq:nC.91}
\eea
so that, for $n\in\N$ and by (\ref{eq:nC.80}),
\bea
&&\hspace{-8mm}
\frac{1}{(1+\eps\chi)^n} < \frac{1}{ \kappa_3^n(\chi_0,\eps_0,d_2)} \; . 
\label{eq:nC.92}
\eea
It follows from (\ref{eq:nC.222ic}),(\ref{eq:nC.92}),
\begin{eqnarray}
&& \hspace{-8mm} 
\Pi_z^2(\theta,\chi,\zeta,\eps,\nu) 
> \check{\Pi}_{z,lb}(\eps_0) \; , 
\label{eq:nC.222if} 
\end{eqnarray}
where
\begin{eqnarray}
&& \hspace{-8mm} 
\check{\Pi}_{z,lb}(\eps):=
1- \eps^2
\frac{1 + K^2  \Pi_{x,ub}^2(\eps)}{{\cal E}\kappa_3^2(\chi_0,\eps,d_2)}
 \; .
\label{eq:nC.222ig} 
\end{eqnarray}
To show that $\check{\Pi}_{z,lb}(\eps_0)>0$ we compute, by using (\ref{eq:3.28}),
\begin{eqnarray}
&& \hspace{-8mm} 
\eps_0^2\frac{1 + K^2  \Pi_{x,ub}^2(\eps_0)}{{\cal E}\kappa_3^2(\chi_0,\eps_0,d_2)}
= \biggl( \frac{1 + \eps_0\chi_{lb}(\eps_0)}
{\kappa_3(\chi_0,\eps_0,d_2)}\biggr)^2
 \; .
\label{eq:nC.222ij} 
\end{eqnarray}
If $\chi_0\leq 0$ then, by (\ref{eq:nC.80}),(\ref{eq:nC.90}),
\begin{eqnarray}
&& \hspace{-8mm} 
\kappa_3(\chi_0,\eps_0,d_2) = 1 + \eps_0(\chi_0-d_2)
> 1 + \eps_0\chi_{lb}(\eps_0) > 0 \; ,
\label{eq:nC.222ik} 
\end{eqnarray}
whence
\begin{eqnarray}
&& \hspace{-8mm} 
 0 < \frac{1 + \eps_0\chi_{lb}(\eps_0)}
         {\kappa_3(\chi_0,\eps_0,d_2)}  < 1 \; ,
\label{eq:nC.222ika} 
\end{eqnarray}
so that, by (\ref{eq:nC.222ij}),
\begin{eqnarray}
&& \hspace{-8mm} 
\eps_0^2\frac{1 + K^2  \Pi_{x,ub}^2(\eps_0)}
{{\cal E}\kappa_3^2(\chi_0,\eps_0,d_2)} < 1 \; .
\label{eq:nC.222ija} 
\end{eqnarray}
If $\chi_0 > 0$ then, by (\ref{eq:3.28}),(\ref{eq:nC.8aa}),(\ref{eq:nC.80}),
\begin{eqnarray}
&& \hspace{-8mm} 
  \kappa_3(\chi_0,\eps_0,d_2) = 1 > 1 - \eps_0\chi_M 
> 1 + \eps_0\chi_{lb}(\eps_0) > 0 \; ,
\label{eq:nC.222il} 
\end{eqnarray}
whence again (\ref{eq:nC.222ika}) holds which entails (\ref{eq:nC.222ija})
by (\ref{eq:nC.222ij}).
Having thus proven (\ref{eq:nC.222ija}) we conclude from
(\ref{eq:nC.222ig}) that
\begin{eqnarray}
&& \hspace{-8mm} 
\check{\Pi}_{z,lb}(\eps_0) > 0 \; ,
\label{eq:nC.222ih} 
\end{eqnarray}
whence, by (\ref{eq:xnC.17n}),(\ref{eq:nC.222if}),
\begin{eqnarray}
&& \hspace{-8mm} 
\Pi_z(\theta,\chi,\zeta,\eps,\nu) 
> \Pi_{z,lb}(\eps_0) \; , 
\label{eq:nC.222ifn} 
\end{eqnarray}
where
\begin{eqnarray}
&& \hspace{-8mm} 
 \Pi_{z,lb}(\eps):=\sqrt{ \check{\Pi}_{z,lb}(\eps) } 
=\sqrt{ 1- \eps^2
\frac{1 + K^2  \Pi_{x,ub}^2(\eps)}{{\cal E}\kappa_3^2(\chi_0,\eps,d_2)}}
\; .
\label{eq:nC.222ign} 
\end{eqnarray}
Of course since $\Pi_z,\Pi_{z,lb}>0$ we conclude from (\ref{eq:nC.222ifn})
that
\begin{eqnarray}
&& \hspace{-8mm} 
\frac{1}{\Pi_z(\theta,\chi,\zeta,\eps,\nu)} < \frac{1}{\Pi_{z,lb}(\eps_0)} \; .
\label{eq:nC.222ii} 
\end{eqnarray}
Inserting (\ref{eq:xnC.17n}),(\ref{eq:nC.92}),(\ref{eq:nC.222ii})
into (\ref{eq:xnC.10nn}) yields to
\bea
&&\hspace{-8mm}
|g_1|
\leq \frac{q}{\bar{q}\kappa_3^2(\chi_0,\eps_0,d_2)}
\biggl(  \frac{3q}{{\cal E}\Pi_{z,lb}(\eps_0)\kappa_3^2(\chi_0,\eps_0,d_2)}
+  3\chi^2 + 2\eps_0|\chi|^3  \biggr)
\nonumber\\
&&\quad +\frac{|\kappa_1|}{\bar{q}\Pi_{z,lb}(\eps_0)
\kappa_3^2(\chi_0,\eps_0,d_2)}
\biggl( 2  + \frac{ 3\eps_0^2q}{{\cal E}\kappa_3^2(\chi_0,\eps_0,d_2)}
\biggr) \; .
\label{eq:xnC.10nnn}
\eea
Furthermore, by (\ref{eq:2.59}),
(\ref{eq:nB.61}), (\ref{eq:nC.8b}),(\ref{eq:xnC.17n}),
\bea
&&\hspace{-8mm}
|\chi|=|\chi-\chi_0+\chi_0|\leq |\chi-\chi_0|+| \chi_0| < d_2 + |\chi_0| \; ,
\nonumber\\
&&\hspace{-8mm}
|\kappa_1(\theta,\zeta,\eps,\nu)|\leq
\frac{K^2 \bar{q}}{\nu}\biggl ( 2 + 2|\Delta P_{x0}|
+\frac{\eps^2 \bar{q}}{\nu} \biggr) 
\leq
\frac{K^2 \bar{q}}{\nu}\biggl ( 2 + 2|\Delta P_{x0}|
+\frac{\eps_0^2 \bar{q}}{\nu} \biggr) \; ,
\nonumber\\
&&\hspace{-8mm}
q(\zeta) \leq 1 + K^2(1 + |\Delta P_{x0}|)^2 =:q_{ub} \; .
\nonumber\\
\label{eq:nC.66}
\eea
Inserting (\ref{eq:nC.66}) into (\ref{eq:xnC.10nnn}) yields to
\bea
&&\hspace{-15mm}
|g_1( \theta,\chi,\zeta;\eps,\nu) |
\leq \frac{q_{ub}}{\bar{q}\kappa_3^2(\chi_0,\eps_0,d_2)}
\nonumber\\
&& \hspace{-5mm}
\times\biggl(  \frac{3q_{ub}}{{\cal E}\Pi_{z,lb}(\eps_0)
\kappa_3^2(\chi_0,\eps_0,d_2)}
+  3(d_2 + |\chi_0|)^2 + 2\eps_0(d_2 + |\chi_0|)^3  \biggr)
\nonumber\\
&& +\frac{K^2}{\nu\Pi_{z,lb}(\eps_0)
\kappa_3^2(\chi_0,\eps_0,d_2)}
\biggl ( 2 + 2|\Delta P_{x0}|
+\frac{\eps_0^2 \bar{q}}{\nu} \biggr)
\biggl( 2  + \frac{ 3\eps_0^2q_{ub}}{{\cal E}\kappa_3^2(\chi_0,\eps_0,d_2)}
\biggr) 
\nonumber\\
&&\quad 
=:C_1(\chi_0,\eps_0,\nu,d_2) \; .
\label{eq:nC.43n}
\eea
We now consider $g_2$ and we obtain from 
(\ref{eq:xnB.10n2}),(\ref{eq:xnC.17n})
\bea
&&\hspace{-8mm}
|g_2| 
\leq \frac{K^2}{ \Pi_z(\Pi_z + 1)(1+\eps\chi)^3} \biggl( 
(1 + |\Delta P_{x0}|)[
\frac{\eps_0}{{\cal E}}
( q + \eps_0^2|\kappa_1| )
\nonumber\\
&&\quad
+|\chi|\Pi_z(\Pi_z + 1)(1+\eps\chi)^2]
+\eps_0|\kappa_2|(\Pi_z + 1)(1+\eps\chi)^2 \biggr) 
\nonumber\\
&&
= K^2 \Biggl( 
\frac{\eps_0(1 + |\Delta P_{x0}|)}{ {\cal E}\Pi_z(\Pi_z + 1)(1+\eps\chi)^3}
( q + \eps_0^2|\kappa_1| )
+ \frac{|\chi|(1 + |\Delta P_{x0}|)}{1+\eps\chi}
\nonumber\\
&&\quad
+\frac{\eps_0|\kappa_2|}{ \Pi_z(1+\eps\chi)} \Biggr) \; .
\label{eq:xnC.10n2}
\eea
Note that, by (\ref{eq:nB.612}),(\ref{eq:xnC.17n}),
\bea
&&\hspace{-8mm}
|\kappa_2(\theta,\zeta,\nu)|\leq
\frac{\bar{q}}{\nu} \; .
\label{eq:nC.612} 
\eea
Inserting 
(\ref{eq:xnC.17n}),(\ref{eq:nC.92}),(\ref{eq:nC.222ii}),
(\ref{eq:nC.66}),(\ref{eq:nC.612})  into (\ref{eq:xnC.10n2}) yields to
\bea
&&\hspace{-8mm}
|g_2( \theta,\chi,\zeta;\eps,\nu)| 
\nonumber\\
&&\hspace{-5mm}
\leq K^2 \Biggl( 
\frac{\eps_0(1 + |\Delta P_{x0}|)}{ {\cal E}\Pi_{z,lb}(\eps_0)  
\kappa_3^3(\chi_0,\eps_0,d_2)}
\biggl(  q_{ub} + \eps_0^2
\frac{K^2 \bar{q}}{\nu}( 2 + 2|\Delta P_{x0}|
+\frac{\eps_0^2 \bar{q}}{\nu})\biggr)
\nonumber\\
&&
+ \frac{   (d_2 + |\chi_0|)(1 + |\Delta P_{x0}|)}{\kappa_3(\chi_0,\eps_0,d_2)}
+\frac{\eps_0\bar{q}}{\nu\Pi_{z,lb}(\eps_0)\kappa_3(\chi_0,\eps_0,d_2)} 
\Biggr)
\nonumber\\
&&
=:C_2(\chi_0,\eps_0,\nu,d_2)\; ,
\label{eq:nC.95n}
\eea
where $\kappa_3,\Pi_{z,lb},q_{ub}$ are given by
(\ref{eq:nC.80}),(\ref{eq:nC.222ign}),(\ref{eq:nC.66}).
With (\ref{eq:nC.43n}),(\ref{eq:nC.95n}) we have shown that 
$g_1(\cdot,\nu)$ and $g_2(\cdot,\nu)$ are bounded for $\nu\geq 1/2$ 
for the points
\begin{eqnarray}
&& \hspace{-8mm}
(\theta,\chi,\zeta,\eps)\in \R\times(\chi_0-d_2,\chi_0+d_2)\times\R \times
(0,\eps_0] \; .
\end{eqnarray}
\section{Limit of $g_1^R,g_2^R$}
\label{D}
\setcounter{equation}{0}
Let $\eps\in(0,\eps_0]$ with $\eps_0\in(0,1]$ and
$k\in\N,a\in[-1/2,1/2]$ and let $(\theta,\chi,\zeta)\in W(\eps_0)\times\R$. 
In this appendix we will prove the properties 
(\ref{eq:xnD.10n}),(\ref{eq:nD.20}),(\ref{eq:nD.32}),(\ref{eq:nD.46}),
(\ref{eq:nD.36a}),(\ref{eq:nD.60}) of $g_1^R$ and $g_2^R$.
The properties (\ref{eq:nD.20}),(\ref{eq:nD.60})
are used in the proof of Proposition \ref{P2}. Furthermore
the properties (\ref{eq:xnD.10n}),(\ref{eq:nD.32}),(\ref{eq:nD.46}),
(\ref{eq:nD.36a}) will be used in Appendix \ref{E}.
Since all assumptions of this
appendix are also satisfied in Appendix \ref{B}, we can apply the results
of Appendix \ref{B}.

We first consider $g_1$ and we obtain from (\ref{eq:n2.101aaca}),
(\ref{eq:xnB.10n}) that
\bea
&&\hspace{-8mm}
g^R_1(\theta,\chi,\zeta,\eps,k,a) 
=g_1(\theta,\chi,\zeta;\eps,k+\eps a) 
\nonumber\\
&&=  -\frac{q}{\bar{q} \Pi_z(\Pi_z + 1)^2 (1 + \eps\chi)^4}
\biggl(  \frac{q}{{\cal E}} (\Pi_z + 2) 
\nonumber\\
&& 
+  \Pi_z(\Pi_z + 1)^2
(3\chi^2 + 2\eps\chi^3)(1 + \eps\chi)^2  \biggr)
\nonumber\\
&&\quad -\frac{\kappa_1}{\bar{q} \Pi_z(\Pi_z + 1)^2 (1 + \eps\chi)^4}
\biggl( 2(1 + \eps\chi)^2  + \frac{\eps^2q}{{\cal E} }(\Pi_z + 2)\biggr) \; ,
\label{eq:xnD.10n}
\eea
where $\Pi_z=\Pi_z(\theta,\chi,\zeta,\eps,k+\eps a)$ and 
$\kappa_1=\kappa_1(\theta,\zeta,\eps,k+\eps a)$
whence, by (\ref{eq:xnB.10n}),(\ref{eq:nB.63}),
\bea
&&\hspace{-8mm}
\lim_{\eps\rightarrow 0+}\;[g_1^R(\theta,\chi,\zeta;\eps,k,a)] 
= \lim_{\eps\rightarrow 0+}\;[g_1( \theta,\chi,\zeta;\eps,k)] 
=  -\frac{q(\zeta)}{4\bar{q}} (   \frac{3}{{\cal E}} q(\zeta)  +  12\chi^2)
\nonumber\\
&&\quad - \frac{K^2}{2k}
\biggl( \sin(k[\theta - Q(\zeta)]) -\sin(k\theta_0) \biggr)
(\cos\zeta + \Delta P_{x0}) \; .
\label{eq:nD.20}
\eea
We now consider $g_2^R$ and we conclude from (\ref{eq:nnn2.101aada})
that
\bea
&&\hspace{-8mm}
g^R_2(\theta,\chi,\zeta,\eps,k,a) 
=g^R_{2,1}(\theta,\chi,\zeta;\eps,k,a) 
+ g^R_{2,2}(\theta,\chi,\zeta;\eps,k,a) \; ,
\label{eq:nD.32}
\eea
where
\bea
&&\hspace{-8mm}
g^R_{2,1}(\theta,\chi,\zeta;\eps,k,a) 
:=g_2(\theta,\chi,\zeta;\eps,k+\eps a) \; ,
\label{eq:nD.35}
\\
&&\hspace{-8mm}
g^R_{2,2}(\theta,\chi,\zeta;\eps,k,a):= 
-\frac{K^2}{\eps}
(\cos\zeta + \Delta P_{x0}) 
\biggl( \cos( \kappa_4 + \kappa_5) -  \cos( \kappa_4)\biggr) 
\nonumber\\
&& =-\frac{K^2}{\eps}
(\cos\zeta + \Delta P_{x0}) 
\biggl( \cos(\kappa_4) [\cos(\kappa_5)-1]
-\sin(\kappa_4)\sin(\kappa_5)\biggr) 
\nonumber\\
&& = -\frac{K^2}{\eps}
(\cos\zeta + \Delta P_{x0}) 
\biggl( -2\cos(\kappa_4) \sin^2(\kappa_5/2)
\nonumber\\
&& \quad
-2\cos(\kappa_5/2)
\sin(\kappa_5/2)\sin(\kappa_4)\biggr)
\nonumber\\
&& = \frac{2K^2}{\eps}
(\cos\zeta + \Delta P_{x0}) \sin(\kappa_5/2)
\biggl( \cos(\kappa_4) \sin(\kappa_5/2)
+\cos(\kappa_5/2)\sin(\kappa_4)\biggr) \; ,
\nonumber\\
\label{eq:nD.46}
\eea
with
\begin{eqnarray}
&&  \hspace{-12mm}  \kappa_4(\theta,\zeta,\eps,k,a):=
k(\theta-\zeta-\Upsilon_0\sin\zeta - \Upsilon_1\sin 2\zeta) - \eps a\zeta \; ,
\nonumber\\
&& \hspace{-12mm}  \kappa_5(\theta,\zeta,\eps,a):=
\eps a (\theta-\Upsilon_0\sin\zeta - \Upsilon_1\sin 2\zeta) \; .
\nonumber\\
\label{eq:nD.36b}
\end{eqnarray}
We obtain from (\ref{eq:xnB.10n2}),(\ref{eq:nD.35})
\bea
&&\hspace{-8mm}
g^R_{2,1}(\theta,\chi,\zeta;\eps,k,a) 
=g_2(\theta,\chi,\zeta;\eps,k+\eps a) 
\nonumber\\
&&= \frac{K^2}{ \Pi_z(\Pi_z + 1)(1+\eps\chi)^3}
\cos( \nu[\theta - Q(\zeta)])\biggl( 
(\cos\zeta + \Delta P_{x0})[
-\frac{\eps}{{\cal E}}
( q + \eps^2\kappa_1 )
\nonumber\\
&&\quad
+\chi\Pi_z(\Pi_z + 1)(1+\eps\chi)^2]
- \eps\kappa_2(\Pi_z + 1)(1+\eps\chi)^2 \biggr) \; ,
\label{eq:nD.36a}
\eea
where $\Pi_z=\Pi_z(\theta,\chi,\zeta,\eps,k+\eps a)$ and
$\kappa_2=\kappa_2(\theta,\zeta,k+\eps a)$ whence,
by (\ref{eq:xnB.10n2}),(\ref{eq:nB.632}),
\bea
&&\hspace{-8mm}
\lim_{\eps\rightarrow 0+}\;[g_{2,1}^R(\theta,\chi,\zeta;\eps,k,a)] 
= \lim_{\eps\rightarrow 0+}\;[g_2( \theta,\chi,\zeta;\eps,k)] 
\nonumber\\
&&\quad 
= \chi K^2\cos( k[\theta - Q(\zeta)])
(\cos\zeta + \Delta P_{x0}) \; .
\label{eq:nD.45}
\eea
Clearly, by (\ref{eq:nD.36b}),
\bea
&&\hspace{-8mm}
\lim_{\eps\rightarrow 0+}\;[\frac{\sin(\kappa_5(\theta,\zeta,\eps,a)/2)}{\eps}] 
= \frac{a}{2} (\theta-\Upsilon_0\sin\zeta - \Upsilon_1\sin 2\zeta) \; ,
\nonumber\\
&&\hspace{-8mm}
\lim_{\eps\rightarrow 0+}\;[\kappa_5(\theta,\zeta,\eps,a)] = 0 \; ,
\nonumber\\
&&\hspace{-8mm}
\lim_{\eps\rightarrow 0+}\;[\kappa_4(\theta,\zeta,\eps,k,a)]
= k(\theta-\zeta-\Upsilon_0\sin\zeta - \Upsilon_1\sin 2\zeta) \; ,
\nonumber\\
\label{eq:nD.48} 
\eea
whence, by (\ref{eq:nD.46}),
\bea
&&\hspace{-8mm}
\lim_{\eps\rightarrow 0+}\;[g^R_{2,2}( \theta,\chi,\zeta;\eps,k,a)] 
= K^2 a (\theta-\Upsilon_0\sin\zeta - \Upsilon_1\sin 2\zeta)
\nonumber\\
&&\times\sin(k[\theta-\zeta-\Upsilon_0\sin\zeta - \Upsilon_1\sin 2\zeta])
(\cos\zeta + \Delta P_{x0}) \; ,
\label{eq:nD.50}
\eea
so that, by (\ref{eq:nD.32}),(\ref{eq:nD.45}),
\bea
&&\hspace{-8mm}
\lim_{\eps\rightarrow 0+}\;[g^R_2(\theta,\chi,\zeta,\eps,k,a)] 
= \chi K^2\cos( k[\theta - Q(\zeta)])
(\cos\zeta + \Delta P_{x0}) 
\nonumber\\
&& 
+ K^2  a (\theta-\Upsilon_0\sin\zeta - \Upsilon_1\sin 2\zeta)
\nonumber\\
&&\quad \times\sin(k[\theta-\zeta-\Upsilon_0\sin\zeta - \Upsilon_1\sin 2\zeta])
(\cos\zeta + \Delta P_{x0}) \; .
\label{eq:nD.60}
\eea
\section{Bounds on $g_1^R,g_2^R$}
\label{E}
\setcounter{equation}{0}
Let $\eps\in(0,\eps_0]$ with $\eps_0\in(0,1]$ and let
$k\in\N,a\in[-1/2,1/2]$. Let also 
$(\theta_0,\chi_0)\in W(\eps_0)$. Moreover let
$\chi_{lb}(\eps_0)$ satisfy the restriction (\ref{eq:nC.8aa})
where $\chi_M$ is the positive constant from Theorem \ref{T2} (see item 2
of the setup list for Theorem \ref{T2}).
Furthermore we assume that
\bea
&&\hspace{-8mm}
(\theta,\chi,\zeta)\in
(\theta_0-d_1,\theta_0+d_1)\times (\chi_0-d_2,\chi_0+d_2)
\times \R \; ,
\label{eq:nE.7} 
\eea
%
where $\chi_0,d_1,d_2$ satisfy 
\bea
&&\hspace{-8mm}
0 < d_1 \; , \quad 0 < d_2 < \chi_0 - \chi_{lb}(\eps_0) \; .
\label{eq:nE.8a} 
\eea
In this appendix we will prove the properties
(\ref{eq:nE.10}),(\ref{eq:nE.80}) of $g_1^R$ and $g_2^R$.
We thus show in this appendix that the
properties (\ref{eq:nE.10}),(\ref{eq:nE.80})
hold in the situation of Theorem \ref{T2}
(see item 8 of the setup of Theorem \ref{T2}). 
Since all assumptions of this
appendix are also satisfied in Appendix \ref{C} and Appendix \ref{D}, 
we can apply the results of those appendices.

We first consider $g_1^R$ and we obtain from (\ref{eq:n2.101aaca}) that
\bea
&&\hspace{-8mm}
|g^R_1(\theta,\chi,\zeta,\eps,k,a)| 
=|g_1(\theta,\chi,\zeta;\eps,k+\eps a)| \; , 
\label{eq:xE.10n}
\eea
whence, by (\ref{eq:nC.43n}),
\bea
&&\hspace{-8mm}
|g^R_1(\theta,\chi,\zeta,\eps,k,a)| \leq
C_1(\chi_0,\eps_0,k+\eps a,d_2) \; ,
\label{eq:nE.10a}
\eea
where $C_1$ is given by (\ref{eq:nC.43n}).
Note that, by (\ref{eq:nC.43n}), $C_1(\chi_0,\eps_0,\nu,d_2)$
is decreasing w.r.t. $\nu$ whence 
\bea
&&\hspace{-8mm}
C_1(\chi_0,\eps_0,k+\eps a,d_2)\leq C_1(\chi_0,\eps_0,1/2,d_2)
=: C_1^R(\chi_0,\eps_0,d_2) \; ,
\label{eq:nE.10b}
\eea
so that, by (\ref{eq:nE.10a}),
\bea
&&\hspace{-8mm}
|g^R_1(\theta,\chi,\zeta,\eps,k,a)| \leq
C_1^R(\chi_0,\eps_0,d_2) \; ,
\label{eq:nE.10}
\eea
where $C_1^R$ is given by (\ref{eq:nE.10b}).

We now consider $g_2^R$ and we obtain from (\ref{eq:nD.32}) that
\bea
&&\hspace{-8mm}
|g^R_2(\theta,\chi,\zeta,\eps,k,a)| 
\leq |g^R_{2,1}(\theta,\chi,\zeta;\eps,k,a)| 
+ |g^R_{2,2}(\theta,\chi,\zeta;\eps,k,a)| \; .
\label{eq:nE.32}
\eea
Note that, by (\ref{eq:nC.95n}),(\ref{eq:nD.35}),
\bea
&&\hspace{-12mm}
|g^R_{2,1}(\theta,\chi,\zeta;\eps,k,a)| 
= |g_2(\theta,\chi,\zeta;\eps,k+\eps a)| 
\leq C_2(\chi_0,\eps_0,k+\eps a,d_2)
\; ,
\label{eq:nE.36a}
\eea
where $C_2$ is given by (\ref{eq:nC.95n}).
Note that, by (\ref{eq:nC.95n}), $C_2(\chi_0,\eps_0,\nu,d_2)$
is decreasing w.r.t. $\nu$ whence 
\bea
&&\hspace{-8mm}
C_2(\chi_0,\eps_0,k+\eps a,d_2)\leq C_2(\chi_0,\eps_0,1/2,d_2)
=: C_{2,1}^R(\chi_0,\eps_0,d_2) \; ,
\label{eq:nE.36c}
\eea
so that, by (\ref{eq:nE.36a}),
\bea
&&\hspace{-8mm}
|g^R_{2,1}(\theta,\chi,\zeta,\eps,k,a)| \leq
C_{2,1}^R(\chi_0,\eps_0,d_2) \; ,
\label{eq:nE.36}
\eea
where $C_{2,1}^R$ is given by (\ref{eq:nE.36c}).
We also have, by (\ref{eq:nD.46}),
\bea
&&\hspace{-8mm}
|g^R_{2,2}(\theta,\chi,\zeta;\eps,k,a)| 
\nonumber\\
&& = \Big{|}\frac{2K^2}{\eps}
(\cos\zeta + \Delta P_{x0}) \sin(\kappa_5/2)
\biggl( \cos(\kappa_4) \sin(\kappa_5/2)
+\cos(\kappa_5/2)\sin(\kappa_4)\biggr)\Big{|} 
\nonumber\\
&& \leq \frac{4K^2}{\eps}|\sin(\kappa_5/2)|
(1 + |\Delta P_{x0}|)
\; .
\label{eq:nE.46}
\eea
Of course, by (\ref{eq:nD.36b}),(\ref{eq:nE.7}),
\begin{eqnarray}
&&  \hspace{-12mm}  
\frac{|\sin(\kappa_5/2(\theta,\zeta,\eps,a))|}{\eps}
= \frac{1}{\eps}
|\sin( \frac{\eps a}{2}[\theta-\Upsilon_0\sin\zeta - \Upsilon_1\sin 2\zeta])|
\nonumber\\
&&\leq
\frac{|a|}{2}
\Big{|}\theta-\Upsilon_0\sin\zeta - \Upsilon_1\sin 2\zeta\Big{|}
\leq
\frac{|a|}{2} (|\theta| +|\Upsilon_0| +|\Upsilon_1|) 
\nonumber\\
&&\leq
\frac{|a|}{2} (|\theta_0| + d_1 +|\Upsilon_0| +|\Upsilon_1|) 
\; ,
\label{eq:nE.36b}
\end{eqnarray}
whence, by (\ref{eq:nE.46}),
\bea
&&\hspace{-8mm}
|g^R_{2,2}(\theta,\chi,\zeta;\eps,k,a)| \leq  2K^2|a| 
(1 + |\Delta P_{x0}|)(|\theta_0| + d_1 +|\Upsilon_0| +|\Upsilon_1|) 
\nonumber\\
&&=: C_{2,2}^R(\theta_0,a,d_1)
\; .
\label{eq:nE.47}
\eea
We conclude from (\ref{eq:nE.32}),(\ref{eq:nE.36}),(\ref{eq:nE.47}) that
\bea
&&\hspace{-8mm}
|g^R_2(\theta,\chi,\zeta,\eps,k,a)| 
\leq C_{2,1}^R(\chi_0,\eps_0,d_2) +
C_{2,2}^R(\theta_0,a,d_1) 
\nonumber\\
&&=:  C_2^R(\theta_0,\chi_0,\eps_0,a,d_1,d_2) \; ,
\label{eq:nE.80}
\eea
where $C_{2,1}^R$ is given by (\ref{eq:nE.36c}) and 
$C_{2,2}^R$ is given by (\ref{eq:nE.47}).

With (\ref{eq:nE.10}),(\ref{eq:nE.80})
we have shown that 
$g_1^R(\cdot,k,a)$ and $g_2^R(\cdot,k,a)$ are bounded for $k\in\N,|a|\leq 1/2$ 
for the points
\begin{eqnarray}
&& \hspace{-8mm}
(\theta,\chi,\zeta,\eps)\in 
(\theta_0-d_1,\theta_0+d_1)\times (\chi_0-d_2,\chi_0+d_2)
\times \R \times
(0,\eps_0] \; .
\end{eqnarray}
\section{Error bounds in a regular perturbation problem}
\label{F}
\setcounter{equation}{0}
Here we outline a derivation of error bounds in a regular perturbation
problem of relevance for \S \ref{3.4.4}. This could be made into a
theorem and proof at the level of \S\ref{3.5} and \S \ref{4} but we leave this to the interested
reader (see \cite[\S 2]{SSC} for a detailed discussion of regular perturbation theory relevant here, complete with a theorem and proof). We write the IVP in (\ref{eq:n3.16}) as
\bea
&&\hspace{-8mm} x_1'= x_2  \; , \quad x_1(0)=\xi \; ,
\label{eq:F.10}\\
&&\hspace{-8mm} x_2'= -\epsilon \sin x_1  \; , \quad x_2(0)=1 \; .
\label{eq:F.11} 
\eea
Then the zeroth-order approximation is
\bea
&&\hspace{-8mm} u_1'= u_2  \; , \quad u_1(0)=\xi \; ,
\label{eq:F.10a}\\
&&\hspace{-8mm} u_2'= 0  \; , \quad u_2(0)=1 \; ,
\label{eq:F.11a}
\eea
with solutions
\bea
&&\hspace{-8mm} u_1(s) = s + \xi \; , \quad
u_2(s) =1 \; .
\label{eq:F.11b}
\eea
Subtracting and integrating we obtain
\begin{eqnarray}
&&\hspace{-5mm} e_1:=|x_1(s)-u_1(s)| \leq \int_0^s\;|x_2(\tau)-u_2(\tau)|d\tau 
\; , \label{eq:F.15} \\
&&\hspace{-5mm} e_2:=|x_2(s)-u_2(s)| =\epsilon
|\int_0^s\;[\sin(x_1(\tau)) - \sin(u_1(\tau)) + \sin(\tau+\xi)]d\tau|
\nonumber\\
&&\leq \epsilon 
\int_0^s\;|x_1(\tau) - u_1(\tau)| d\tau 
+ \epsilon |-\cos(s+\xi) +\cos(\xi)| 
\nonumber\\
&&
\leq \epsilon 
\int_0^s\;|x_1(\tau) - u_1(\tau)| d\tau 
+ 2\epsilon \; . 
\label{eq:F.16} 
\end{eqnarray}
Introducing $R_1$ and $R_2$ as in \S\ref{4}, we have
\begin{eqnarray}
&& \hspace{-10mm} e_1(s)\leq \int_0^s e_2(\tau)d\tau =: R_1(s) \; , 
\label{eq:F.20} \\
&& \hspace{-10mm} e_2(s)\leq \epsilon \int_0^s e_1(\tau)d\tau
+2\epsilon =: R_2(s) \; .
\label{eq:F.21} 
\end{eqnarray}
Differentiating gives the differential inequalities
\begin{eqnarray}
&& \hspace{-10mm} R_1' = e_2 \leq R_2 \; , \quad R_1(0)=0 \; ,
\label{eq:F.22} \\
&& \hspace{-10mm} R_2' = \epsilon e_1 \leq \epsilon R_1 \; , 
\quad R_2(0)=2\epsilon \; .
\label{eq:F.23} 
\end{eqnarray}
Let 
\begin{eqnarray}
&& \hspace{-10mm} w_1' = w_2 \; , \quad w_1(0)=0 \; ,
\label{eq:F.25} \\
&& \hspace{-10mm} w_2' = \epsilon w_1  \; , 
\quad w_2(0)=2\epsilon  \; .
\label{eq:F.26} 
\end{eqnarray}
Then
\begin{eqnarray}
&& \hspace{-10mm} w_1 = \sqrt{\epsilon}
2\sinh(\sqrt{\epsilon}s) \; ,
\label{eq:F.30} \\
&& \hspace{-10mm} w_2 = \epsilon 2 \cosh(\sqrt{\epsilon}s) \; .
\label{eq:F.31} 
\end{eqnarray}
Now as shown in 
Appendix \ref{I}, $R_1(s)\leq w_1(s)$ and $R_2(s)\leq w_2(s)$ whence
if $0\leq s\leq T$,
\begin{eqnarray}
&& \hspace{-10mm} e_1(s) \leq \sqrt{\epsilon}2
\sinh(\sqrt{\epsilon}s) 
 \leq \sqrt{\epsilon}2\sinh(\sqrt{\epsilon}T) = O (\epsilon) \; ,
\label{eq:F.35} \\
&& \hspace{-10mm} e_2(s) \leq \epsilon 2 \cosh(\sqrt{\epsilon}s) 
\leq \epsilon 2 \cosh(\sqrt{\epsilon}T) =  O (\epsilon) 
\; .
\label{eq:F.36} 
\end{eqnarray}
In the context of \S\ref{3.4.4} with $\epsilon=1/Y_0^2,\xi=X_0,x_1=\hat{X},
x_2=\hat{Y}$ we obtain from (\ref{eq:F.35}),(\ref{eq:F.36}) that
$\hat{X}(s)= s + X_0 + O(1/Y_0^2),\hat{Y}(s)= 1 + O(1/Y_0^2)$ whence
$X(t)= Y_0t + X_0 + O(1/Y_0^2),Y(t)= Y_0(1 + O(1/Y_0))$.
%
%
\section{Derivatives for Low Gain Problem}
\label{G}
\setcounter{equation}{0}
We here derive (\ref{eq:G.30}) which is needed in \S\ref{3.7}.
By (\ref{eq:nn430015v}) we have
\begin{eqnarray}
&&\hspace{-10mm} v_1'(\cdot,1) = 2v_2(\cdot,1) \; , \quad v_1(0,1)=
\theta_0 \; , \quad
v_2'(\tau,1) = -K_0(k) \cos(kv_1(\tau,1)-a\tau)
\nonumber\\
&& = -\frac{K_0(k)}{2}\exp(u(\tau)) + cc \; , \quad v_2(0,1)=\chi_0 \; ,
\label{eq:G.10} 
\end{eqnarray}
where
\begin{eqnarray}
&&\hspace{-10mm} 
u(\tau):= i [ kv_1(\tau,1) -a\tau] \; .
\label{eq:G.15} 
\end{eqnarray}
It follows from (\ref{eq:G.10}) that
\begin{eqnarray}
&&\hspace{-10mm} 
v_2''(\tau,1) = K_0(k) (kv_1'(\tau,1)-a)\sin(kv_1(\tau,1)-a\tau)
\nonumber\\
&&\hspace{-5mm} 
= K_0(k)(2kv_2(\tau,1)-a)   \sin(kv_1(\tau,1)-a\tau) 
= -\frac{K_0(k)}{2}\exp(u(\tau))u'(\tau) + cc \; ,
\nonumber\\
&&\hspace{-10mm} 
v_2'''(\cdot,1) = -\frac{K_0(k)}{2}\exp(u)[ u'' + (u')^2]
+ cc \; ,
\nonumber\\
&&\hspace{-10mm} 
v_2''''(\cdot,1) = -\frac{K_0(k)}{2}\exp(u)[ u''' 
+ 3 u' u'' + (u')^3]
+ cc \; ,
\nonumber\\
\label{eq:G.20} 
\end{eqnarray}
and from (\ref{eq:G.10}),(\ref{eq:G.15}),
(\ref{eq:G.20}) that
\begin{eqnarray}
&&\hspace{-10mm} 
u'(\tau)= i [ kv_1'(\cdot,1) -a] = i [ 2kv_2(\cdot,1) -a] \; ,
\nonumber\\
&&\hspace{-10mm} 
u''(\tau) = i 2kv_2'(\tau,1) = -i2kK_0(k) \cos(kv_1(\tau,1)-a\tau) \; ,
\nonumber\\
&&\hspace{-10mm} 
u'''(\tau) = i 2kv_2''(\tau,1) = i2kK_0(k) 
(2kv_2(\tau,1)-a)\sin(kv_1(\tau,1)-a\tau)
\; .
\nonumber\\
\label{eq:G.24} 
\end{eqnarray}
We conclude from (\ref{eq:G.10}),(\ref{eq:G.15}),
(\ref{eq:G.24}) that
\begin{eqnarray}
&&\hspace{-10mm} 
u(0) = i kv_1(0,1) = ik\theta_0  \; ,
\nonumber\\
&&\hspace{-10mm} 
u'(0) = i [ 2kv_2(0,1) -a] = i [ 2k\chi_0 -a] 
 \; ,
\nonumber\\
&&\hspace{-10mm} 
u''(0) = -i2kK_0(k) \cos(kv_1(0,1)) = -i2kK_0(k) \cos(k\theta_0) \; ,
\nonumber\\
&&\hspace{-10mm} 
u'''(0) = i2kK_0(k)(2kv_2(0,1)-a)\sin(kv_1(0,1))
\nonumber\\
&&
= i2kK_0(k)(2k\chi_0-a)\sin(k\theta_0) \; ,
\nonumber\\
\label{eq:G.25} 
\end{eqnarray}
whence, by (\ref{eq:G.10}),(\ref{eq:G.20}),
\begin{eqnarray}
&&\hspace{-10mm} v_2'(0,1) = -K_0(k) \cos(kv_1(0,1))
 =-K_0(k) \cos(k\theta_0) \; ,
\nonumber\\
&&\hspace{-10mm} 
v_2''(0,1) = K_0(k)(2kv_2(0,1)-a)   \sin(kv_1(0,1)) 
 =K_0(k)(2k\chi_0-a)   \sin(k\theta_0) \; , 
\nonumber\\
&&\hspace{-10mm} 
v_2'''(0,1) = -\frac{K_0(k)}{2}\exp(u(0))[ u''(0) + (u'(0))^2]
+ cc 
\nonumber\\
&&\hspace{-5mm} 
= -\frac{K_0(k)}{2}\exp(ik\theta_0)\biggl(
-i2kK_0(k) \cos(k\theta_0) - [ 2k\chi_0 -a]^2\biggr)
+ cc
\nonumber\\
&&
= -K_0(k)\biggl( 2kK_0(k) \sin(k\theta_0)\cos(k\theta_0) 
- [ 2k\chi_0 -a]^2  \cos(k\theta_0)\biggr)
\nonumber\\
&& \qquad
= K_0(k)\biggl( -kK_0(k) \sin(2k\theta_0) 
+ [ 2k\chi_0 -a]^2  \cos(k\theta_0)\biggr)
\; ,
\nonumber\\
&&\hspace{-10mm} 
v_2''''(0,1) = -\frac{K_0(k)}{2}\exp(u(0))[ u'''(0) 
+ 3 u'(0) u''(0) + (u'(0))^3] + cc 
\nonumber\\
&&\hspace{-5mm} 
= -\frac{K_0(k)}{2}\exp(ik\theta_0)\biggl(
i2kK_0(k)(2k\chi_0-a)\sin(k\theta_0) 
\nonumber\\
&& +6kK_0(k) [ 2k\chi_0 -a] \cos(k\theta_0)
- i[ 2k\chi_0 -a]^3\biggr) + cc \;
\nonumber\\
&&\hspace{-5mm} 
= -\frac{K_0(k)}{2}\biggl(
-4kK_0(k)(2k\chi_0-a)\sin^2(k\theta_0) 
\nonumber\\
&& +12kK_0(k) [ 2k\chi_0 -a] \cos^2(k\theta_0)
+2[ 2k\chi_0 -a]^3\sin(k\theta_0)\biggr) 
\nonumber\\
&&\hspace{-5mm} 
= K_0(k)\biggl( 2kK_0(k)(2k\chi_0-a)\sin^2(k\theta_0) 
\nonumber\\
&& -6kK_0(k) [ 2k\chi_0 -a] \cos^2(k\theta_0)
-[ 2k\chi_0 -a]^3\sin(k\theta_0)\biggr) \; .
\nonumber\\
\label{eq:G.30} 
\end{eqnarray}
\section{Calculation of $E_r/cB_u$ in high gain regime}
\label{H}
\setcounter{equation}{0}
In this appendix we aim to estimate the magnitude of the electric field.
The basic field equation is
\begin{eqnarray}
&&\hspace{-10mm} 
(\frac{\partial^2}{\partial t^2} -c^2\frac{\partial^2}{\partial z^2})
E_x(z,t) = -cZ_{vac} \frac{\partial j}{\partial t}(z,t) \; ,
\label{eq:H.10} 
\end{eqnarray}
where $Z_{vac}=1/c\epsilon_0$ is the free space impedance and 
\begin{eqnarray}
&&\hspace{-10mm} 
j(z,t):=-\frac{ecK}{\Sigma_\perp}\cos(k_uz)\sum_{n=1}^N\; \frac{1}{\gamma_n(t)}
\delta(z-z_n(t)) 
\nonumber\\
&&\approx
-\frac{ecKN}{\gamma_c\Sigma_\perp}\cos(k_uz)\frac{1}{N}
\sum_{n=1}^N\; \delta(z-z_n(t))
 \; ,
\label{eq:H.15} 
\end{eqnarray}
with $\Sigma_\perp$ being the transverse emittance, 
see \cite{KHL} and \cite{Bob}. We proceed in two ways. In the first we
solve (\ref{eq:H.10}) and (\ref{eq:H.15}) directly and in the second
we use Fourier tranforms.

The unique solution of the homogeneous IVP at $t=0$ is
\begin{eqnarray}
&&\hspace{-10mm} 
E_x(z,t)=-\frac{Z_{vac}}{2}\int_0^t\;ds\int_{z-ct+cs}^{z+ct-cs}
\;dy \frac{\partial j}{\partial s}(y,s) 
\nonumber\\
&&
=-\frac{Z_{vac}}{2}[U_-(z,t) + U_+(z,t)] \; ,
\label{eq:H.20} 
\end{eqnarray}
where
\begin{eqnarray}
&&\hspace{-10mm} 
U_-(z,t):=\int_{z-ct}^z \;dy [j(y,t+\frac{1}{c}(y-z))- j(y,0)] \; ,
\label{eq:H.25} \\
&&\hspace{-10mm} 
U_+(z,t):=\int^{z+ct}_z \;dy [j(y,t-\frac{1}{c}(y-z)) - j(y,0)] \; .
\label{eq:H.26} 
\end{eqnarray}
The first equality in (\ref{eq:H.20})
is often obtained using Duhamel's principle
and d'Alembert's formula and the second equality is obtained after
changing the order of integration. To obtain our estimate we
consider $z_n(t)=\beta_c c t +z_n(0)$ which is quite crude (but may suffice
for a rough estimate)
and where the nonnegative $\beta_c$ is determined by
$\beta_c^2=(\gamma_c^2-1)/\gamma_c^2$. 
We obtain \cite{EH} $U_+\ll U_-$ and
\begin{eqnarray}
&&\hspace{-10mm} 
U_-(z,t) \approx -\frac{2ecK\gamma_c N}{\Sigma_\perp}\frac{1}{N}
\sum_{n=1}^N\; I_n(z,t)
\cos(2k_u\gamma_c^2[z -ct-z_n(0)]) \; ,
\label{eq:H.30} 
\end{eqnarray}
where
\begin{eqnarray}
&& I_n(z,t):=\left\{ \begin{array}{ll}  
 1  & \;\; {\rm if\;} z_n(t)< z < z_n(0)+ct \\ 
 0  &\;\;{\rm if\;otherwise} \; . \end{array} 
                  \right.
 \label{eq:H.31}
\end{eqnarray}
%
So if all the particles contributed
at $z$, which they don't, then
$U_-(z,t) =O(\frac{2ecK\gamma_c N}{\Sigma_\perp})$
and $E_{r1}=\frac{Z_{vac}ecK\gamma_c N}{\Sigma_\perp}$ would be a typical
value of the field $E_x$ at $(z,t)$. 

We now give a second estimate, $E_{r2}$, of $E_r$.
Following \cite{Bob} which is based on \cite{KHL} we Fourier transform
(\ref{eq:H.10}) by defining
\begin{eqnarray}
&&\hspace{-10mm} 
\hat{E}_x(z,\omega):=\frac{1}{2\pi}\int_{-\infty}^\infty\;ds
E_x(z,\frac{z}{c}-\frac{s}{c k_r}) \exp(-i\omega s) \; .
\label{eq:H.35} 
\end{eqnarray}
The Fourier inversion theorem gives
\begin{eqnarray}
&&\hspace{-10mm} 
E_x(z,t) = \int_{-\infty}^\infty\;d\omega \hat{E}_x(z,\omega)
\exp(i\omega k_r[z-ct]) \; .
\label{eq:H.40} 
\end{eqnarray}
We define $\hat{j}(z,\omega)$ in the same way as $\hat{E}_x(z,\omega)$
whence, in the slowly varying approximation,
(\ref{eq:H.10}) reduces to
\begin{eqnarray}
&&\hspace{-10mm} 
\frac{\partial \hat{E}_x}{\partial z}(z,\omega)
= -\frac{Z_{vac}}{2}  \hat{j}(z,\omega) \; ,
\label{eq:H.45} 
\end{eqnarray}
and from (\ref{eq:H.15}) we obtain
\begin{eqnarray}
&&\hspace{-10mm} 
\hat{j}(z,\omega) = -\frac{ec K N k_r}{2\pi\beta_c\gamma_c\Sigma_\perp}
\check{j}(z,\omega) \; ,
\label{eq:H.50} 
\end{eqnarray}
where
\begin{eqnarray}
&&\hspace{-10mm} 
\check{j}(z,\omega):= \cos(k_uz)\exp(-i\omega k_r z)\frac{1}{N}
\sum_{n=1}^N\; \exp(i\omega c k_r T_n(z)) \; .
\label{eq:H.51} 
\end{eqnarray}
Here the function $T_n$ is the inverse of the function $z_n$.
To obtain our estimate we note that $|\check{j}|$ is bounded by $1$
and replace it by $1$ which is quite crude but may suffice
for a rough estimate.
Inserting this into (\ref{eq:H.45}) and integrating we obtain
\begin{eqnarray}
&&\hspace{-10mm} 
\hat{E}_x(z,\omega) = O\biggl( \frac{Z_{vac}}{2}
\frac{ec K N k_r}{2\pi\beta_c\gamma_c\Sigma_\perp}\frac{1}{k_u}k_u z\biggr) \; ,
\label{eq:H.55} 
\end{eqnarray}
and, for $k_u z=O(1)$, 
\begin{eqnarray}
&&\hspace{-10mm} 
\hat{E}_x = O(E_{r2}) \; , \quad 
E_{r2}:= \frac{Z_{vac}}{4\pi}
\frac{ec K N}{\Sigma_\perp}\frac{k_r}{k_u\beta_c\gamma_c^2}\gamma_c \; .
\label{eq:H.60} 
\end{eqnarray}
We now have, recalling that $K=3.7$ in LCLS,
\begin{eqnarray}
&&\hspace{-10mm} 
\frac{E_{r1}}{E_{r2}} = 4\pi\frac{k_u\gamma_c^2}{k_r}
= 4\pi/K_r = 2\pi(1+\frac{K^2}{2})
\approx 2\pi(1+(3.7)^2/2)\approx 49 \; ,
\label{eq:H.65} 
\end{eqnarray}
and we calculate $E_{r2}/cB_u$. From (\ref{eq:H.60}) 
\begin{eqnarray}
\frac{E_{r2}}{cB_u}=\frac{Z_{vac}c}{4\pi}\frac{eK}{cB_u}\frac{k_r}{k_u\gamma_c^2}
\gamma_c\frac{N}{\Sigma_\perp} \; .
\end{eqnarray}
Now $K/cB_u=e/mc^2k_u$ and $k_r/k_u\gamma_c^2=2(1+K^2/2)^{-1}$ therefore
\begin{eqnarray}
\frac{E_{r2}}{cB_u}=\frac{Z_{vac}c}{4\pi}\frac{e^2}{mc^2}
\frac{1}{k_u}\frac{2}{(1+K^2/2)} \gamma_c\frac{N}{\Sigma_\perp} 
=r_e \frac{1}{k_u}\frac{2}{(1+K^2/2)} \gamma_c\frac{N}{\Sigma_\perp} 
\; ,
\end{eqnarray}
where $r_e$ denotes the classical electron radius. Furthermore
\begin{eqnarray}
&& r_e \approx 2.82\cdot 10^{-15}m \; , \quad 
\frac{1}{k_u}=\frac{3cm}{2\pi} \; , \quad
\frac{2}{(1+K^2/2)}\approx 0.255 \; , \quad  \gamma_c=10^4 \; ,
\nonumber
\end{eqnarray}
and so 
\begin{eqnarray}
\frac{E_{r2}}{cB_u} \approx 
0.034\cdot 10^{-12} m^2 \frac{N}{\Sigma_\perp} \approx 34 \; ,
\quad \frac{E_{r1}}{cB_u} = \frac{E_{r2}}{cB_u} \frac{E_{r1}}{E_{r2}} 
\approx 34\cdot 49 \approx 1700 \; ,
\nonumber
\end{eqnarray}
for $N=10^9$ and $\Sigma_\perp=1mm^2$.
\section{IVP for a system of differential inequalities}
\label{I}
\setcounter{equation}{0}
Here we present and verify a solution of the IVP for a system of differential inequalities which is used
in \S\ref{4.1},\S\ref{4.2} and Appendix \ref{F}.
Consider the IVP for
\begin{eqnarray}
&& \hspace{-5mm} R_1'(\zeta) \leq a_1 R_2(\zeta) \; , 
\label{eq:I.1} 
\\
&& \hspace{-5mm} R_2'(\zeta) \leq a_2  R_1(\zeta) \; , 
\label{eq:I.2} 
\end{eqnarray}
%
where $a_1,a_2>0$ and $R_1,R_2$ are of class $C^1$. We
want to show, for $\zeta \geq 0$, that
\begin{eqnarray}
&& \hspace{-5mm} R_1(\zeta)\leq r_1(\zeta) \; , \quad
 R_2(\zeta)\leq r_2(\zeta) \; , 
\label{eq:I.3} 
\end{eqnarray}
where 
\begin{eqnarray}
&& \hspace{-5mm} r_1' =  a_1 r_2  \; , \quad
r_1(0) =  R_1(0) \; , 
\label{eq:I.4} 
\\
&& \hspace{-5mm} r_2' =  a_2 r_1  \; , \quad
r_2(0) =  R_2(0) \; .
\label{eq:I.5} 
\end{eqnarray}
We do this in two ways.
First we define
$\hat{r}_j(\zeta):=R_j(\zeta)- r_j(\zeta)$ for $j=1,2,\zeta\geq 0$
whence, by (\ref{eq:I.1}),(\ref{eq:I.2}),(\ref{eq:I.4}),\\(\ref{eq:I.5}),
\begin{eqnarray}
&& \hspace{-12mm} \hat{r}_1'(\zeta) \leq 
a_1  \hat{r}_2(\zeta) \; , \quad
 \hat{r}_2'(\zeta) \leq 
a_2  \hat{r}_1(\zeta) \; , \quad
\hat{r}_1(0)=\hat{r}_2(0)= 0 \; .
\label{eq:I.10} 
\end{eqnarray}
Clearly we have to show that, for $j=1,2,\zeta\geq 0$,
\begin{eqnarray}
&& \hspace{-12mm} \hat{r}_j(\zeta) \leq 0 \; . 
\label{eq:I.15} 
\end{eqnarray}
It follows from (\ref{eq:I.10}) that
\begin{eqnarray}
&& \hspace{-12mm} \hat{r}_1'(\zeta) \leq a_1\int_0^\zeta\;ds  \hat{r}_2'(s) 
\leq a_1 a_2\int_0^\zeta\;ds  \hat{r}_1(s) \; ,
\nonumber\\
&& \hspace{-12mm} \hat{r}_2'(\zeta) \leq a_2\int_0^\zeta\;ds  \hat{r}_1'(s) 
\leq a_1 a_2\int_0^\zeta\;ds  \hat{r}_2(s) \; ,
\nonumber
\end{eqnarray}
i.e.,
\begin{eqnarray}
&& \hspace{-12mm} \hat{r}_j'(\zeta) \leq a_0^2 
\int_0^\zeta\;ds  \hat{r}_j(s) \; ,
\label{eq:I.25} 
\end{eqnarray}
where $a_0:=\sqrt{a_1 a_2}$.
It follows from (\ref{eq:I.25}) and by partial integration that
\begin{eqnarray}
&& \hspace{-12mm} \exp(-a_0\zeta)\hat{r}_j(\zeta) + a_0\int_0^\zeta\;ds  
\exp(-a_0 s)\hat{r}_j(s) 
= \int_0^\zeta\;ds  
\exp(-a_0 s)\hat{r}_j'(s) 
\nonumber\\
&&\leq  a_0^2 \int_0^\zeta\;ds  
\exp(-a_0 s) \int_0^s\;d\tilde{s} \hat{r}_j(\tilde{s}) 
\nonumber\\
&&\quad = -a_0\exp(-a_0 \zeta) \int_0^\zeta\;ds  \hat{r}_j(s)
+a_0 \int_0^\zeta\;ds  \exp(-a_0 s) \hat{r}_j(s) \; ,
\label{eq:I.30} 
\end{eqnarray}
which entails
\begin{eqnarray}
&& \hspace{-12mm} \hat{r}_j(\zeta) 
\leq  -a_0\int_0^\zeta\;ds  \hat{r}_j(s) \; .
\label{eq:I.35} 
\end{eqnarray}
Abbreviating
\begin{eqnarray}
&& \hspace{-12mm} \check{r}_j(\zeta) :=\int_0^{\zeta/a_0}\;ds  \hat{r}_j(s) \; ,
\label{eq:I.40} 
\end{eqnarray}
we obtain from (\ref{eq:I.35})
\begin{eqnarray}
&& \hspace{-12mm} \check{r}_j'(\zeta) =\frac{1}{a_0}\hat{r}_j(\zeta/a_0) 
\leq -\int_0^{\zeta/a_0}\;ds  \hat{r}_j(s) 
=- \check{r}_j(\zeta) 
\; ,
\label{eq:I.45} 
\end{eqnarray}
whence
\begin{eqnarray}
&& \hspace{-12mm} 
0\geq \exp(\zeta)[\check{r}_j(\zeta) + \check{r}_j'(\zeta) ]
 =  [\exp(\zeta)\check{r}_j(\zeta)]' \; , 
\label{eq:I.50} 
\end{eqnarray}
so that $\exp(\zeta)\check{r}_j(\zeta)$ is decreasing w.r.t. $\zeta$
which entails, by (\ref{eq:I.40}), that
\begin{eqnarray}
&& \hspace{-12mm} 
0 = \exp(0)\check{r}_j(0) \geq \exp(\zeta)\check{r}_j(\zeta) \; ,
\label{eq:I.55} 
\end{eqnarray}
i.e.,
\begin{eqnarray}
&& \hspace{-12mm} 
\check{r}_j(\zeta) \leq 0  \; .
\label{eq:I.60} 
\end{eqnarray}
We conclude from (\ref{eq:I.25}),(\ref{eq:I.40}),(\ref{eq:I.60}) that
\begin{eqnarray}
&& \hspace{-12mm} \hat{r}_j'(\zeta) \leq 
a_0^2\int_0^\zeta\;ds  \hat{r}_j(s) =  \check{r}_j(a_0\zeta) \leq 0
\; ,
\label{eq:I.65} 
\end{eqnarray}
whence $\hat{r}_j(\zeta)$ is decreasing w.r.t. $\zeta$
so that (\ref{eq:I.15}) follows from
(\ref{eq:I.10}).

The result in (\ref{eq:I.3}) is a special case of a much more 
general theorem on pages 112-113 of \cite{Walter}. That proof simplifies 
in the special case here and we present it for the
interested reader. The proof proceeds 
by cleverly introducing a comparison function $\hb$. Here
\begin{eqnarray}
&&\hspace{-10mm} 
\hb(\zeta) = \left( \begin{array}{c} h_1(\zeta) \\ h_2(\zeta)
\end{array}\right)
: = a_4\exp(2a_3\zeta)
\left( \begin{array}{c} 1 \\ 1
\end{array}\right) \; ,
\label{eq:I.66} 
\end{eqnarray}
where
$a_3:=max(a_1,a_2),a_4>0$. Then
\begin{eqnarray}
&& \hspace{-5mm} h_1' =  2a_3  h_1 =2a_3 h_2 > a_1 h_2  \; , 
\label{eq:I.70} 
\\
&& \hspace{-5mm} h_2' =  2a_3  h_2 =2a_3 h_1 > a_2 h_1  \; , 
\label{eq:I.71} 
\end{eqnarray}
and we have, by (\ref{eq:I.10}),
\begin{eqnarray}
&& \hspace{-5mm} \hat{r}_1' - a_1 \hat{r}_2 \leq 0 < h_1' - a_1 h_2  \; , 
\label{eq:I.72} \\ 
&& \hspace{-5mm} \hat{r}_2' - a_2 \hat{r}_1 \leq 0 < h_2' - a_2 h_1  \; .
\label{eq:I.73} 
\end{eqnarray}
We now show that, for $j=1,2,\zeta\geq 0$,
\begin{eqnarray}
&& \hspace{-5mm} \hat{r}_j(\zeta)\leq h_j(\zeta) \; .
\label{eq:I.74} 
\end{eqnarray}
Suppose that (\ref{eq:I.74}) is wrong then there exists a smallest 
$\zeta_0>0$ such that an index $j_0$ exists with
\begin{eqnarray}
&& \hspace{-5mm} \hat{r}_{j_0}(\zeta_0)= h_{j_0}(\zeta_0) \; ,
\label{eq:I.75} 
\end{eqnarray}
where we used that, by (\ref{eq:I.10}),(\ref{eq:I.66}) and 
for $j=1,2$,
\begin{eqnarray}
&& \hspace{-5mm} \hat{r}_j(0)= 0 < a_4 = h_j(0) \; .
\label{eq:I.67} 
\end{eqnarray}
Clearly, for $j=1,2,0\leq \zeta <\zeta_0$, 
\begin{eqnarray}
&& \hspace{-5mm} \hat{r}_j(\zeta) < h_j(\zeta) \; .
\label{eq:I.68} 
\end{eqnarray}
Without loss of generality we take $j_0=1$ whence,
for $0\leq \zeta \leq\zeta_0$, 
\begin{eqnarray}
&& \hspace{-5mm} \hat{r}_2(\zeta) \leq h_2(\zeta) \; .
\label{eq:I.78} 
\end{eqnarray}
It follows from (\ref{eq:I.68}) that at the first intersection 
\begin{eqnarray}
&& \hspace{-5mm} \hat{r}_1'(\zeta_0)\geq  h_1'(\zeta_0) \; .
\label{eq:I.79} 
\end{eqnarray}
But by (\ref{eq:I.72}),(\ref{eq:I.78})
\begin{eqnarray}
&& \hspace{-5mm} \hat{r}_1'(\zeta_0)-h_1'(\zeta_0) 
< a_1( \hat{r}_2(\zeta_0)-h_2(\zeta_0)) \leq 0 \; ,
\label{eq:I.80} 
\end{eqnarray}
which is a contradiction.
\newpage
\addcontentsline{toc}{section}{Bibliography}


\begin{thebibliography}{00}
\bibitem{SPPPP} The well known phase plane portrait for the pendulum equation can be found in many places, e.g., see Fig. 6.4, p. 248 of \cite{BN}.
\bibitem{PP}  The potential plane is simply a plot of the ``potential''
$1-\cos X$ versus $X$ placed above the phase plane portrait. This is nicely illustrated in \cite{BN}, Figs. 6.5 and 6.6 on p. 249. See also Fig. 95, p. 142 of
\cite{Ar}.
\bibitem{Period} See e.g., \cite{BN} Section 6.2 for a derivation of the period as a function of amplitude as well a proof that the integral is well defined as an improper integral (note that the integrand is singular). The formula for the pendulum equation is given explicitly on p.244.
\end{thebibliography}
\end{document}